\documentclass[numbers=noenddot]{jfm}
\usepackage[latin9]{inputenc}

\usepackage{array, ragged2e}
\usepackage{float}
\usepackage{mathtools}
\usepackage{graphicx}
\usepackage{esint}
\usepackage{verbatim}
\usepackage{caption}
\usepackage{subcaption}
\renewcommand\[{\begin{equation}}
\renewcommand\]{\end{equation}}
\newcommand{\un}[1]{\,\mathrm{#1}}
\numberwithin{equation}{section}
\usepackage{amsmath} 
\newcommand{\beq}{\begin{equation}}
\newcommand{\eeq}{\end{equation}}
\newcommand{\lb}{\left(}
\newcommand{\rb}{\right)}
\usepackage{tikz}
\usepackage{pgfplots}
\usepackage{overpic}
\usepackage{hyperref}
\usepackage{todonotes}
\author{Graham P. Benham \aff{1,2}\corresp{\email{gpb35@cam.ac.uk}}, 
Mike J. Bickle \aff{1},  
Jerome A. Neufeld \aff{1,2,3}}
\affiliation{\aff{1} Department of Earth Sciences, University of Cambridge, Cambridge CB3 0EZ, UK
\aff{2} BP Institute, University of Cambridge, Cambridge CB3 0EZ, UK
\aff{3} Department of Applied Mathematics and Theoretical Physics, University of Cambridge, Cambridge CB3 0WA, UK
}

\begin{document}

\title{Two-phase gravity currents in layered porous media}

\maketitle

\abstract{
We examine the effects of horizontally layered heterogeneities on the spreading of two-phase gravity currents in a porous medium, with application to numerous environmental flows, most notably geological carbon sequestration. 
Geological heterogeneities, which are omnipresent within all reservoirs, affect the large-scale propagation of such flows through the action of small-scale capillary forces, yet the relationship between these small and large scales is poorly understood.  
Here, we derive a simple upscaled model for a gravity current under an impermeable cap rock, which we use to investigate the effect of a wide range of heterogeneities on the macroscopic flow.  
By parameterising in terms of different types of archetypal layering, we assess the sensitivity of the gravity current to the distribution and magnitude of these heterogeneities.
Furthermore, since field measurements of heterogeneities are often sparse or incomplete, we quantify how uncertainty in such measurements is manifest as uncertainty in the macroscale flow predictions.
Finally we apply our model to the Sleipner case study in the North Sea, positing how heterogeneities may have played a role in the observed migration of CO$_2$. 
}

\section{Introduction}

Injection of CO$_2$ into underground reservoirs to reduce greenhouse gas emissions, also known as geological carbon sequestration, is one of the major proposed technological solutions to meet future global temperature targets \citep{bickle2009geological,bui2018carbon}. During this process, the injected CO$_2$ rises as a buoyant plume within the porous aquifer, encountering impermeable cap rocks which force it to spread laterally as a gravity current. 
As the flow spreads out, capillary forces play a key role in determining the saturation distribution and consequent flow properties via the relative permeability and capillary pressure \citep{nordbotten2011geological}. 
Heterogeneities in rock properties at the $10-100\un{cm}$ scales substantially amplify and complicate the effect of variations in pore-scale capillary forces, and are manifest in the large-scale saturation distributions within the CO$_2$ current.
Hence, in order to ensure safe and efficient sequestration, it is imperative to be able to model how small-scale heterogeneities, which are ubiquitous in all subsurface reservoirs, affect spreading rates at the macroscale \citep{benham2020upscaling,jackson2020small}.

The only previous attempts to model such flows in heterogeneous media involve using upscaled relative permeability data, often acquired using numerical calculations or core flooding experiments \citep{jackson2018characterizing}, and applying these to reservoir simulators \citep{braun2005macro,cavanagh2014sleipner,li2015influence,trevisan2017impact}. However, these studies are often computationally demanding and focus on a specific type of heterogeneity (e.g. over some horizontal/vertical length scale, as investigated by \citet{jackson2020small}).
In particular, it is not currently understood how generic small-scale heterogeneities affect the propagation of such large-scale flows. 
Furthermore, since heterogeneities are usually measured through isolated and sparsely distributed bore hole samples, such measurements often come with a large degree of uncertainty, yet there are very few studies which discuss how such uncertainty at small-scales translates to large-scale predictions. 
Indeed, whilst the studies of \citet{hinton2020shear2,hinton2020shear1} investigated how variations in permeability cause shear dispersion for miscible flows, the corresponding effects due to capillary forces within immiscible flows have not yet been addressed. However, as discussed by \citet{jackson2020small}, these capillary forces associated with the heterogeneities play a critical role during CO$_2$ migration, and therefore require careful modelling.

In this study, we derive a simple upscaled model for the evolution of an axisymmetric two-phase gravity current beneath an impermeable cap rock, where the structure and distribution of layered heterogeneities is a model input. 
This simplified approach not only greatly reduces the computational demand of modelling such small-scale details, but also allows us to study a large range of heterogeneities by parameterising them as different types of archetypal sedimentary layering. 
In this way, we assess how the properties of the heterogeneities affect the macroscale flow, as well as how uncertainty in the measurements is manifest in such flow predictions.
Our simple model provides other key insights, such as the self-similar nature of the upscaled gravity current, scaling laws for the speed of propagation and an understanding of where and when the flow transitions between the so-called capillary and viscous limiting behaviours.

In modelling subsurface migration of CO$_2$, one of the key difficulties is resolving the complex heterogeneous rock properties of the porous medium. This is a two-fold challenge, since on the one hand a resolution which captures all the details of the rock geometry is very computationally intensive, and on the other hand it is almost impossible to obtain such fine-scaled resolution over field scales from field measurements.
Hence, there is a strong motivation for an \textit{upscaled} modelling approach which describes the bulk, or average effect of heterogeneities on the large-scale flow features. There are many possible levels of upscaling (from the pore scale upwards, as discussed by \citet{krevor2015capillary}), and this depends on the desired amount of detail as well as the available data, but here we focus on length scales between the size of the heterogeneities and the size of the aquifer. Therefore, in the context of this study small-scale heterogeneities refer to variations on the scale of $10^{-2}$ - $1\un{m}$ (as opposed to pore-scale heterogeneities which occur on the scale of $10^{-6}$ - $10^{-3}\un{m}$), and the large-scale flow refers to a gravity current which is typically around $1$ - $10\un{m}$ thick \citep{cowton2016inverse}.

Heterogeneity types range from variation within the pore structure of a rock to variation in the rock type itself. 
Owing to the complexity of geological processes, these heterogeneities arise from many different causes, including sedimentary layering, subsequent diagenetic changes in the mineral fabrics and tectonic fracturing and faulting.
Each type of heterogeneity affects the flow in a different way, via the action of small-scale capillary forces, thereby presenting a significant challenge for generic upscaling approaches. 
However, the low computational cost of our approach allows us to investigate a wide variety of heterogeneity types via simple parameterisations of archetypal cases. 
Among these, we study the effects of lithostatic compaction as well as sedimentary strata with permeability sampled from a probability distribution. The latter case is particularly useful since it captures how the uncertainty in field studies, due to a lack of measurements, is manifest in the uncertainty of modelling predictions.

When upscaling the effect of heterogeneities, a key parameter is the capillary number \citep{jackson2018characterizing}, which is defined as the ratio between horizontal flow-driving pressure gradients (associated with Darcy flow) and vertical capillary pressure gradients (associated with the capillary forces). Hence, in the limit of small capillary number, known as the \textit{capillary limit}, the capillary forces due to heterogeneities dominate the flow behaviour, while in the limit of large capillary number, known as the \textit{viscous limit}, they have a negligible effect.
Many previous studies focus on each of these cases separately, though recently semi-analytical approaches have been derived by \citet{benham2020upscaling} and \citet{boon2021} that capture the transition between the viscous and capillary regimes, demonstrating which regions of a confined aquifer are in each of these limits, and which regions are in between the limits. 
However, gravity was neglected in that study, restricting the applicability to very thin aquifers.

For CO$_2$ sequestration sites in large aquifers, gravity plays a dominant role in the rise and spreading of the buoyant plume of injected fluid \citep{nordbotten2011geological}.
The role of buoyancy is characterised by the ratio between the strength of gravitational forces and capillary forces, and may be characterised by a dimensionless  Bond number \citep{golding2013effects}, which we define later in Section \ref{sechet}. 
As discussed by \citet{benham2020upscaling}, the Bond number is greater than unity for aquifers larger than around $\sim1\un{m}$ thick, in which case gravity alters the upscaled flow properties significantly.
Hence, in this study we focus on the upscaled modelling of such gravity currents, so that more general injection scenarios in larger aquifers can be addressed.

There is a long history of studying gravity currents in porous media, from early work which explored the first fundamentals \citep{huppert1995gravity}, to later studies which investigated the effect of confinement \citep{pegler2014fluid}, permeability variations \citep{hinton2018buoyancy}, and capillary forces \citep{golding2013effects}. 
Recently, \citet{jackson2020small} showed that small-scale capillary heterogeneities can significantly modify the large-scale migration of a buoyant plume within an aquifer. 
However, their numerical study is both computationally intensive, and does not provide general scalings for different types of heterogeneity.

The aim of the present study is to quantify the macroscopic effect of a wide range of heterogeneities on the axisymmetric injection of CO$_2$ beneath an impermeable cap rock. 
The low computational cost of our simple approach allows us to explore different parameterisations of heterogeneities, providing insights into the dominant controls on the evolution of the gravity current. 
Similar to \citet{benham2020upscaling} (though focussing on gravitational effects), we investigate both the viscous limit, the capillary limit and the transition between these limits using a locally defined capillary number that determines where and when heterogeneities play an important role. 
We show that away from this transition zone the upscaled gravity current is self-similar, where the front moves like the square root of time (like the homogeneous case discussed by \citet{golding2013effects}) and the prefactor varies significantly depending on the type and strength of the heterogeneity, as well as the Bond number.
In addition, we provide a framework for managing real permeability data with uncertainty in the measurements, illustrating how this uncertainty is manifest in modelling predictions. 
Finally, we use our upscaled approach to investigate how heterogeneities may have affected the injection of CO$_2$ at the Sleipner site in the North Sea \citep{bickle2007modelling}.

Our paper is laid out as follows. In Section \ref{sec1} we derive a simplified model for the upscaled gravity current, discussing different types of heterogeneities.
Section \ref{sec2} presents numerical and analytical results in the viscous and capillary limits, as well as numerical results accounting for the dynamic transition between these limits.
In Section \ref{sleipsec} we apply our results to the case study of the Sleipner project, and finally we close with some concluding remarks in Section \ref{concsec}.

\section{Upscaled modelling of two-phase gravity currents}\label{sec1}

In this section, we outline the assumptions used to model an upscaled two-phase gravity current in a heterogeneous porous medium, making note of how the saturation of phases varies within the current. Then, we derive the upscaled governing equations and boundary conditions which describe the macroscopic dynamics. Subsequently, we discuss a variety of different types of heterogeneity and how these are manifest in the upscaled properties. Finally, despite the added complexity of the heterogeneities, we demonstrate the self-similar nature of the gravity current, thereby greatly reducing the complexity of the problem.

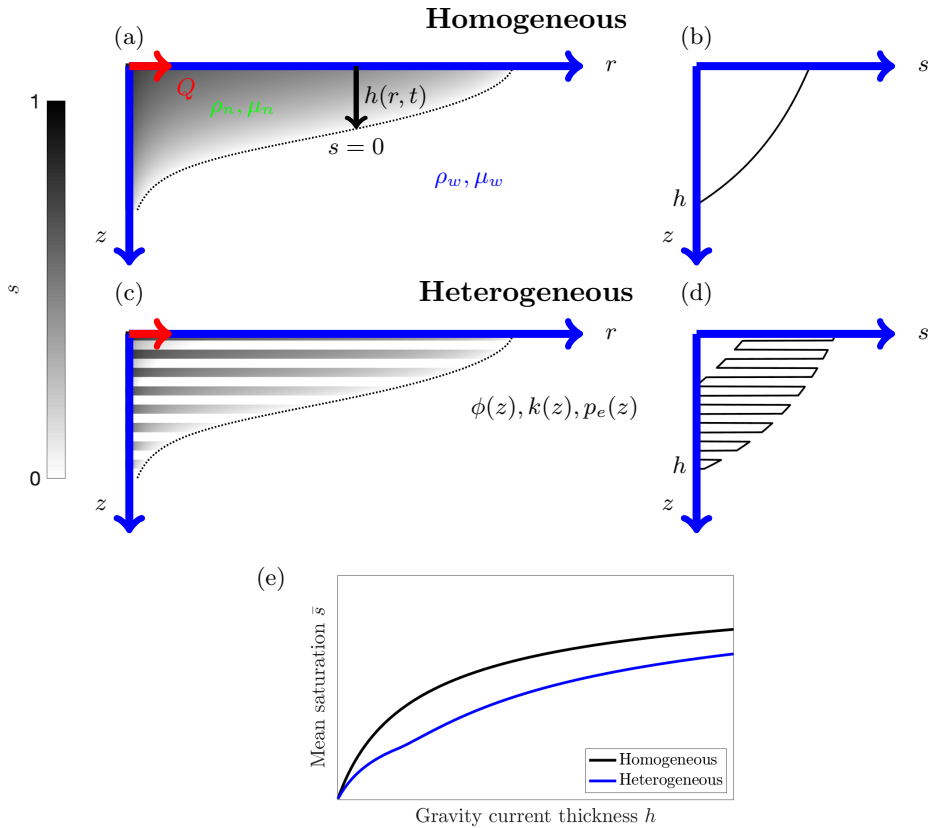
\begin{figure}
\centering
\begin{tikzpicture}[scale=0.75]
\node at (3.2,-5) {\includegraphics[height=0.175\textwidth]{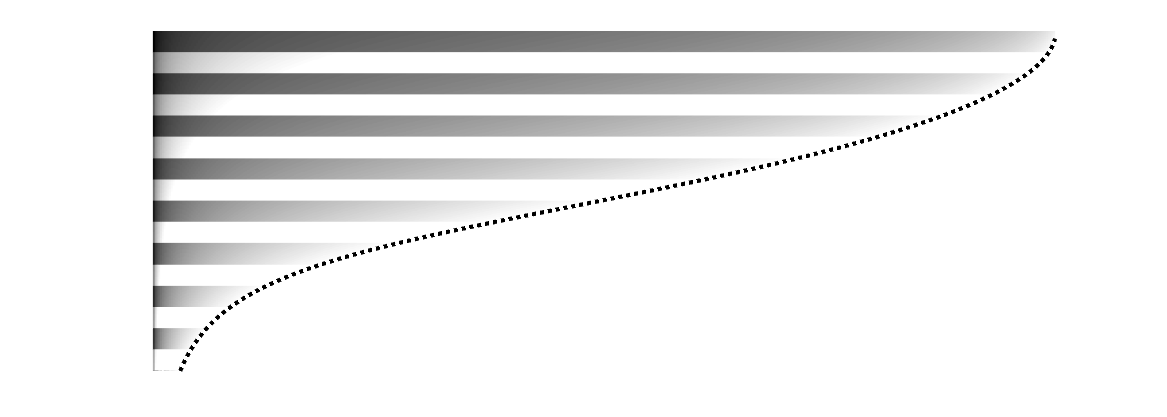}};
\node at (3.2,-0.3) {\includegraphics[height=0.175\textwidth]{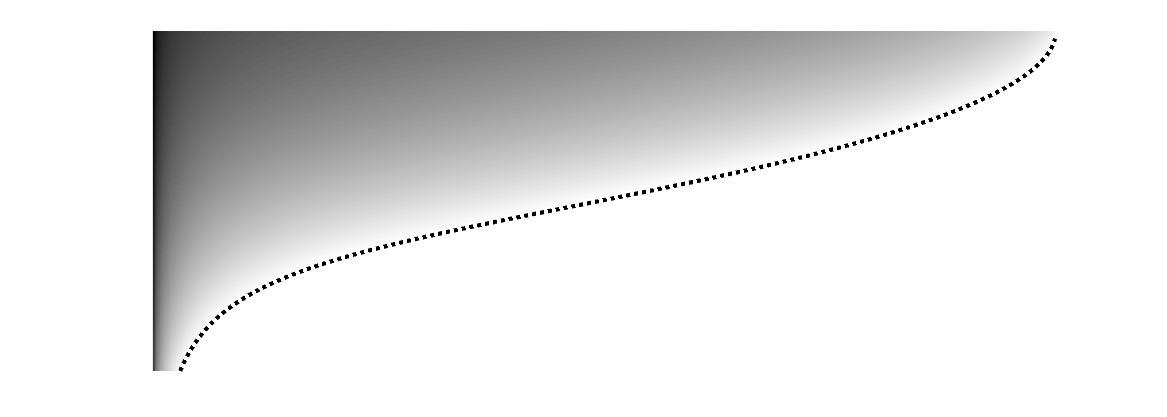}};
\node at (11.2,-5) {\includegraphics[height=0.175\textwidth]{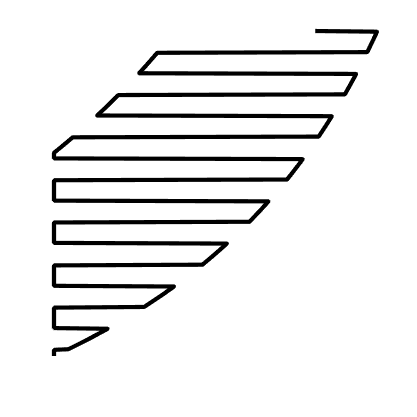}};
\node at (11.2,-0.3) {\includegraphics[height=0.175\textwidth]{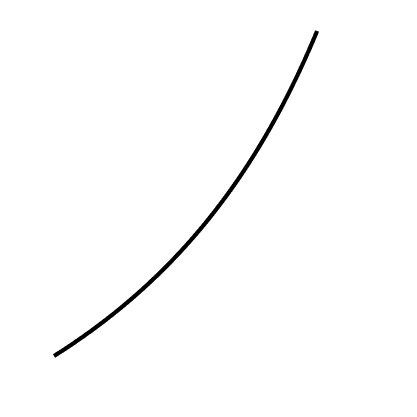}};
\node at (7,-3) {\bf \large  Heterogeneous};
\node at (7,1.8) {\bf \large  Homogeneous};
\draw[line width=3,->,blue] (0,1) -- (0,-2.5);
\draw[line width=3,->,blue] (0,1) -- (8,1);
\draw[line width=3,->,blue] (0,-3.7) -- (0,-7.2);
\draw[line width=3,->,blue] (0,-3.7) -- (8,-3.7);
\draw[line width=3,->,blue] (10,1) -- (10,-2.5);
\draw[line width=3,->,blue] (10,1) -- (13.5,1);
\draw[line width=3,->,blue] (10,-3.7) -- (10,-7.2);
\draw[line width=3,->,blue] (10,-3.7) -- (13.5,-3.7);
\node at (8.5,1) {$r$};
\node at (8.5,-3.7) {$r$};
\node at (-0.5,-2) {$z$};
\node at (-0.5,-6.7) {$z$};
\node at (9.5,-2) {$z$};
\node at (9.5,-6.7) {$z$};
\node at (9.7,-1.3) {$h$};
\node at (9.7,-6.) {$h$};
\node at (14,1) {$s$};
\node at (14,-3.7) {$s$};
\node[green] at (2,0.25) {$\rho_n,\mu_n$};
\node[blue] at (6,-1.) {$\rho_w,\mu_w$};
\node[black] at (7.5,-5) {$\phi(z),k(z),p_e(z)$};
\node[red] at (1,0.6) {$Q$};
\node[black] at (4,-0.4) {$s=0$};
\draw[line width=2,->,black] (4,1) -- (4,-0.1);
\node[black] at (4.7,0.45) {$h(r,t)$};
\draw[line width=3,->,red] (0,1) -- (0.75,1);
\draw[line width=3,->,red] (0,-3.7) -- (0.75,-3.7);
\node at (7,-10) {\includegraphics[width=0.5\textwidth]{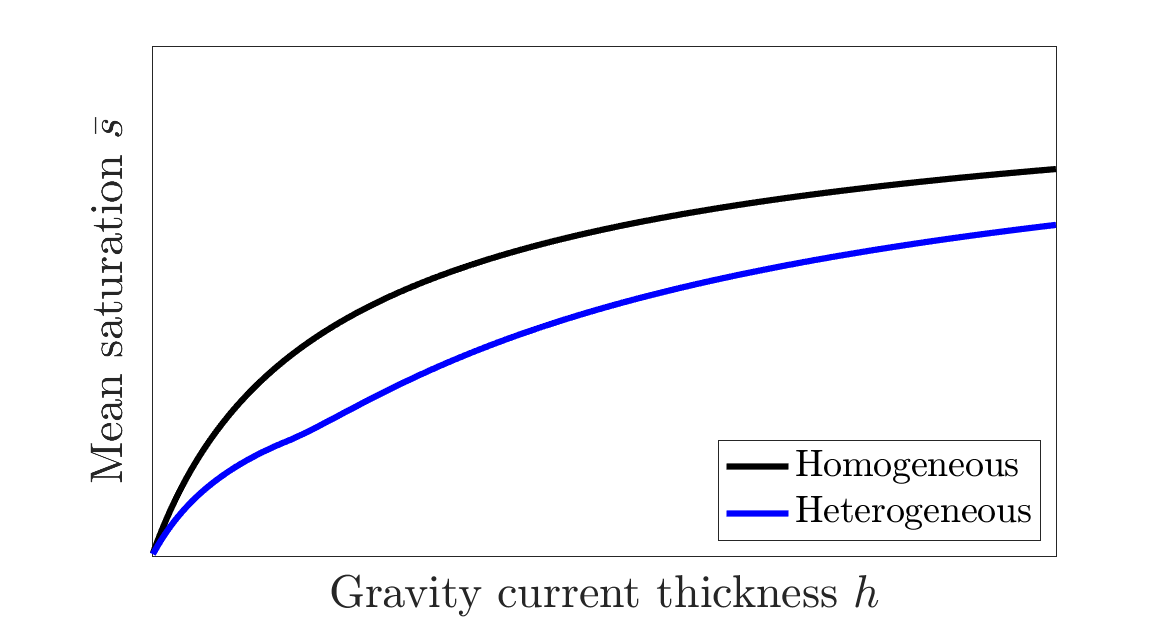}};
\node at (-1.5,-3) {\includegraphics[width=0.075\textwidth]{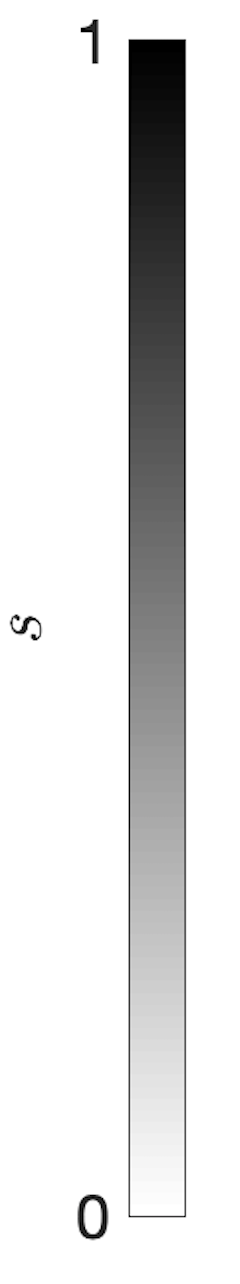}};
\node at (0,1.5) {(a)};
\node at (10,1.5) {(b)};
\node at (0,-3) {(c)};
\node at (10,-3) {(d)};
\node at (2.5,-8) {(e)};
\end{tikzpicture}
\caption{Schematic diagram of an axisymmetric gravity current (with constant injection $Q$) in both the homogeneous case (a) and the heterogeneous case (with sedimentary strata) (c), also illustrating the corresponding vertical non-wetting saturation profiles (b,d), given by \eqref{sateq},\eqref{sateq2} (note the heterogeneity wavelength is exaggerated for illustration purposes). (e) Relationship between mean non-wetting saturation $\bar{s}$ \eqref{meansatdef} and gravity current thickness $h$. \label{fig1}}
\end{figure}

\subsection{Fundamentals of two-phase flow in heterogeneous porous media}\label{secfunds}

The flow scenario we consider is illustrated in figure \ref{fig1} with a radial coordinate system $(r,z)$. A buoyant non-wetting phase (e.g. CO$_2$) is injected at a point source at the origin with flow rate $Q$ into a surrounding porous medium saturated with a denser wetting phase (e.g. water). The resulting current spreads out radially under gravity with thickness $z=h(r,t)$ beneath a horizontal impermeable cap rock located at $z=0$. Motivated by the dominant heterogeneity arising from sedimentary layering, we consider a porous medium which has vertically varying permeability $k(z)$ and porosity $\phi(z)$. 

We model this scenario using conservation of mass and Darcy's law for two-phase flow \citep{bear2013dynamics}. Hence, the governing equations are
\begin{align}
\phi(z) \frac{\partial S_i}{\partial t}+{\nabla}\cdot\boldsymbol{u}_i&=0,\quad &i=n,w\label{ge1},\\
\boldsymbol{u}_i&=-\frac{k(z)k_{ri}(S_i)}{\mu_i}{\nabla}\, \lb p_i -\rho_i g \boldsymbol{z} \rb,\quad &i=n,w\label{ge2},
\end{align}
where subscripts $i=n,w$ indicate non-wetting and wetting phases, and $S_i$, $p_i$, $\boldsymbol{u}_i$, $\rho_i$, $\mu_i$, $k_{ri}(S_i)$ are the saturations, pressures, Darcy velocities, densities, viscosities, and relative permeabilities of the two phases. We assume that the pore spaces are filled, such that $S_n+S_w=1$. Furthermore, due to capillary forces, the pressure difference between phases satisfies
\beq
p_n-p_w=p_c(S_n),\label{prepc}
\eeq  
where $p_c(S_n)$ is the capillary pressure. As is often done, we assume that both $k_{ri}$ and $p_c$ depend on the saturation only for simplicity (though in general they may have more complex dependencies). These are usually approximated with empirical parameterised formulae, such as those proposed by \citet{corey1954interrelation,brooks1964hydrau,chierici1984novel}. Here we use the \textit{Brooks-Corey} and \textit{Corey} relationships, which are given by
\begin{align}
p_c=p_e(z)(1-s)^{-1/\lambda},\label{pceq}\\
k_{rn}=k_{rn0}s^\alpha,\label{krneq}
\end{align}
where $p_e(z)$ is the pore entry pressure, $s=S_n/(1-S_{w0})$ is the reduced saturation of non-wetting phase, $\lambda$ represents the pore size distribution,  $k_{rn0}$ is the end-point relative permeability, and  $\alpha$ is a fitting parameter. The irreducible wetting saturation $S_{w0}$ is the amount of wetting phase that is permanently stored in the pores during drainage flows, and consequently the end-point relative permeability corresponds to $k_{rn0}=k_{rn}(1)$. In this new formulation, the reduced saturation $s$ conveniently varies between 0 and 1.

The pore entry pressure $p_e(z)$ is the minimum pressure difference required to allow the non-wetting phase to enter the largest pore spaces at a given position. Likewise, as the pressure difference between phases increases, the non-wetting phase is able to enter smaller and smaller pore spaces.  
Hence, clearly the pore entry pressure depends on the size and geometry of the pores (and hence varies vertically), and the same is true for the permeability and porosity. 
However, whilst this dependence has been measured for specific rock types, it is not fully understood in general. Hence, as is often done, we use power laws to relate these different quantities, such that 
\begin{align}
\phi\propto k^a,\\
p_e\propto k^{-b},
\end{align}
for some constants $a, b>0$, which we take to be positive since large pore size corresponds to large porosity, large permeability, and small pore entry pressure. As discussed by \cite{cloud1941effects,nelson1994permeability}, we do not expect these constants to be the same for different rock types. Therefore, we keep them in general form for this analysis. However, we note a commonly used scaling proposed by \citet{leverett1941capillary}, $p_e\sim(\phi/k)^{1/2}$ which implies $b=1/2(1-a)$.

Motivated by field observations of gravity currents (e.g. see \citet{cowton2016inverse}, where the aspect ratio of the gravity current at Sleipner was calculated to be less than $\sim1/1000$) and following \citet{golding2011two,golding2013effects}, we assume that the gravity current is long and thin, such that the vertical velocity is much smaller than the horizontal velocity $w_i\ll u_i$. In this case, the pressure within each phase is approximately hydrostatic
\beq
\frac{\partial p_i}{\partial z}=\rho_i g,\quad i=n,w,\label{hydrostat}
\eeq
and consequently, \eqref{hydrostat} is integrated to match the capillary pressure \eqref{prepc}, giving
\beq
p_c=-\Delta \rho  g (z-h) + p_0,\label{pceq2}
\eeq
where $p_0$ is the pressure at the edge of the gravity current $(z=h)$ and $\Delta \rho=\rho_w-\rho_n$. 
The saturation is calculated by combining \eqref{pceq} and \eqref{pceq2}, enforcing the physical lower bound on $s$, such that
\beq
s=\max \left\{ 1-\left[ \frac{p_0}{p_e(z)}+\frac{\Delta \rho g (h-z)}{p_e(z)}\right]^{-\lambda},\quad 0 \right\} .\label{sateq}
\eeq
To determine the value of $p_0$ we consider that the edge of the gravity current is defined as the boundary below which no saturation of non-wetting phase exists.  
Hence, from \eqref{pceq2},\eqref{sateq}, it is sufficient to ensure that $s=0$ for all $z>h$ if we choose $p_0=\min p_e(z)$.
In other words, by setting the capillary pressure at the edge of the gravity current as the smallest required pressure difference to invade any pores in the aquifer, we guarantee that anywhere below the edge of the gravity current $(p_c<p_0)$ no saturation will be found. Therefore, even though there may be disconnected regions of non-wetting phase within $0\leq z\leq h$ (e.g. see figure \ref{fig1}c,d), there will never be such regions for $z>h$.

The saturation distribution \eqref{sateq} represents a balance between capillary forces (due to heterogeneities) and gravitational forces. However, this is only valid for situations where capillary forces are large enough to drive the saturation into regions of larger pore space, or equivalently when the capillary number is small. Therefore, in general the saturation distribution must depend on the local capillary number N$_c$, which is given as the ratio between the horizontal flow-driving pressure gradient and the typical vertical gradient in pore entry pressure  \citep{benham2020upscaling}. For the former, we use the pressure gradient of non-wetting phase ${\partial p_n}/{\partial r}$, and for the latter we use ${\Delta p_e}/{h}$, where $\Delta p_e=\max p_e(z) - \min p_e(z)$ is the maximum difference in pore entry pressure across the aquifer (constant), and the gravity current thickness $h$ is used as the vertical length scale.
Hence, the capillary number is given by
\beq
\mathrm{N}_c=\left|\frac{h}{\Delta p_e}\frac{\partial p_n}{\partial r}\right|.\label{capdef}
\eeq
In the limit of very small capillary number N$_c\ll1$, also known as the \textit{capillary limit}, the saturation distribution \eqref{sateq} remains accurate. However, when the capillary number is very large, also known as the \textit{viscous limit}, capillary forces due to heterogeneities are effectively negligible (i.e. we can ignore pore entry pressure variations $p_e(z)=p_0$), and the saturation distribution becomes
\beq
s= 1-\left[ 1+\frac{\Delta \rho g (h-z)}{p_0}\right]^{-\lambda} ,\label{sateq2}
\eeq
which is identical to the homogeneous case addressed by \citet{golding2013effects}\footnote{We note that although the upscaled description of the viscous limit is mathematically identical to the homogeneous case, the model would still have to account for flow variations due to permeability gradients through an effective permeability. The saturation would, however, be identical to the homogeneous case.}. 

For intermediate capillary number (i.e. when the flow is neither in the viscous limit nor the capillary limit), the saturation distribution lies in between \eqref{sateq} and \eqref{sateq2}, and therefore the expression for the saturation must contain the capillary number itself $s=s(z,h,\mathrm{N}_c)$. Typically, the dependence of the saturation on the capillary number is logarithmic \citep{benham2020upscaling}, as with the upscaled flow properties, and we will return to address this later in Section \ref{transsec}.

In figure \ref{fig1}c we illustrate a radially symmetric gravity current in a heterogeneous layered medium composed of sedimentary strata. We contrast this to the classic homogeneous case in figure \ref{fig1}a (as studied by \citet{golding2013effects}), which is equivalent to the upscaled viscous limit (N$_c\gg1$) for a heterogeneous medium (see earlier footnote). For each case we plot typical vertical saturation profiles in figure \ref{fig1}b,d. In the homogeneous case the saturation distribution satisfies a balance between capillary and buoyancy forces, so that lighter regions (of high saturation) are pushed towards the cap rock. In the heterogeneous case, the same overall balance is sustained, but within that balance capillary forces push the saturation into layers where the pores are larger. Hence, significant oscillatory behaviour is observed within the vertical saturation profile, including patches where the saturation drops to zero. This corresponds with regions where the pore spaces are too small to allow any non-wetting phase (i.e. the zero value is chosen in \eqref{sateq}). 
Hence, one interesting consequence of heterogeneities is that they modify the mean saturation value in the gravity current. In figure \ref{fig1}e we plot the mean saturation, defined as 
\beq
\bar{s}(h)=\frac{1}{h}\int_0^h s\,\mathrm{d}z,\label{meansatdef}
\eeq
whilst varying the gravity current thickness $h$ for both the homogeneous and heterogeneous cases. In both cases $\bar{s}$ is an increasing function of $h$, but the heterogeneous case always has a lower mean value. This is due to the substantial fraction of the gravity current with zero saturation.

\subsection{Upscaled model: Governing equation and boundary conditions}

Having discussed the flow scenario and laid down the key assumptions, now we outline the upscaling procedure, deriving a single governing equation and accompanying boundary conditions for the macroscopic description of the gravity current.

To do so, \eqref{hydrostat} is integrated to obtain the pressure, and then the conservation of mass equation for the non-wetting phase 
\eqref{ge1} is integrated between $z=0$ and $z=h(r,t)$, such that
\beq
\varphi\frac{\partial}{\partial t} \int_0^h \hat{\phi}(z) s \,\mathrm{d}z  -\frac{u_b}{r}\frac{\partial}{\partial r}\lb \frac{r}{k_0k_{rn0}}\frac{\partial h}{\partial r} \int_0^h k(z) k_{rn}(s) \,\mathrm{d}z \rb=0, \label{goveheq}
\eeq
where $\varphi=(1-S_{w0})\phi_0$ is the reduced porosity scaling, $\phi_0$ and $k_0$ are typical scalings for the porosity and permeability (where $\hat{\phi}=\phi/\phi_0$), and $u_b={k_0k_{rn0} \Delta \rho g}/{\mu_n}$ is the buoyancy velocity. 

We note that the flow of wetting phase is ignored in this analysis under the long-thin approximation. Essentially, the flow of non-wetting phase within the gravity current decouples from the flow of wetting phase, which is not present at leading order. Nevertheless, multiphase effects are still manifest at leading order via the multiphase properties, such as the relative permeability and capillary pressure. However, as we discuss later in Section \ref{sechet}, this assumption breaks down if the permeability ratio between layers becomes very small. In particular, if there are regions of very low permeability, these will act as a vertical obstruction to the flow. In such situations, as discussed by \citet{pegler2014fluid}, the flow must be treated as confined, where the flow of the ambient fluid plays an important role on the dynamics and therefore can no longer be ignored. We give more details of this consideration in the next section.

We also note that \eqref{goveheq} is already an upscaled description of the flow, since the heterogeneities are only manifest within the integrals.
Hence, \eqref{goveheq} represents how the heterogeneities affect the evolution of the gravity current in a spatially averaged sense. 
Such an upscaling approach is desirable, since we wish to avoid having to resolve all the heterogeneities, both to reduce computation time, and also because realistically the low resolution of field measurements means that such details are uncertain anyway.

It is convenient to write \eqref{goveheq} in a more standard \textit{diffusion equation} form to render it amenable to conventional analysis. Therefore, by switching variables to the integrated saturation $\mathcal{S}(h,\mathrm{N}_c)$, which is defined as
\beq
\mathcal{S}=\varphi\int_0^h \hat{\phi}(z) s\,\mathrm{d}z\label{Seqn},
\eeq 
\eqref{goveheq} is rewritten as 
\beq
\frac{\partial \mathcal{S}}{\partial t} =\frac{1}{r}\frac{\partial}{\partial r}\left[ r \mathcal{F}(\mathcal{S},\mathrm{N}_c) \frac{\partial \mathcal{S}}{\partial r}  \right],\label{Geqn}
\eeq
where the flux is given by
\beq
\mathcal{F}=\frac{u_b}{k_0k_{rn0}}\left[\frac{\mathcal{K}(h,\mathrm{N}_c)}{\mathcal{S}_h(h,\mathrm{N}_c)}\right],\label{Feqn}
\eeq
and the two functions $\mathcal{K}(h,\mathrm{N}_c)$ and $\mathcal{S}_h(h,\mathrm{N}_c)$ are defined as
\begin{align}
\mathcal{K}&=\int_0^h k(z) k_{rn}(s)\,\mathrm{d}z,\label{Feqn1}\\
\mathcal{S}_h&=\varphi\int_0^h \hat{\phi}(z) {\partial s}/{\partial h}\,\mathrm{d}z=\frac{\partial\mathcal{S}}{\partial h}.\label{Feqn2}
\end{align}
Further details of this coordinate transformation are presented in Appendix \ref{appB}. We note that $\mathcal{S}$ has dimensions of length, and $\mathcal{F}$ has dimensions of length squared over time. Therefore, \eqref{Geqn} is just a standard diffusion equation for the total volume of non-wetting phase (per unit area), where the flux is a non-linear function that represents how capillary forces modify the flow. Hence, there is an interesting analogy between our scenario and a viscous gravity current, where the flux function represents how viscous forces modify the flow (e.g. plug flow, Poiseuille flow, etc...).
As we will find out later, $\mathcal{F}$ is sometimes well-approximated by a power law of $\mathcal{S}$, and the solutions to such equations are detailed by a large  historical body of literature (see \citet{huppert1982propagation} for example).

In general, \eqref{Geqn} must be solved in tandem with the equation for the capillary number \eqref{capdef}. Therefore, writing \eqref{capdef} in terms of the integrated saturation $\mathcal{S}$, we arrive at the transcendental equation for the capillary number
\beq
\mathrm{N}_c=\frac{\Delta\rho g}{\Delta p_e}\left|\frac{h(\mathcal{S},\mathrm{N}_c)}{\mathcal{S}_h(h(\mathcal{S},\mathrm{N}_c),\mathrm{N}_c)}\frac{\partial \mathcal{S}}{\partial r}\right|,\label{capdef2}
\eeq
where $h$ is written in terms of $\mathcal{S}$ under the assumption that \eqref{Seqn} has a uniquely-defined inverse (which we later find to be the case).

For the remainder of this study (up until Section \ref{transsec}) we restrict our attention to the two limiting cases of small and large capillary number (capillary and viscous limits), where the saturation is given by \eqref{sateq} or \eqref{sateq2} and the flux is just given by $\mathcal{F}=\mathcal{F}(\mathcal{S})$, thereby decoupling \eqref{Geqn} and \eqref{capdef2}. However, later in Section \ref{transsec} we address the case of intermediate capillary number, for which the equations must be solved in tandem.

\begin{figure}
\centering
\begin{tikzpicture}[scale=0.6]
\node at (0,-6) {\includegraphics[width=0.25\textwidth]{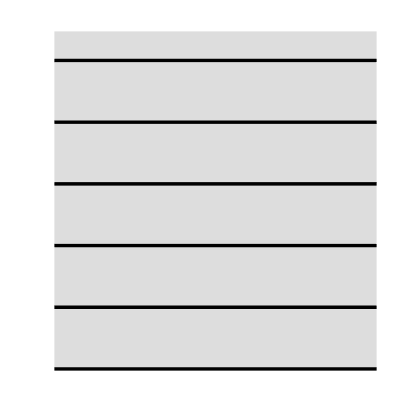}};
\node at (7,-6) {\includegraphics[width=0.25\textwidth]{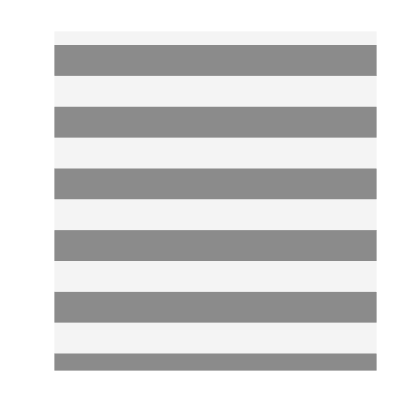}};
\node at (14,-6) {\includegraphics[width=0.25\textwidth]{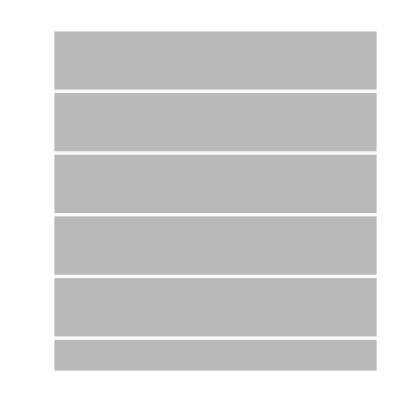}};
\draw[line width=2,->,black] (-2,-3.6) -- (2.75,-3.6);
\draw[line width=2,->,black] (-2,-3.6) -- (-2,-9);
\draw[line width=2,->,black] (5,-3.6) -- (9.75,-3.6);
\draw[line width=2,->,black] (5,-3.6) -- (5,-9);
\draw[line width=2,->,black] (12,-3.6) -- (16.75,-3.6);
\draw[line width=2,->,black] (12,-3.6) -- (12,-9);
\node at (0,-3) {\bf $\boldsymbol{{H}_\mathrm{low}/{H}_\mathrm{high}=0.1}$};
\node at (7,-3) {\bf$\boldsymbol{{H}_\mathrm{low}/{H}_\mathrm{high}=1}$};
\node at (14,-3) {\bf $\boldsymbol{{H}_\mathrm{low}/{H}_\mathrm{high}=10}$};
\node at (-3,-3) {(a)};
\node at (4,-3) {(b)};
\node at (11,-3) {(c)};
\node at (3,-4) {\large $r$};
\node at (-2.5,-8) {\large $z$};
\node at (10,-4) {\large $r$};
\node at (4.5,-8) {\large $z$};
\node at (17,-4) {\large $r$};
\node at (11.5,-8) {\large $z$};
\node[red] at (6.5,-6.1) {\bf High};
\node[blue] at (6.5,-6.5) {\bf Low};
\end{tikzpicture}
\begin{tikzpicture}[scale=0.9]
\node at (1.0,-0.4) {\includegraphics[height=0.21\textwidth]{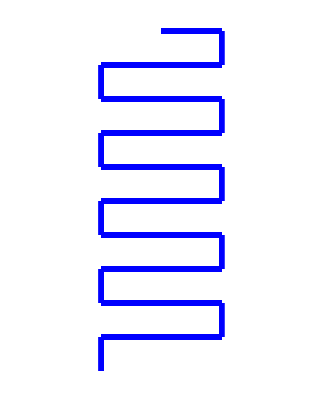}};
\draw[line width=2,->] (0,1) -- (0,-1.5);
\draw[line width=2,->] (0,1) -- (2.5,1);
\node at (-0.5,-1.25) {$z$};
\node at (2,1.4) {$k$};
\node at (0,1.5) {(d)};
\node at (5.0,-0.4) {\includegraphics[height=0.21\textwidth]{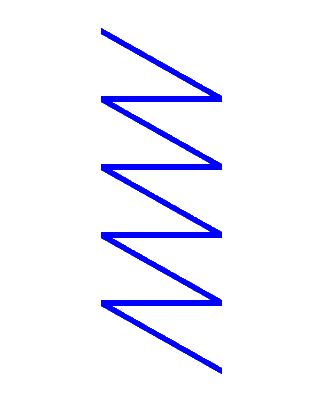}};
\draw[line width=2,->] (4,1) -- (4,-1.5);
\draw[line width=2,->] (4,1) -- (6.5,1);
\node at (3.5,-1.25) {$z$};
\node at (6,1.4) {$k$};
\node at (4,1.5) {(e)};
\node at (8.0,-0.4) {\includegraphics[height=0.21\textwidth]{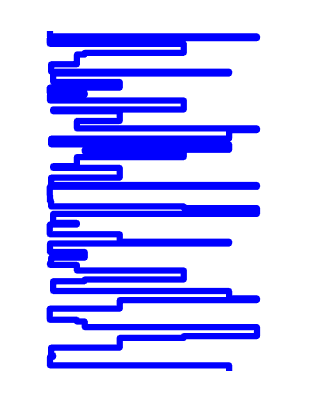}};
\draw[line width=2,->] (7,1) -- (7,-1.5);
\draw[line width=2,->] (7,1) -- (9.5,1);
\draw[blue,line width=2,->] (9.5,-1) -- (9.5,0.5);
\draw[blue,line width=2,->] (9.5,-1) -- (11.25,-1) ;
\draw[black,line width=2,dotted] (9.7,-1) -- (9.7,0) -- (10.8,0) -- (10.8,-1);
\node[blue] at (10.25,0.5) {\bf PDF};
\node[blue] at (10.25,-1.3) {\bf $\boldsymbol{\log k}$};
\node at (6.5,-1.25) {$z$};
\node at (9,1.4) {$k$};
\node at (7,1.5) {(f)};
\node at (13.0,-0.4) {\includegraphics[height=0.21\textwidth]{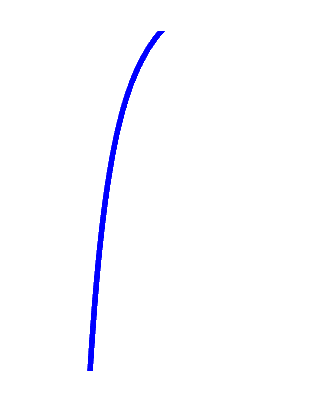}};
\draw[line width=2,->] (12,1) -- (12,-1.5);
\draw[line width=2,->] (12,1) -- (14.5,1);
\node at (11.5,-1.25) {$z$};
\node at (14,1.4) {$k$};
\node at (12,1.5) {(g)};
\node[red] at (1,-2) {\bf Sedimentary strata};
\node[red] at (1,-2.5) {\bf $(\boldsymbol{H_\mathrm{low}/H_\mathrm{high}=1})$};
\node[red] at (5,-2) {\bf Turbidites};
\node[red] at (13,-2) {\bf Compacted};
\node[red] at (9,-2) {\bf Spectrum};
\end{tikzpicture}
\caption{Illustrations of the different types of heterogeneity we consider, where the heterogeneity is characterised by variation of the permeability with depth. (a-f) represent the deposition of sediments through various geological mechanisms, whereas (g) represents compaction due to lithostatic pressure. In (a,b,c) we illustrate the case of sedimentary strata with greyscale permeability maps for three different values of the width ratio between low/high permeability regions $(H_\mathrm{low}/H_\mathrm{high})$. In the spectrum case (f) we display the probability density function (PDF) of the permeability which is randomly sampled from a uniform distribution on a logarithmic scale. \label{hettypes}}
\end{figure}
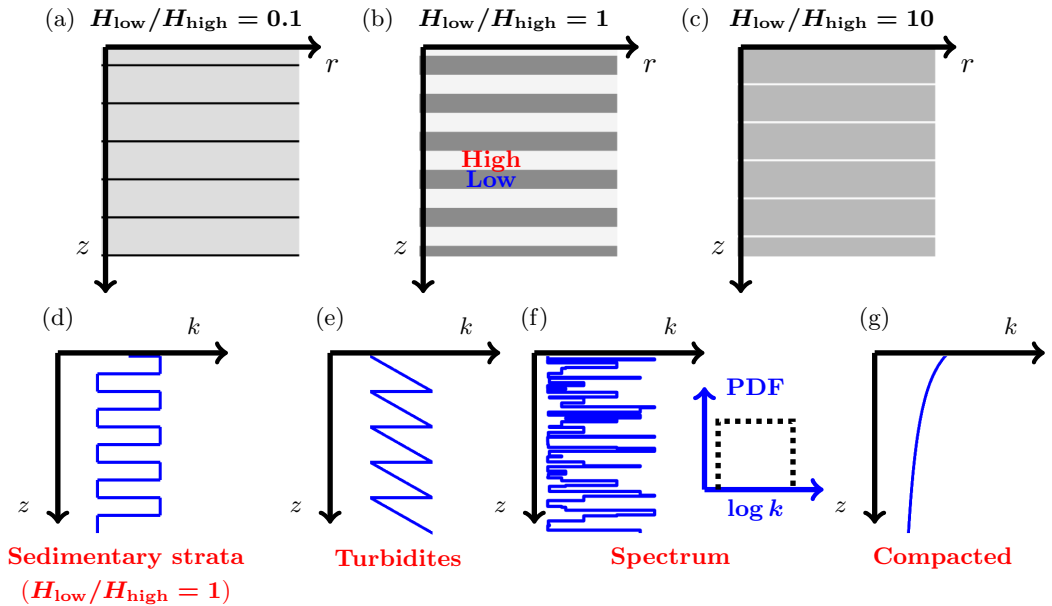

The governing equation \eqref{Geqn} must be accompanied by some initial and boundary conditions to create a well-posed system. Firstly, we define the nose of the gravity current at position $r=r_N(t)$, at which the thickness is zero, such that
\beq
\left.\mathcal{S}\right|_{r=r_N(t)}=0.\label{bc1}
\eeq
Secondly, following \cite{golding2013effects}, we impose global conservation of mass of the non-wetting phase, such that
\beq
2\pi\int_0^{r_N(t)} r\mathcal{S}\,\mathrm{d}r=Qt,\label{bc2}
\eeq
or equivalently, we impose the input flux at the origin and zero flux through the nose of the current,
\begin{align}
-2\pi \left[r\mathcal{F}\frac{\partial \mathcal{S}}{\partial r}\right]_{r=0}&=Q,\label{fluxbc}\\
2\pi \left[r\mathcal{F}\frac{\partial \mathcal{S}}{\partial r}\right]_{r=r_N(t)}&=0.\label{zerofluxbc}
\end{align}
The finite flux value $Q$ in \eqref{fluxbc} indicates that the gradient ${\partial \mathcal{S}}/{\partial r}$ must become infinite as $r\rightarrow 0$. Therefore, it is expected that the long-thin approximation made earlier may become invalid very close to injection. Furthermore, near the nose of the gravity current $r=r_N$, where the gravity current becomes thinner than the heterogeneity length scale, we do not expect our upscaled approximation to be accurate.

\subsection{Incorporating heterogeneity}\label{sechet}

\begin{table}
\centering
\begin{tabular}{|c|c|c|}
Heterogeneity type & Functional form &  Parameters  \\
\hline
Sedimentary strata   & $k=\begin{cases}k_\mathrm{low}  \\k_\mathrm{high} \end{cases}$ with ratio $H_\mathrm{low}/H_\mathrm{high}$  & $k_\mathrm{low}/k_\mathrm{high}$, $H_\mathrm{low}/H_\mathrm{high}$ \\
\\
Turbidites   & $k=k_0\left[1-({2A}/{\pi}) \tan^{-1} \cot \lb 2 n\pi {z
}\rb\right] $ & $A=\frac{1-k_\mathrm{low}/k_\mathrm{high}}{1+k_\mathrm{low}/k_\mathrm{high}}$\\
\\
Spectrum &  $k=\exp\lb\mathcal{U}\left[\log k_\mathrm{low},\log k_\mathrm{high}\right]\rb $ &   $k_\mathrm{low}/k_\mathrm{high}$  \\
& &  \\
 Compacted   & $k=k_0(1+z)^{-\beta} $ &  $\beta \geq 0$\\
\hline
\end{tabular}
\caption{Definitions of the different types of heterogeneity (characterised by the permeability), as displayed in figure \ref{hettypes}. Sedimentary strata take binary permeability values $k_\mathrm{low}/k_\mathrm{high}$ with the width ratio of low/high regions given by $H_\mathrm{low}/H_\mathrm{high}$. Turbidites, the deposits of turbidity currents, consist of a periodic array of layers with linearly varying permeability, where the wavenumber $n$, is considered in the limit $nh\rightarrow\infty$.
In the spectrum case, permeability is a series of strata, where each layer has permeability taken from a uniform random distribution, distributed logarithmically across range $[k_\mathrm{low}, k_\mathrm{high}]$. Likewise, the widths of the layers are taken from a uniform random distribution on a linear scale.
Compacted rock corresponds to a permeability profile which decreases with depth under a power law $\beta$, starting with a finite value at $z=0$. 
\label{hettypestab}}
\end{table}

To close the system, we must choose a type of vertical heterogeneity. Since we have used power laws $a,b$ to relate the porosity and pore entry pressure to the permeability, we need only choose a functional form for $k(z)$. 
Whilst in general this function may vary in three dimensions (i.e. $k(\mathbf{x})$), here we restrict our attention to pure vertical variation since, not only does this capture the leading order behaviour for sedimentary layering, but also because this is consistent with the long-thin approximation of a gravity current made earlier.

To model the permeability we have a variety of different physically-motivated choices which we list in table \ref{hettypestab} and plot in figure \ref{hettypes}. Firstly, \textit{sedimentary strata} represent a periodic deposition of two different types of sediments, such that the permeability alternates between two values, $k_\mathrm{low}$ and $k_\mathrm{high}$, in a periodic array of layers, where the width ratio of each of these is given by $H_\mathrm{low}/H_\mathrm{high}$ (see figure \ref{hettypes}a,b,c). 
Unlike the sedimentary strata, which are uniformly deposited in each layer, \textit{turbidites} represent the deposition of sediments from the continuous arrival of turbidity currents, such that within each layer the permeability varies linearly. The sign of the linear slope indicates that layers become more permeable as one descends deeper, since this corresponds to the early/late arrival of large/small particles in a turbidity current. 
Thirdly, we consider a permeability profile which is generated by randomly sampling from a distribution, or \textit{spectrum}, of permeability values, spread out logarithmically.
This case is motivated by realistic measurements of sedimentary strata which are  often noisy and aperiodic.
Finally, we consider a \textit{compacted} rock, where the permeability decreases with depth due to the buildup of lithostatic pressure over time.

Although there are many other possible choices for the permeability, we restrict our investigation to these four examples since they are canonical cases from which we may learn about the fundamental effects of heterogeneities. Each case is parameterised, either by the ratio of the permeabilities and widths of the lowest-highest permeability regions $k_\mathrm{low}/k_\mathrm{high}$, $H_\mathrm{low}/H_\mathrm{high}$, or by the compaction power law $\beta$, which represents the strength of the compaction effect. 

It is important to note the possible limitations on these parameters. In particular, sufficiently low permeability layers may cause a vertical obstruction, such that an unconfined description of the gravity current is no longer applicable.
To investigate the limitations on the permeability ratio $k_\mathrm{low}/k_\mathrm{high}$ we have performed a set of numerical simulations of the two-dimensional miscible Darcy equations using the \textit{DarcyLite} finite element package in \textit{Matlab}, adapted to account for gravity \citep{liu2016darcylite,harper2021coupling}. The miscible Darcy equations are equivalent to the immiscible equations \eqref{ge1}-\eqref{ge2} in the limit where the relative permeabilities become independent of the phases, $k_{rn},k_{rw}\rightarrow 1$, and the phase pressures equalise such that $p_c\rightarrow0$. Studying the miscible flow problem allows us to investigate the applicability of upscaling for small values of the permeability ratio without accounting for the more complex effects of immiscible phase saturations. We do not display the numerical results here, since a rigorous analysis of this query is outside the scope of this paper, but we describe our findings here in writing instead.

For very small values of the permeability ratio (e.g. $k_\mathrm{low}/k_\mathrm{high}=0.001$) there are several important observations from these numerical simulations. At early times, due to the effective obstruction from the low permeability layers, the injection is focussed within the nearest high permeability layers instead, and behaves approximately like a confined flow in which the pressure has significant streamwise gradients (i.e. deviating away from the hydrostatic condition \eqref{hydrostat}). As a result, the shape of the gravity current is highly distorted and loses its self-similar structure. At later times, once the gravity current has invaded a sufficient number of vertical layers, it begins to assume self-similar dynamics and the pressure becomes hydrostatic to good approximation. Therefore, there is no strict lower bound on the permeability ratio $k_\mathrm{low}/k_\mathrm{high}$ for an upscaling procedure, but rather this becomes a question of temporal and spatial scales. In other words, for any positive permeability ratio, given enough time and spatial extent, such an injection will eventually resemble a self-similar gravity current and is therefore amenable to upscaling. However, the overall aspect ratio of the gravity current is significantly reduced (i.e. longer and thinner) due to the delay in the vertical flow caused by the obstruction of the low permeability layers. 
Therefore, to avoid dealing with the prolonged transient effects that precede self-similarity in the case of very low permeability ratios, for the remainder of this study we restrict our attention to $k_\mathrm{low}/k_\mathrm{high}\geq0.01$. 

Continuing our upscaling analysis we note that, given a particular type of heterogeneity $k(z)$ and power laws $a,b$ for the porosity and pore entry pressure, the integrals \eqref{Seqn},\eqref{Feqn1}-\eqref{Feqn2} must first be calculated before we can solve \eqref{Geqn}. For general values of $a,b$, these integrals must be calculated numerically, using a trapezoidal integration rule, for example. 
In the layered cases, we wish to remove the dependence of these integrals on the heterogeneity wavelength, since it is undesirable to have upscaled properties like $\mathcal{F}$ that oscillate depending on the gravity current thickness. Therefore, instead of resolving all of the details of the flow, we build a macroscopic picture of the gravity current, which is consistent with other upscaling approaches (see \citet{rabinovich2016analytical} for example).

In the case of sedimentary strata, since $k$ (and therefore $p_e$ and $\phi$) takes either one of two possible values, integrals can be simply decomposed into bulk fractions
\beq
\int\,\cdot\,\mathrm{dz}=\frac{H_\mathrm{low}\int_{k_\mathrm{low}}\,\cdot\,\mathrm{dz}+H_\mathrm{high}\int_{k_\mathrm{high}}\,\cdot\,\mathrm{dz}}{H_\mathrm{low}+H_\mathrm{high}},\label{intsum}
\eeq 
thereby removing the need to resolve individual layers. 
A similar approach can be taken in the case of the permeability spectrum, although in that case \eqref{intsum} is replaced by a sum over the number permeability values sampled from the random distribution\footnote{Note that in the case of the permeability spectrum, we sample $N$ pairs of values $\left\{k_i,H_i\right\}$ ($i=1,\ldots,N$) from a random distribution of permeabilities and layer widths. Once sampled, it does not matter how these values are arranged. Therefore, a bulk decomposition like \eqref{intsum} is still possible.}.
However, in the case of the turbidites, to remove the dependence on the wavenumber $n$ (as defined in table \ref{hettypestab}), the integrals must be calculated with a fine numerical scheme for a large but finite value of $nh\gg1$.

\begin{figure}
\centering
\begin{tikzpicture}[scale=0.8]
\node at (0,0) {\includegraphics[width=0.48\textwidth]{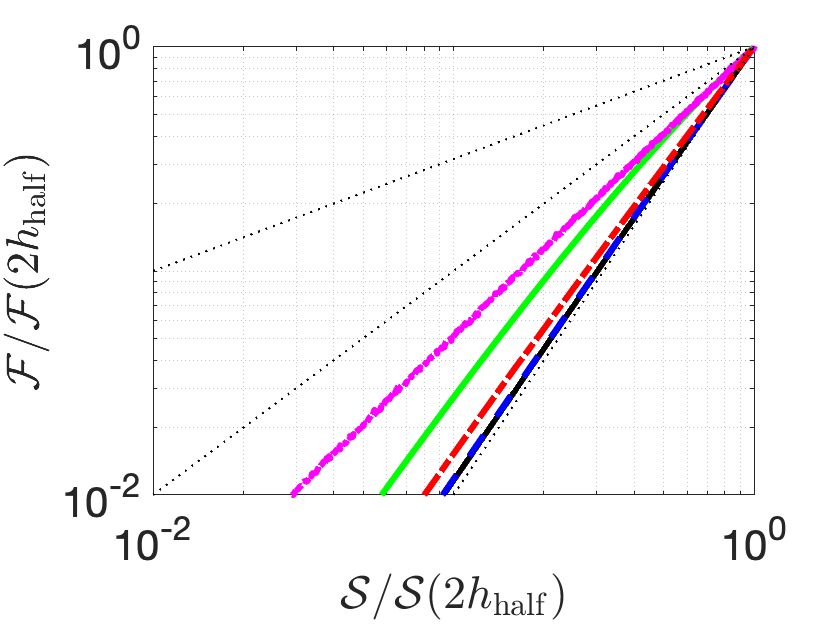}};
\node at (8,0) {\includegraphics[width=0.48\textwidth]{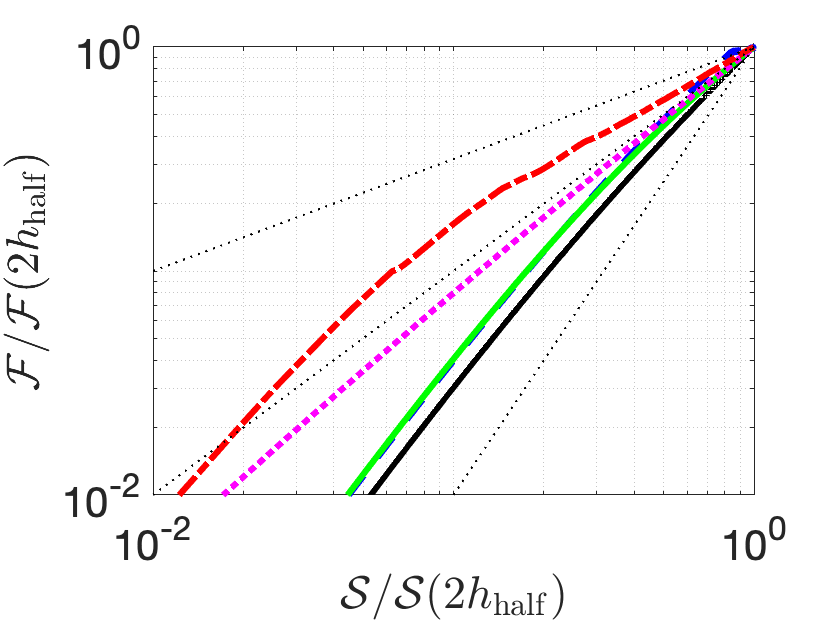}};
\node at (0,-7) {\includegraphics[width=0.48\textwidth]{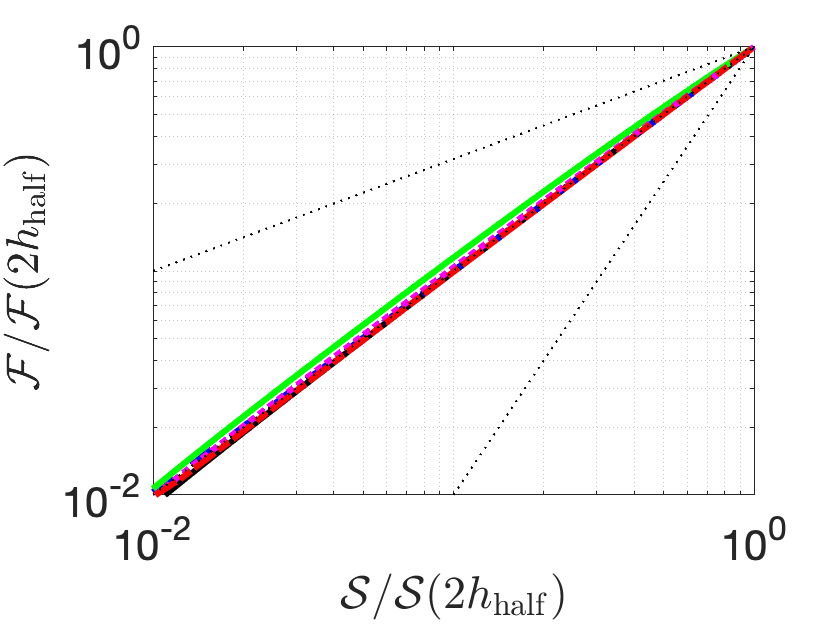}};
\node at (8,-7) {\includegraphics[width=0.48\textwidth]{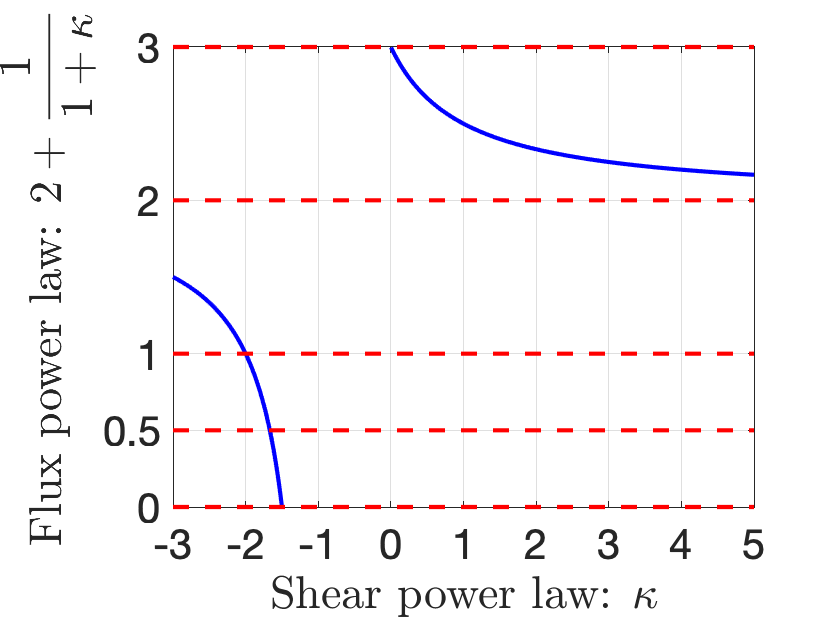}};
\node at (4,-10.5) {\includegraphics[width=0.9\textwidth]{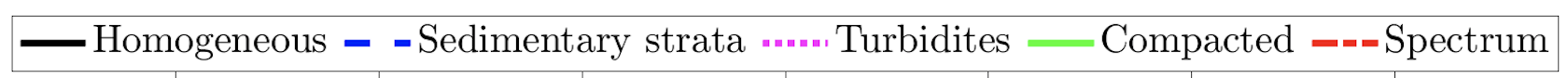}};
\node at (-4,3) {(a)};
\node at (0.5,3) {\large \bf Bo$\boldsymbol{=10^{-3}}$};
\node at (4,3) {(b)};
\node at (8.5,3) {\large \bf Bo$\boldsymbol{=1}$};
\node at (-4,-3.8) {(c)};
\node at (4,-3.8) {(d)};
\node at (0.5,-4) {\large \bf Bo$\boldsymbol{=10^{3}}$};
\node[red] at (0.5,-5) {\large \bf 1/2};
\node[red] at (-0.5,-6.7) {\large \bf 1};
\node[red] at (1.9,-7) {\large \bf 2};
\end{tikzpicture}
\caption{(a,b,c) Variation of the flux $\mathcal{F}$ \eqref{Feqn} of the integrated saturation $\mathcal{S}$ in \eqref{Geqn} for different values of the Bond number Bo. Both $\mathcal{F}$ and $\mathcal{S}$ are normalised by reference values (measured at twice the mid range value of the gravity current thickness, $h_\mathrm{half}=h(r_N(t)/2,t)$) for illustration purposes. In each plot we indicate power law values of $1/2$, $1$ and $2$ with dotted lines for comparison. (d) Analogy between a two-phase gravity current in a heterogeneous porous medium and a non-Newtonian viscous gravity current with viscosity power law $\mu\propto (\partial u/\partial z)^\kappa$. The resultant flux power law is given by $\int_0^h u\,\mathrm{d}z\propto h^{2+1/(1+\kappa)}$, as indicated with the blue curve. Red dashed lines illustrate particular power law values of interest. \label{allfluxes0}}
\end{figure}
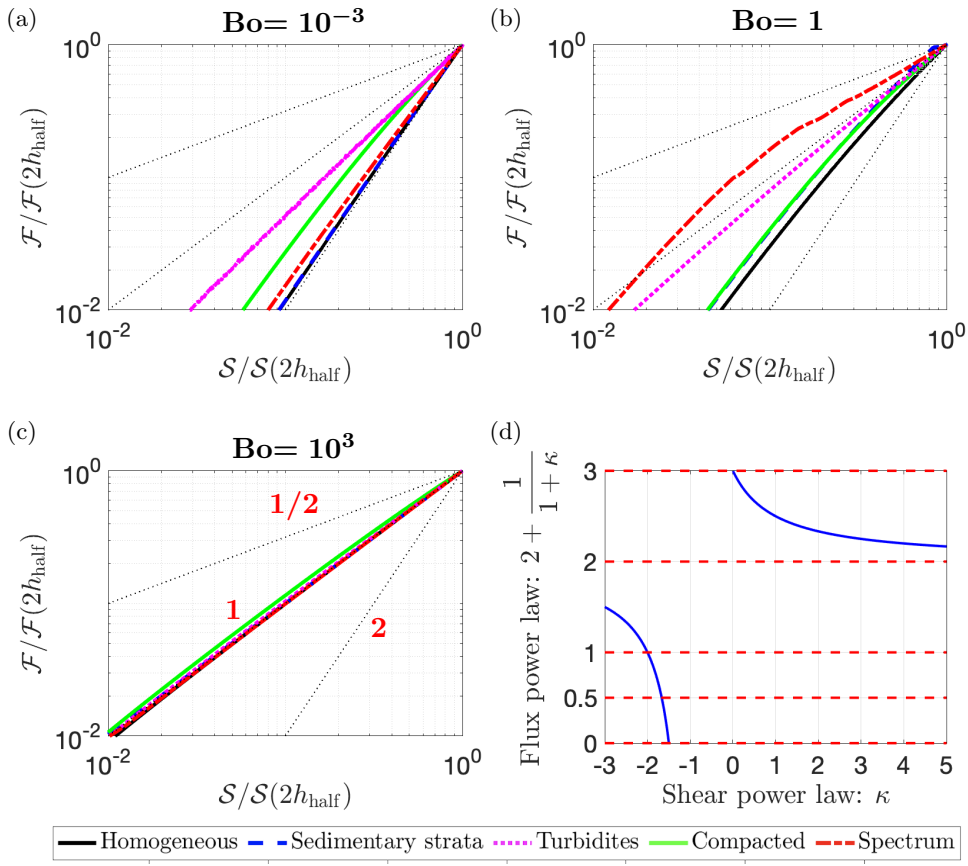

The most salient features of this analysis are the integrated saturation $\mathcal{S}(h)$ and the flux $\mathcal{F}({h})$, since $\mathcal{F}$ allows us to solve the diffusion equation \eqref{Geqn}, and $\mathcal{S}$ allows us to calculate the gravity current thickness by way of inversion. These both depend on a number of non-dimensional parameters. Ignoring the capillary number (since for now we restrict our attention to N$_c\ll1$ or N$_c\gg1$) there are a total of 8 non-dimensional parameters which govern the problem. These consist of the heterogeneity parameters $k_\mathrm{low}/k_\mathrm{high}$, $H_\mathrm{low}/H_\mathrm{high}$, $\beta$ (if compaction present); the power laws relating porosity and pore entry pressure to the permeability $a$, $b$; the Brooks-Corey parameters $\lambda$, $\alpha$; and finally the Bond number, which is defined as
\beq
\mathrm{Bo}=\lb \frac{Q \Delta \rho g \mu_n}{k_0 k_{rn0} p_{0}^2}\rb^{1/2}.\label{Bodef}
\eeq
The Bond number, which can alternatively be written as Bo$={\Delta \rho g H}/{p_0}$, where $H=\sqrt{Q/u_b}$ is the buoyancy length scale, is interpreted as the ratio between buoyancy forces and capillary forces. This quantity largely controls the saturation distribution \eqref{sateq},\eqref{sateq2}, which is evident upon dimensional analysis. For example, when Bo$\gg1$ the saturation, written in dimensionless form, approximates to
\beq
s= 1-\left[ \frac{1}{{p}_e({z})/{p}_0}+\frac{\mathrm{Bo} ({h}/H-{z}/H)}{{p}_e({z})/p_0}\right]^{-\lambda}\approx 1,\label{sateqdimless}
\eeq
regardless of which type of heterogeneity ${p}_e({z})$ we consider (so long as $\mathrm{Bo}p_0/p_e\gg1$).

We note that some of the above parameters have already been studied by other authors. For example, the power laws $a,b$ were already addressed by \citet{benham2020upscaling} and the Brooks-Corey parameter $\lambda$ was studied by \cite{golding2013effects}. Therefore, for the remainder of the current study we focus on the heterogeneity parameters $k_\mathrm{low}/k_\mathrm{high}$, $H_\mathrm{low}/H_\mathrm{high}$, $\beta$, and the Bond number as the key parameters of interest. We use the homogeneous case $k_\mathrm{low}/k_\mathrm{high}=1$ as a proxy to study the viscous limit behaviour N$_c\gg1$, since they are equivalent (see footnote in Section \ref{secfunds}). For $k_\mathrm{low}/k_\mathrm{high}<1$, however, we assume capillary limit behaviour N$_c\ll1$.

We fix the remaining parameters at typical values $a=1/7$, $b=3/7$ (using the Leverett scaling), $\lambda=3$ and $\alpha=4$, which we have extracted from a variety of different sources \citep{golding2011two,berg2013displacement,bickle2017rapid}.
For the remainder of this study (up until Section \ref{sleipsec}), we keep the same values for these parameters so that we can focus on the effect of the heterogeneities instead, but we note that our approach is by no means restricted to these values.

Nevertheless, continuing with these parameter values, we illustrate how the flux $\mathcal{F}$ depends on the type of heterogeneity and the Bond number in figure \ref{allfluxes0}a,b,c. 
For each of the layered cases, we use a permeability ratio value of $k_\mathrm{low}/k_\mathrm{high}=0.1$, whereas in the compacted case we use a power law of $\beta=1$.
In all cases (except the spectrum case) the flux is well approximated by a power law $\mathcal{F}\propto \mathcal{S}^\psi$, for some value of $\psi$ between $1/2$ and $2$. In some cases, as we will show later in Section \ref{limcasesec}, these power laws can be derived analytically.

Figure \ref{allfluxes} in Appendix \ref{appA} displays the integrated saturation $\mathcal{S}$, as well as the velocity distribution $u_n\propto \Delta \rho g k(z)k_{rn}(s)/\mu_n$ within the gravity current.
There are several interesting observations to make. Firstly, no matter which type of heterogeneity nor which Bond number we choose, the integrated saturation $\mathcal{S}(h)$ is always a monotone increasing function, such that the inversion $h=\mathcal{S}^{-1}(\mathcal{S})$ is always well-posed. Secondly, we note that there is an interesting interpretation to the value of the flux power law $\psi$, by way of analogy to viscous gravity currents. 
In the governing diffusion equation for a classic viscous gravity current, the flux power law relates to the velocity distribution within that current. 
For example, the velocity distribution for Poiseuille flow, which varies quadratically in the vertical coordinate, when integrated gives a cubic flux power law. 
Likewise, a uniform plug flow, when integrated gives a linear flux power law.

In general, any viscous gravity current flux power law can be achieved by considering a shear thinning/thickening power law viscosity $\mu\propto (\partial u/\partial z)^\kappa$. Then, it is easy to show that the lubrication balance $\mu \partial^2 u/\partial z^2 \sim \partial p/\partial x$ can be integrated to give a flux $F=\int_0^h u\,\mathrm{d}z$ which obeys the power law $F\propto h^{2+1/(1+\kappa)}$. This is illustrated in figure \ref{allfluxes0}d, indicating specific cases with dashed lines. For example, a shear-thinning fluid with power law $\kappa=-5/3$ will produce a flux with power law $F\propto h^{1/2}$. 
Whilst our upscaling problem is very different from a non-Newtonian viscous gravity current, the analogy is nevertheless useful in helping to relate the flux functions observed in figures \ref{allfluxes0}a,b,c, to the velocity distributions within our gravity current (which are displayed in figure \ref{allfluxes}b,d,f, in Appendix \ref{appA}).

\begin{figure}
\centering
\begin{tikzpicture}[scale=0.7]
\node at (1.0,-0.8) {\includegraphics[height=0.21\textwidth]{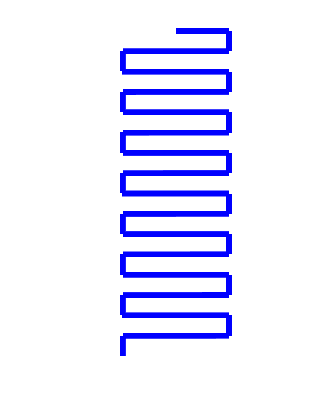}};
\draw[line width=1,->] (0,1) -- (0,-2.75);
\draw[line width=1,->] (0,1) -- (2.5,1);
\node at (-0.5,-2) {\large$h$};
\node at (-0.5,-3) {\large$z$};
\node at (2,1.4) {\large$k$};
\node at (4.85,-0.8) {\includegraphics[height=0.21\textwidth]{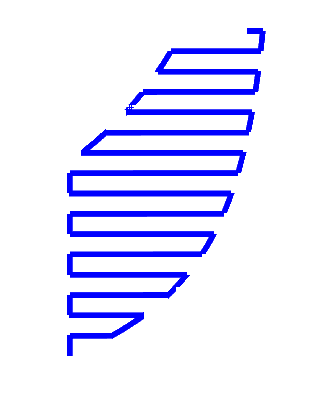}};
\draw[line width=1,->] (4,1) -- (4,-2.75);
\draw[line width=1,->] (4,1) -- (6.5,1);
\node at (3.5,-2) {\large$h$};
\node at (3.5,-3) {\large$z$};
\node at (6,1.4) {\large$s$};
\node at (8.9,-0.8) {\includegraphics[height=0.21\textwidth]{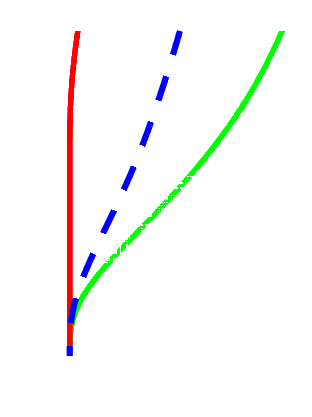}};
\node at (9.8,-2.5) {\includegraphics[height=0.09\textwidth]{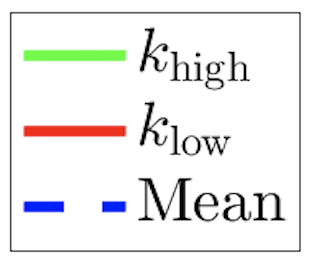}};
\draw[line width=1,->] (8,1) -- (8,-2.75);
\draw[line width=1,->] (8,1) -- (10.5,1);
\node at (7.5,-2) {\large$h$};
\node at (7.5,-3) {\large$z$};
\node at (10,1.4) {\large$u_n$};
\node at (13.1,-1.25) {\includegraphics[height=0.21\textwidth]{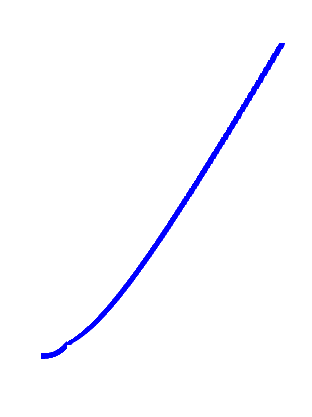}};
\draw[line width=1,->] (12,-2.75) -- (12,1);
\draw[line width=1,->] (12,-2.75) -- (14.5,-2.75);
\node at (11.5,-1.25) {\large$\mathcal{F}$};
\node at (14,-3.2) {\large$h$};
\node at (17.1,-0.7) {\includegraphics[height=0.21\textwidth]{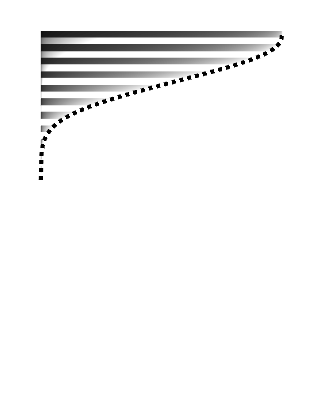}};
\draw[line width=1,->] (16,1) -- (16,-2.75);
\draw[line width=1,->] (16,1) -- (18.5,1);
\node at (15.5,-1.25) {\large$h$};
\node at (18,1.4) {\large$r$};
\node at (0,1.5) {(a)};
\node at (4,1.5) {(b)};
\node at (8,1.5) {(c)};
\node at (12,1.5) {(d)};
\node at (16,1.5) {(e)};
\node[blue] at (1,-4) {\bf  Heterogeneity};
\node[blue] at (5,-4) {\bf  Saturation};
\node[blue] at (9,-4) {\bf  Velocity};
\node[blue] at (13,-4) {\bf  Flux};
\node[blue] at (17,-4) {\bf  Thickness};
\draw[line width=2,->,red] (0,-5.5) -- (18,-5.5) ;
\node[red] at (8,-6.5) {\bf \large Methodology};
\end{tikzpicture}
\caption{Schematic illustration of our methodology, with stages going from left to right (a-e). We start by parameterising the heterogeneity $k(z),p_e(z),\phi(z)$; then we use \eqref{sateq} to determine the saturation distribution $s(z,h)$; then we obtain the velocity distribution $u_n\propto \Delta \rho g k(z)k_{rn}(s)/\mu_n$ (velocities for high and low permeability regions are illustrated as well as the mean); then from this we calculate the integrals comprising the flux $\mathcal{F}(h(\mathcal{S}))$ \eqref{Feqn}; then finally we use \eqref{Geqn} to calculate the gravity current thickness $h$ (via $\mathcal{S}(h)$). \label{panel}}
\end{figure}
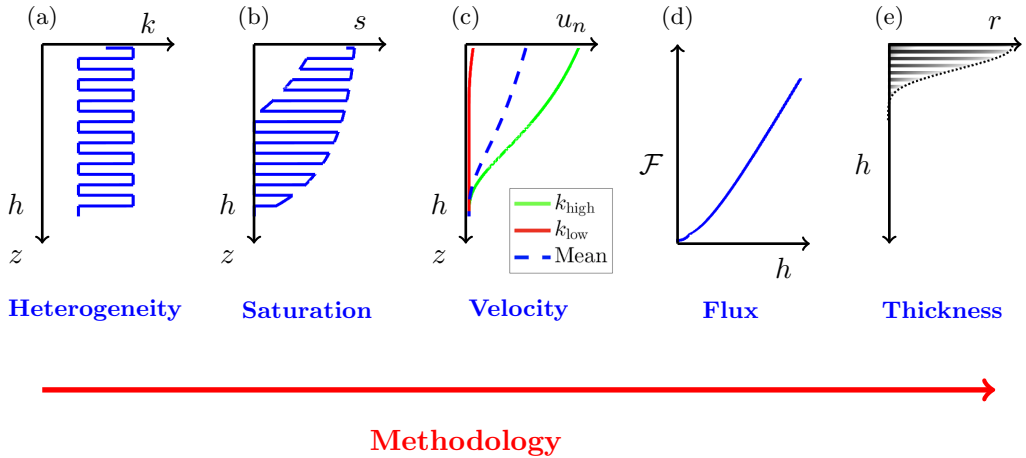

Now that all the steps in our approach have been outlined, we summarise our methodology for analysing the gravity current in figure \ref{panel}.
This illustrates the steps between initially choosing a heterogeneity type and finally solving for the gravity current thickness $h$. We have provided some example code in the supplemental materials to demonstrate these steps in the case of sedimentary strata, including how to numerically calculate the flux functions.

\subsection{Discussion of self-similarity and the numerical scheme}

There is a final simplification that can be made owing to a coordinate invariance, which allows calculation of the solution using a simple numerical scheme. In particular, much like the classic single phase axisymmetric\footnote{Note that the two-dimensional case is not necessarily self-similar. The two-dimensional gravity current thickness scales like $h\sim t^{1/3}$, such that the flux function $F(h)$ cannot be written in a general self-similar form. However, this becomes possible in certain specific cases (e.g. a linear power law $F\propto h$, as discussed by \citet{huppert1995gravity}).} gravity current discussed by \citet{lyle2005axisymmetric}, the heterogeneous case is self-similar. Upon inspection, for constant flux our governing equation \eqref{Geqn} (under the assumption of viscous N$_c\gg1$ or capillary N$_c\ll1$ limits) admits the similarity variables
\begin{align}
\eta&=(\varphi^2/Q u_b)^{1/4} r t^{-1/2},\\
\mathcal{S}&={H}{\varphi}f(\eta),
\end{align}
where the nose of the gravity current is located at $\eta=\eta_N$ for some constant $\eta_N$ which we will determine shortly. To further simplify the equations, and to convert to a unit interval domain, we write our system in terms of the variables $y=\eta/\eta_N$ and $\hat{f}(y)=f(\eta)$. In this way, the governing equation for the integrated saturation \eqref{Geqn} and the boundary conditions \eqref{bc1},\eqref{bc2} become
\begin{align}
\left[y \hat{\mathcal{F}}(\hat{f}) \hat{f}'\right]'+\frac{1}{2}\eta_N^2 y^2 \hat{f}'&=0,\label{simgov}\\
\eta_N^2\left[2\pi\int_0^1 y\hat{f}(y)\,\mathrm{d}y\right]&=1,\label{simint}\\
\hat{f}(1)&=0,\label{simfinal}
\end{align}
where $\hat{\mathcal{F}}=\mathcal{F}\varphi/u_bH$. The system can be solved numerically using a simple finite difference scheme, starting at $y=1$ and marching back towards $y=0$, where the second order ODE \eqref{simgov} is conveniently written as a set of two first order ODE's with boundary conditions
\begin{align}
\mathcal{L}'(y)&=-\frac{\eta_N^2 y\mathcal{L}}{2\hat{\mathcal{F}}(\hat{f})},\label{Leqn1}\\
\hat{f}'(y)&=\frac{\mathcal{L}}{y\hat{\mathcal{F}}(\hat{f})},\\
\mathcal{L}(1)&=\epsilon,\\
\hat{f}(1)&=\epsilon,\label{Leqn4}
\end{align}
where $\mathcal{L}$ is the total dimensionless flux, and $\epsilon\ll1$ is a small but finite numerical value (we cannot choose $\epsilon=0$ since it will generate infinite gradients). To find the constant $\eta_N$, we start with an initial guess $\eta_{N_0}$, and then use Newton's method to iteratively solve the flux condition \eqref{simint} (see our code which we have uploaded in the supplemental material).

We make the key observation that, independent of the form of $\hat{\mathcal{F}}(\hat{f})$, the gravity current is self-similar, with a nose that moves like the square root of time. Hence, the heterogeneities are only capable of modifying the prefactor $\eta_N$ for the nose speed, not the power law (which is always $r\sim t^{1/2}$). However, the heterogeneities may also change the shape of the gravity current via $\mathcal{F}$ and $\mathcal{S}$.  

As discussed later in Section \ref{transsec}, in the case where the capillary number is neither small nor large, the flux must depend on the capillary number itself. In this case, since the capillary number involves derivatives of $\mathcal{S}$ with respect to $r$, this modifies the form of the governing equation \eqref{Geqn}, rendering such similarity variables inadmissible. Over time the solution changes from a viscous limit regime (self-similar) to a capillary limit regime (self-similar), but the transition itself, therefore, cannot be self-similar.

\section{Results: viscous limit, capillary limit and transition}\label{sec2}

Our results comprise the following three different cases. 
\begin{enumerate}
\item The capillary number is large throughout the aquifer (viscous limit), in which case the upscaled problem is equivalent to the homogeneous case.
\item The capillary number is small throughout the aquifer (capillary limit), in which case the upscaled problem is dominated by the effect of the heterogeneities. 
\item The capillary number varies across the aquifer, such that different regions simultaneously lie within the viscous limit and the capillary limit, and other regions lie between these limits. 
\end{enumerate}

We start by addressing the former two cases (viscous and capillary limits), for which the problem is self-similar. 
The homogeneous case is used to study the viscous limit (since they are equivalent in an upscaled sense) and the heterogeneous cases are used to study the capillary limit.
The gravity current thickness must be calculated numerically by solving the ODE system \eqref{simint},\eqref{Leqn1}-\eqref{Leqn4}. 
In specific limiting cases, such as when the Bond number is very large or very small, analytical progress can be made.
We first present our numerical results which we use to understand the broad effect of the heterogeneities across large parameter regimes. Then, we use asymptotic analysis to help interpret the results in certain limiting cases. Finally, we address the transition between the viscous and capillary limits, for which the full PDE system \eqref{Geqn},\eqref{bc1},\eqref{bc2} must be solved in tandem with an algebraic equation for the local capillary number \eqref{capdef2}.

\subsection{Numerical solution in the viscous and capillary limits}

\begin{figure}
\centering
\begin{tikzpicture}[scale=0.78]
\node at (0,0) {\includegraphics[width=0.5\textwidth]{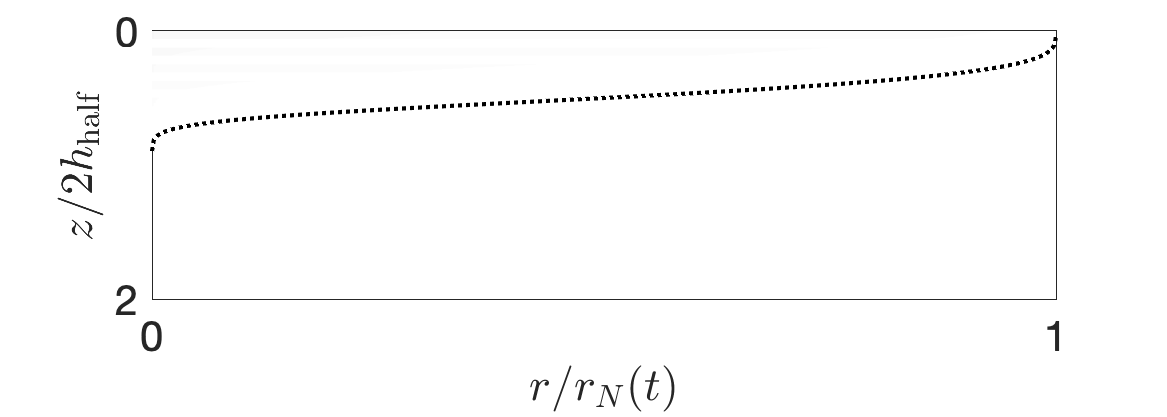}};
\node at (2,0) {\includegraphics[width=0.05\textwidth]{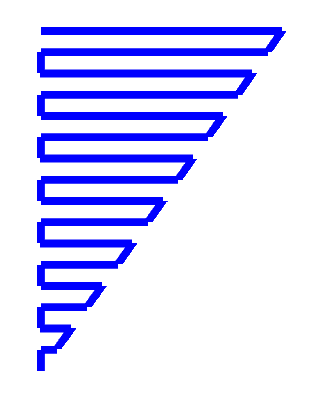}};
\node at (8,0) {\includegraphics[width=0.5\textwidth]{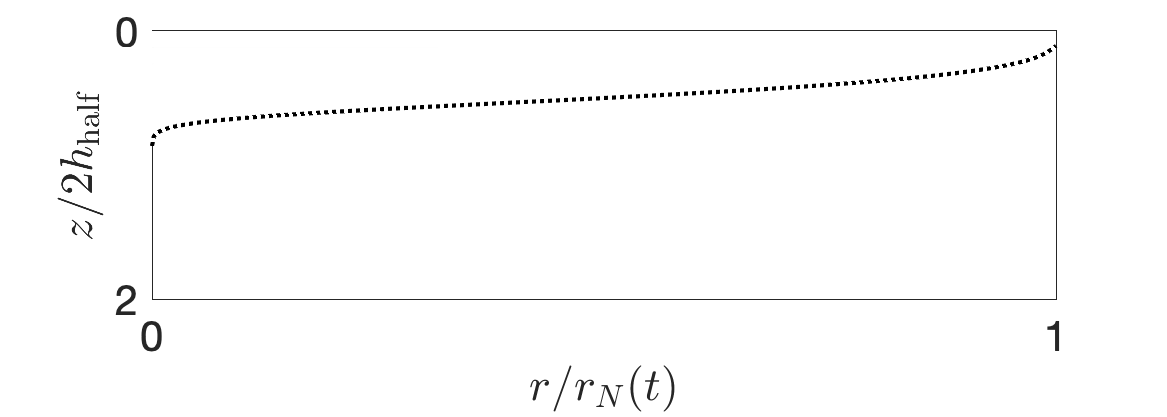}};
\node at (10,0) {\includegraphics[width=0.05\textwidth]{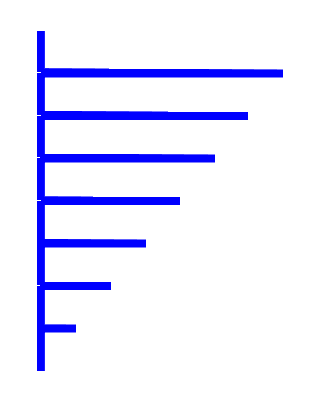}};
\node at (0,-4) {\includegraphics[width=0.5\textwidth]{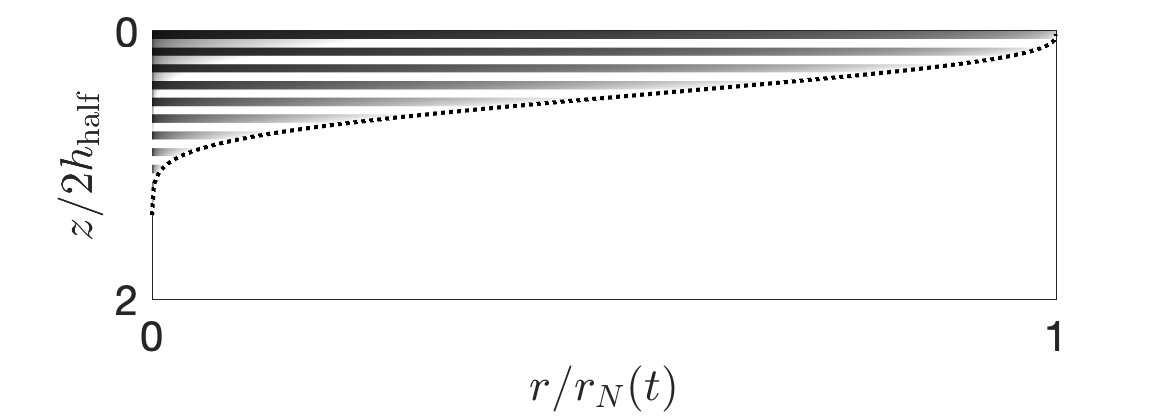}};
\node at (2,-4) {\includegraphics[width=0.05\textwidth]{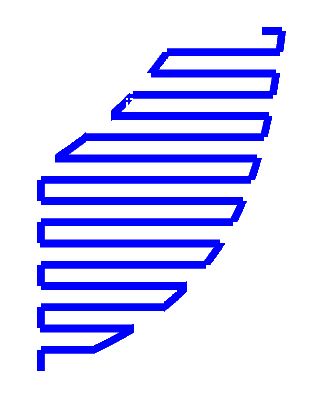}};
\node at (8,-4) {\includegraphics[width=0.5\textwidth]{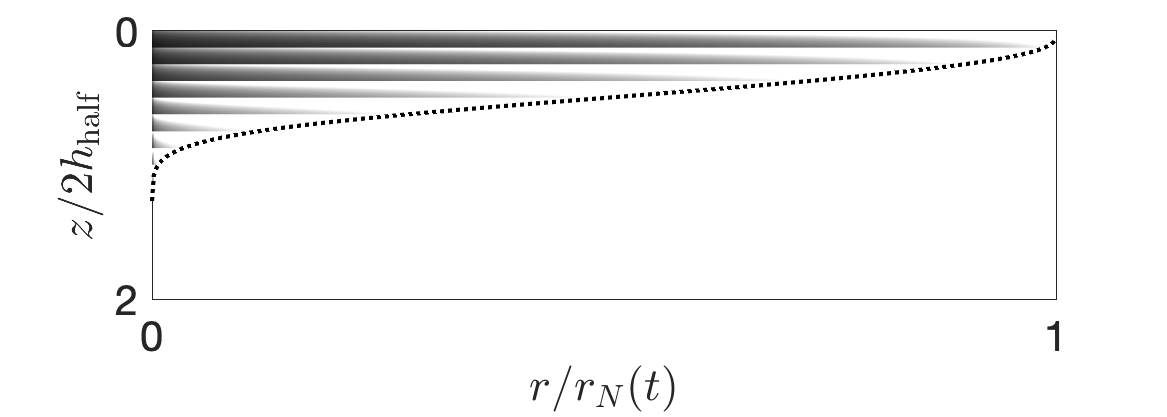}};
\node at (10,-4) {\includegraphics[width=0.05\textwidth]{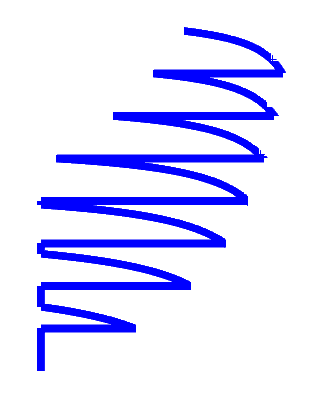}};
\node at (0,-8) {\includegraphics[width=0.5\textwidth]{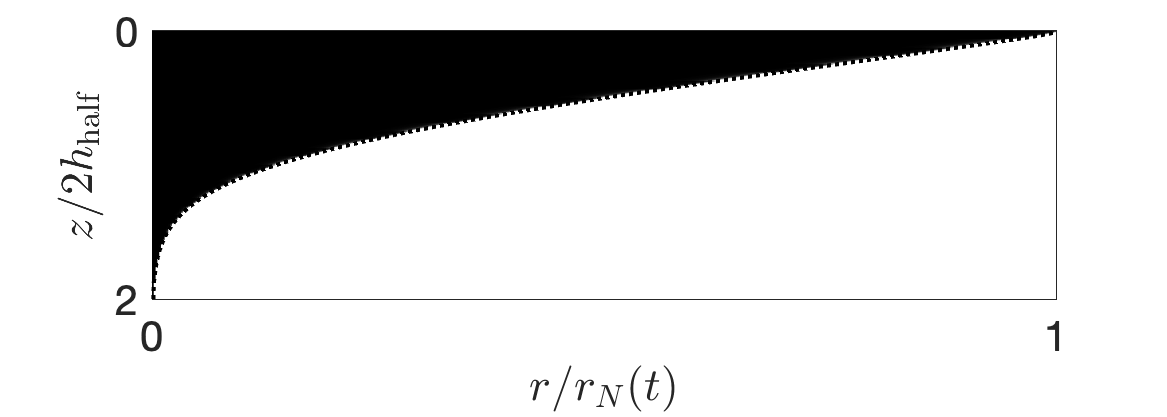}};
\node at (2,-8) {\includegraphics[width=0.05\textwidth]{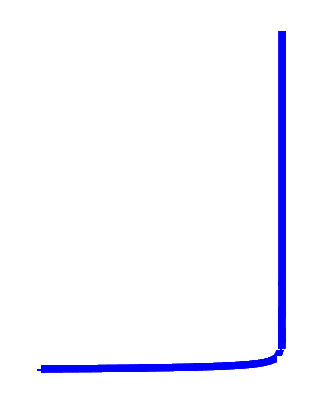}};
\node at (8,-8) {\includegraphics[width=0.5\textwidth]{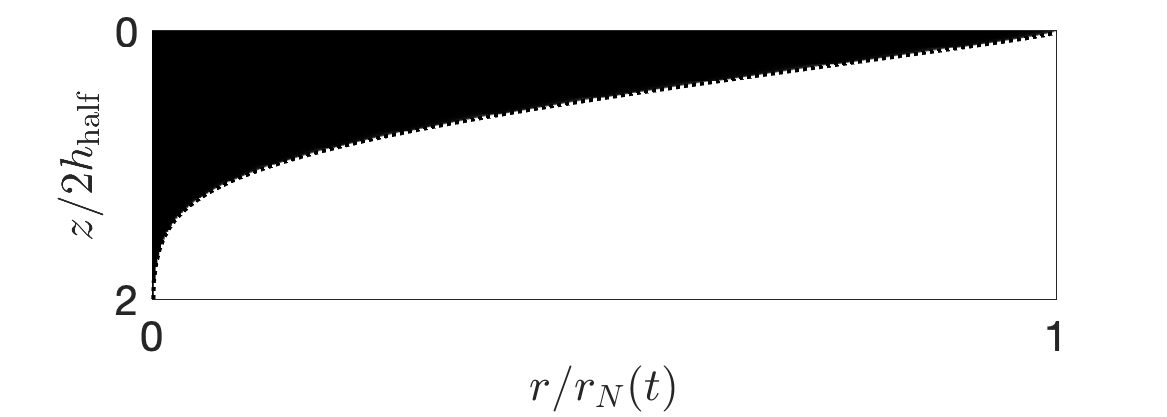}};
\node at (10,-8) {\includegraphics[width=0.05\textwidth]{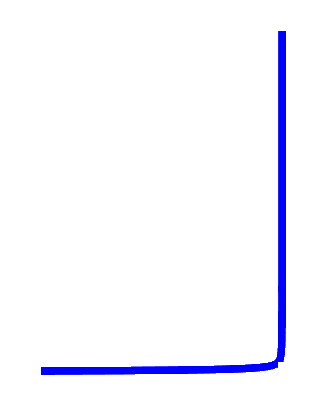}};
\draw[line width=1,->] (1.65,0.5) -- (2.6,0.5);
\draw[line width=1,->] (1.65,0.5) -- (1.65,-0.6); 
\node at (2.1,0.8) {$s/s_0$};
\node at (1.45,0) {$z$};
\draw[line width=1,->] (9.65,0.5) -- (10.6,0.5);
\draw[line width=1,->] (9.65,0.5) -- (9.65,-0.6); 
\node at (10.1,0.8) {$s/s_0$};
\node at (9.45,0) {$z$};
\draw[line width=1,->] (1.65,-3.5) -- (2.6,-3.5);
\draw[line width=1,->] (1.65,-3.5) -- (1.65,-4.6); 
\node at (2.1,-3.2) {$s/s_0$};
\node at (1.45,-4) {$z$};
\draw[line width=1,->] (9.65,-3.5) -- (10.6,-3.5);
\draw[line width=1,->] (9.65,-3.5) -- (9.65,-4.6); 
\node at (10.1,-3.2) {$s/s_0$};
\node at (9.45,-4) {$z$};
\draw[line width=1,->] (1.65,-7.5) -- (2.6,-7.5);
\draw[line width=1,->] (1.65,-7.5) -- (1.65,-8.6); 
\node at (2.1,-7.2) {$s/s_0$};
\node at (1.45,-8) {$z$};
\draw[line width=1,->] (9.65,-7.5) -- (10.6,-7.5);
\draw[line width=1,->] (9.65,-7.5) -- (9.65,-8.6); 
\node at (10.1,-7.2) {$s/s_0$};
\node at (9.45,-8) {$z$};
\node[red] at (0,2.5) {\bf \large Sedimentary strata };
\node[red] at (0,1.8) {\bf \large $(H_\mathrm{low}/H_\mathrm{high}=1)$};
\node[red] at (8,2.5) {\bf \large Turbidites};
\node at (-4.5,0) {\rotatebox{90}{$\boldsymbol{\mathrm{Bo}=10^{-2}}$}};
\node at (-4.5,-4) {\rotatebox{90}{$\boldsymbol{\mathrm{Bo}=10^{0}}$}};
\node at (-4.5,-8) {\rotatebox{90}{$\boldsymbol{\mathrm{Bo}=10^{2}}$}};
\node at (12.5,-3.5) {\includegraphics[width=0.07\textwidth]{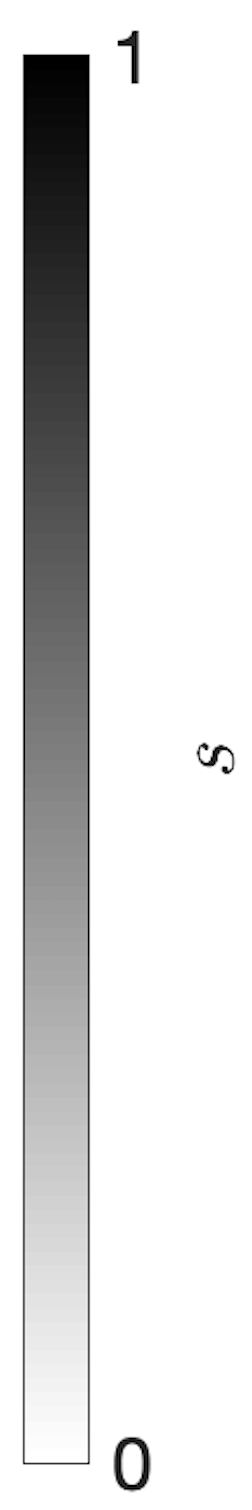}};
\node at (-4,2) {(a)};
\node at (4,2) {(b)};
\node at (-4,-2) {(c)};
\node at (4,-2) {(d)};
\node at (-4,-6) {(e)};
\node at (4,-6) {(f)};
\end{tikzpicture}
\caption{Numerical results for the capillary limit in the case of sedimentary strata (a,c,e) and turbidites (b,d,f) (with $k_\mathrm{low}/k_\mathrm{high}=1/3,H_\mathrm{low}/H_\mathrm{high}=1$). From top to bottom, capillary forces become less important with respect to gravitational forces. The radius $r$ is given in terms of the nose position $r_N(t)$, and the thickness $h$ is normalised by the reference value $2h_\mathrm{half}=2h(r_N(t)/2,t)$ for the sake of comparison.
The heterogeneity wavelength is exaggerated for illustration purposes. In each plot inserts illustrate the vertical saturation profile, normalised by the uppermost value $s_0=s(0)$. \label{het1}}
\end{figure}
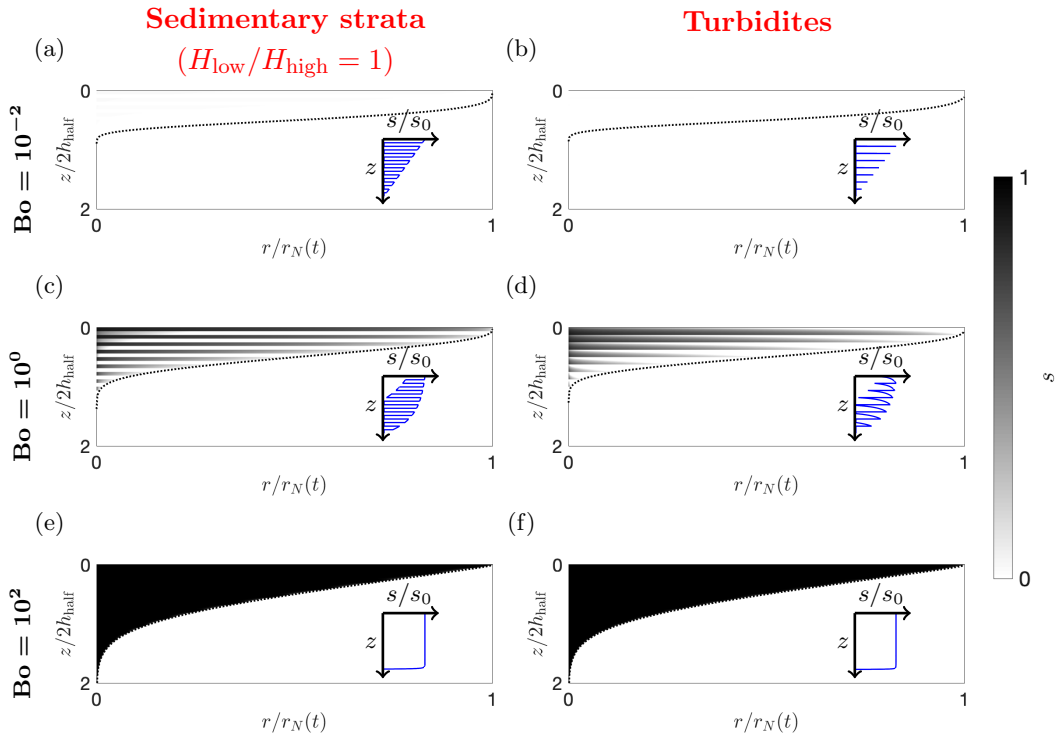

\begin{figure}
\centering
\begin{tikzpicture}[scale=0.78]
\node at (0,0) {\includegraphics[width=0.5\textwidth]{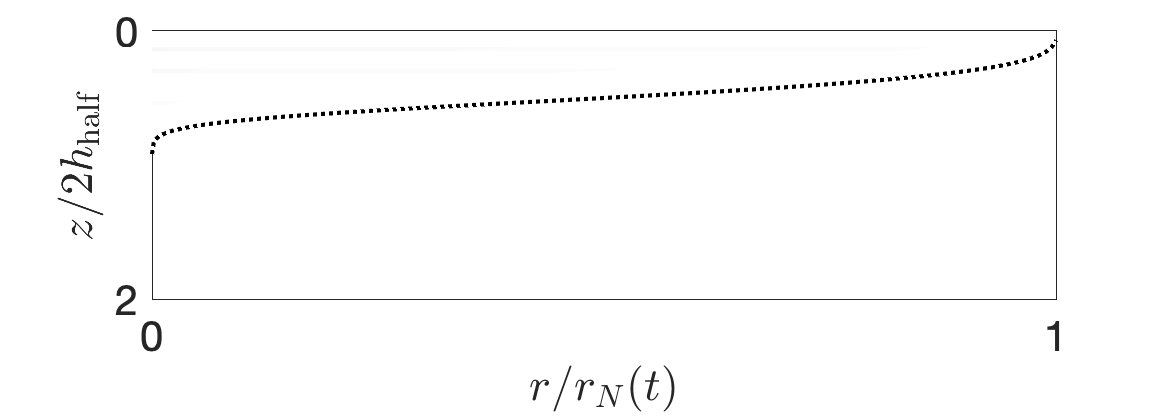}};
\node at (2,0) {\includegraphics[width=0.05\textwidth]{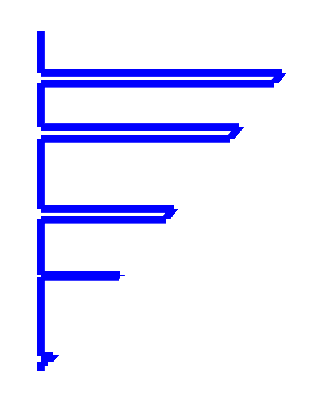}};
\node at (8,0) {\includegraphics[width=0.5\textwidth]{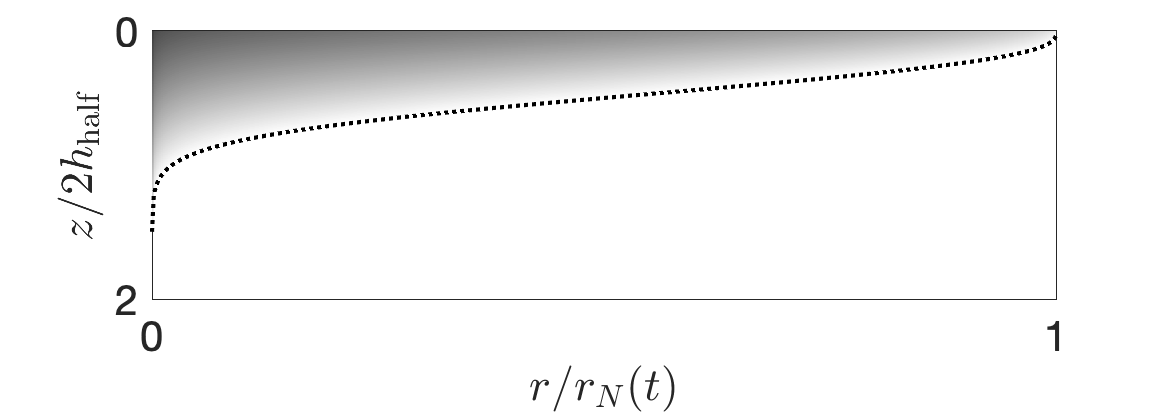}};
\node at (10,0) {\includegraphics[width=0.05\textwidth]{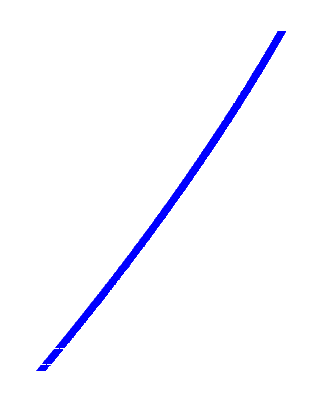}};
\node at (0,-4) {\includegraphics[width=0.5\textwidth]{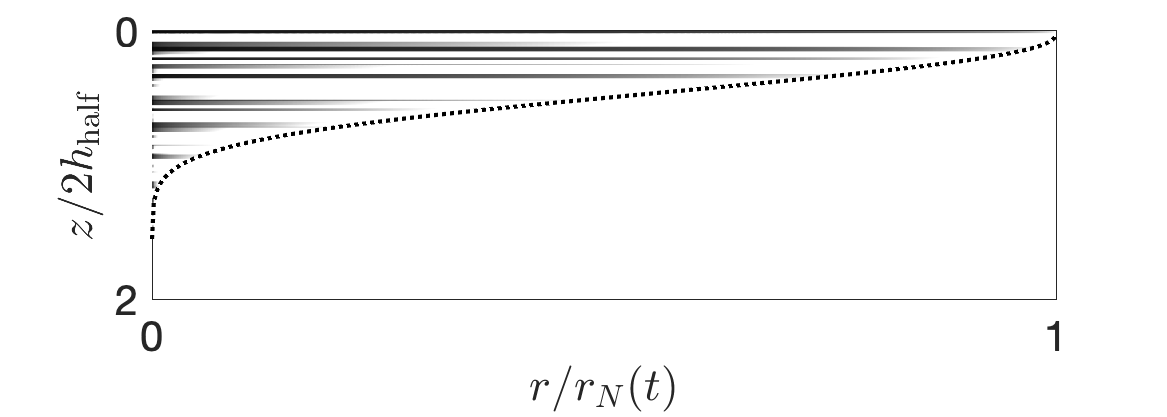}};
\node at (2,-4) {\includegraphics[width=0.05\textwidth]{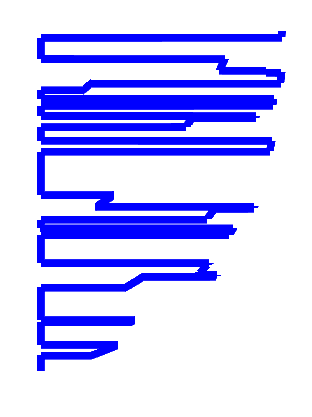}};
\node at (8,-4) {\includegraphics[width=0.5\textwidth]{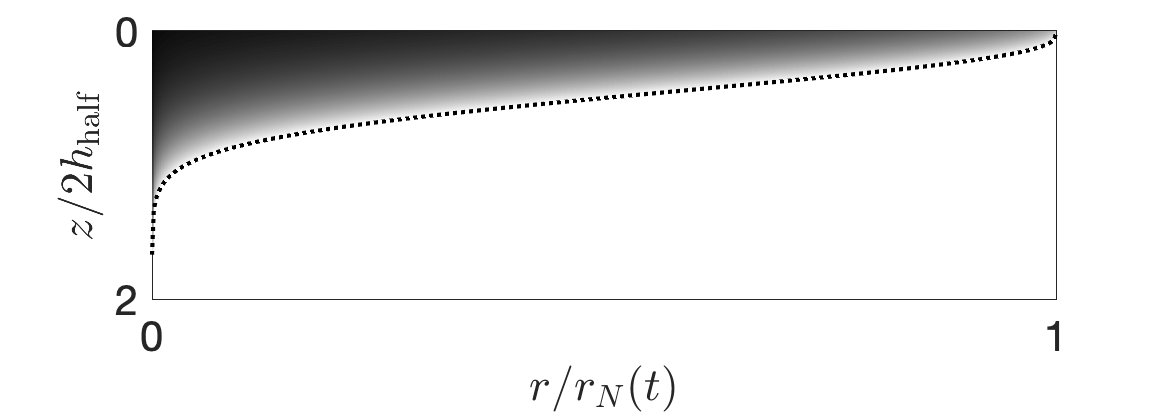}};
\node at (10,-4) {\includegraphics[width=0.05\textwidth]{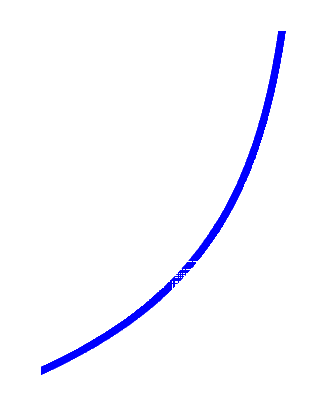}};
\node at (0,-8) {\includegraphics[width=0.5\textwidth]{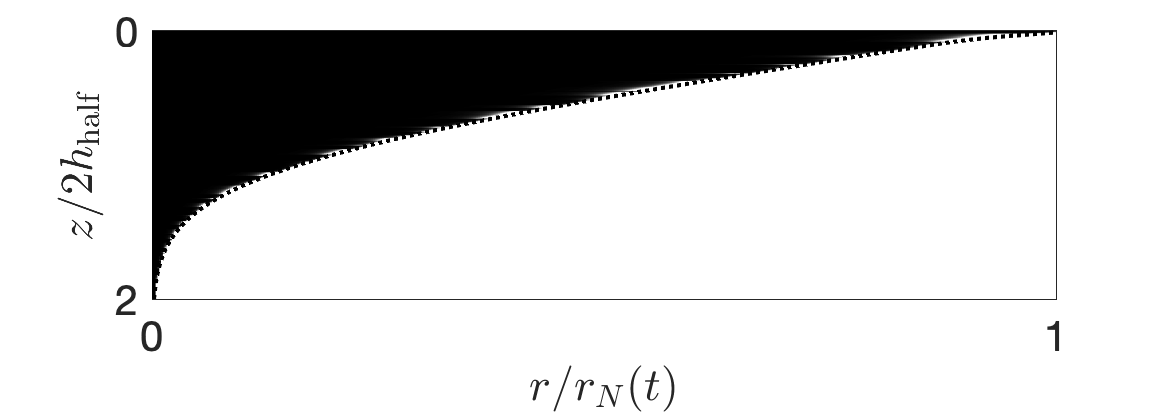}};
\node at (2,-8) {\includegraphics[width=0.05\textwidth]{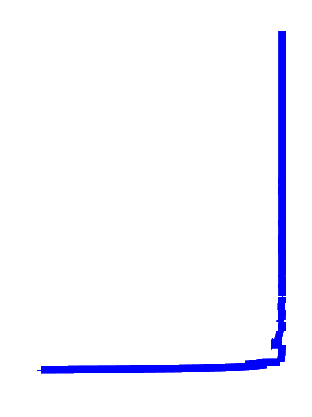}};
\node at (8,-8) {\includegraphics[width=0.5\textwidth]{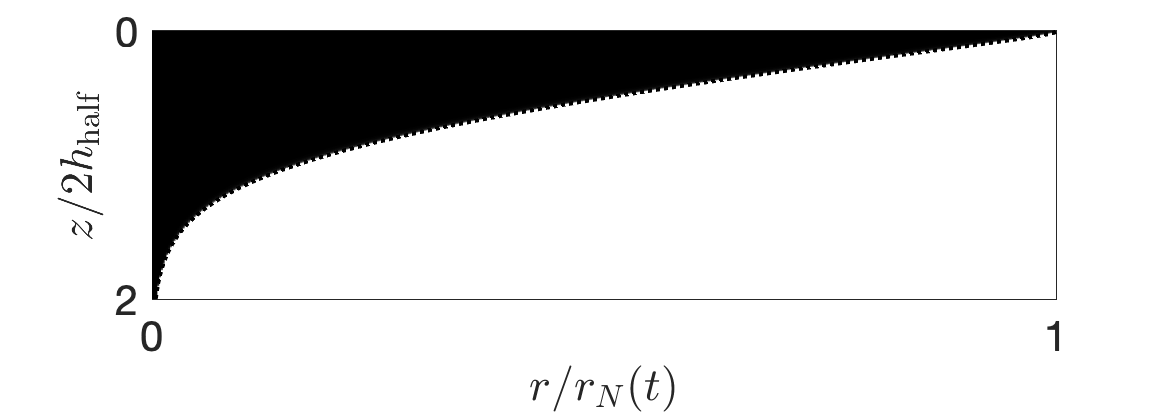}};
\node at (10,-8) {\includegraphics[width=0.05\textwidth]{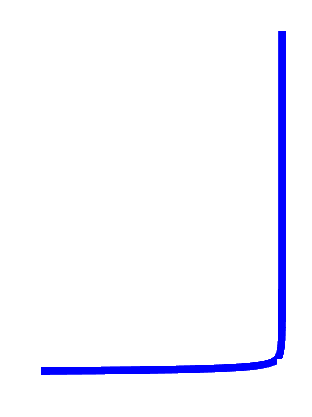}};
\draw[line width=1,->] (1.65,0.5) -- (2.6,0.5);
\draw[line width=1,->] (1.65,0.5) -- (1.65,-0.6); 
\node at (2.1,0.8) {$s/s_0$};
\node at (1.45,0) {$z$};
\draw[line width=1,->] (9.65,0.5) -- (10.6,0.5);
\draw[line width=1,->] (9.65,0.5) -- (9.65,-0.6); 
\node at (10.1,0.8) {$s/s_0$};
\node at (9.45,0) {$z$};
\draw[line width=1,->] (1.65,-3.5) -- (2.6,-3.5);
\draw[line width=1,->] (1.65,-3.5) -- (1.65,-4.6); 
\node at (2.1,-3.2) {$s/s_0$};
\node at (1.45,-4) {$z$};
\draw[line width=1,->] (9.65,-3.5) -- (10.6,-3.5);
\draw[line width=1,->] (9.65,-3.5) -- (9.65,-4.6); 
\node at (10.1,-3.2) {$s/s_0$};
\node at (9.45,-4) {$z$};
\draw[line width=1,->] (1.65,-7.5) -- (2.6,-7.5);
\draw[line width=1,->] (1.65,-7.5) -- (1.65,-8.6); 
\node at (2.1,-7.2) {$s/s_0$};
\node at (1.45,-8) {$z$};
\draw[line width=1,->] (9.65,-7.5) -- (10.6,-7.5);
\draw[line width=1,->] (9.65,-7.5) -- (9.65,-8.6); 
\node at (10.1,-7.2) {$s/s_0$};
\node at (9.45,-8) {$z$};
\node[red] at (8,2) {\bf \large Compacted};
\node[red] at (0,2) {\bf \large Spectrum};
\node at (-4.5,0) {\rotatebox{90}{$\boldsymbol{\mathrm{Bo}=10^{-2}}$}};
\node at (-4.5,-4) {\rotatebox{90}{$\boldsymbol{\mathrm{Bo}=10^{0}}$}};
\node at (-4.5,-8) {\rotatebox{90}{$\boldsymbol{\mathrm{Bo}=10^{2}}$}};
\node at (12.5,-3.5) {\includegraphics[width=0.07\textwidth]{snap}};
\node at (-4,2) {(a)};
\node at (4,2) {(b)};
\node at (-4,-2) {(c)};
\node at (4,-2) {(d)};
\node at (-4,-6) {(e)};
\node at (4,-6) {(f)};
\end{tikzpicture}
\caption{Numerical results  for the capillary limit in the case of spectrum permeability (a,c,e) (with mean permeability ratio $k_\mathrm{low}/k_\mathrm{high}=0.04$) and compacted rock (b,d,f) (with compaction power law $\beta=1$).
In each plot inserts illustrate the vertical saturation profile, normalised by the uppermost value $s_0=s(0)$.
The heterogeneity wavelength is exaggerated for illustration purposes.  \label{het2}}
\end{figure}

The capillary limit numerical solution for different types of heterogeneities is plotted in figures \ref{het1} and \ref{het2}, for Bond number values between Bo $=10^{-2}$ and Bo $=10^2$. Typical saturation profiles are also displayed as inserts in each plot. 
To compare the different profiles, we have normalised the thickness by twice its mid range value $h_\mathrm{half}=h(r_N(t)/2,t)$. The radial coordinate is normalised by the nose position $r_N(t)$ so that the shape remains on a fixed domain for all time.
For now, we do not include plots of the viscous limit numerical solutions, since these are very similar to the study by \citet{golding2013effects}. However, shortly we will use these as a reference when comparing the different types of heterogeneities.

\begin{figure}
\centering
\begin{tikzpicture}[scale=0.75]
\node at (0,0) {\includegraphics[width=0.5\textwidth]{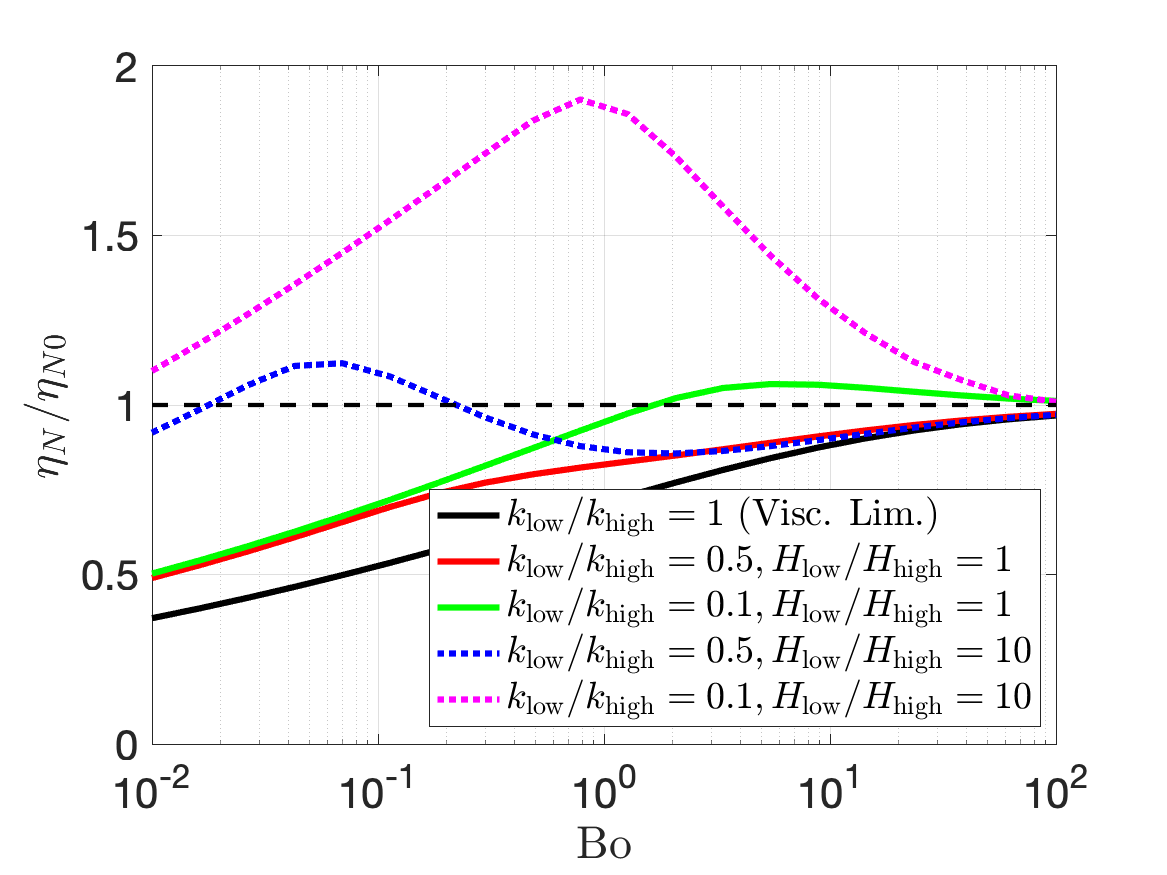}};
\node at (9,0) {\includegraphics[width=0.5\textwidth]{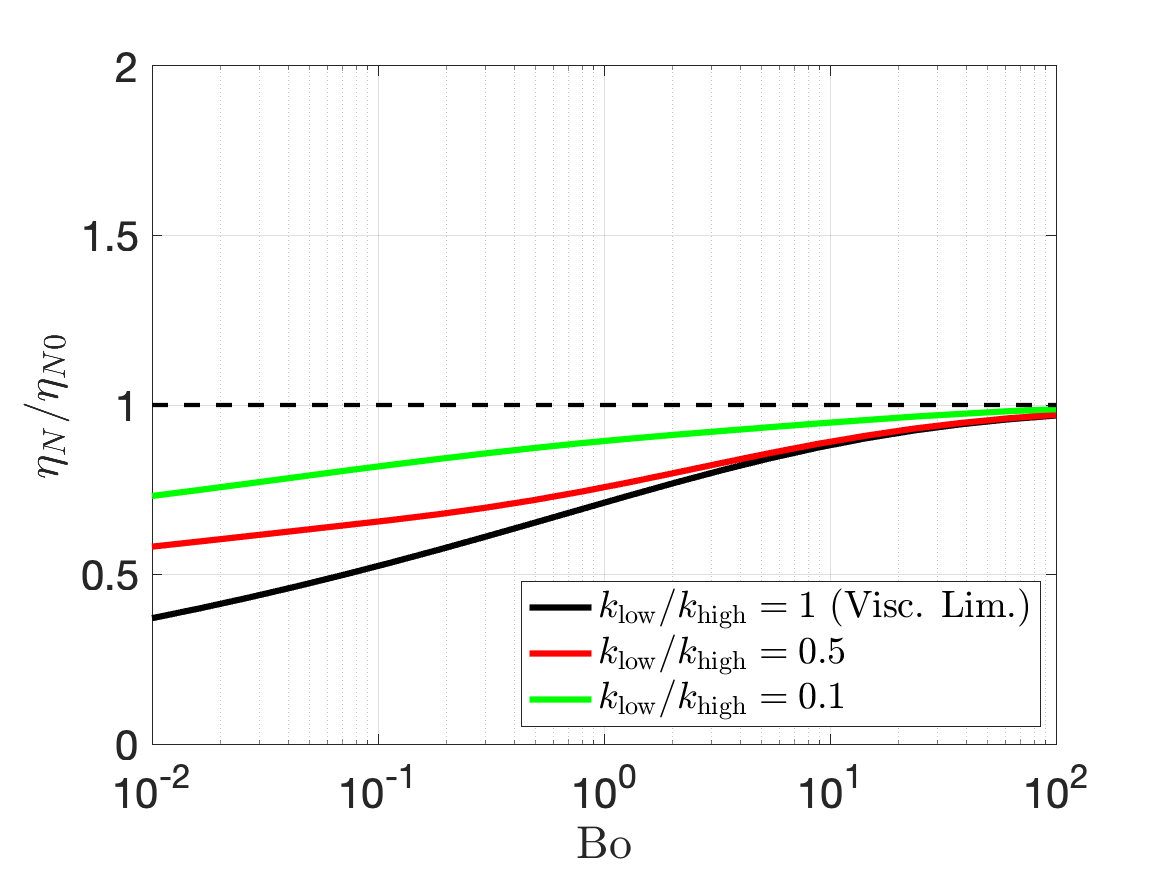}};
\node at (-4.5,3.2) {(a)};
\node at (5,3.2) {(b)};
\node at (0,3.5) {\bf \large Sedimentary strata};
\node at (9,3.5) {\bf \large Turbidites};
\node at (2.5,1.8) {\includegraphics[height=0.09\textwidth]{perm1}};
\draw[line width=1,->] (2,2.5) -- (2,1.0);
\draw[line width=1,->] (2,2.5) -- (3.3,2.5);
\node at (11.0,1.8) {\includegraphics[height=0.09\textwidth]{perm2}};
\draw[line width=1,->] (10.5,2.5) -- (10.5,1.0);
\draw[line width=1,->] (10.5,2.5) -- (11.8,2.5);
\end{tikzpicture}
\begin{tikzpicture}[scale=0.75]
\node at (0,0) {\includegraphics[width=0.5\textwidth]{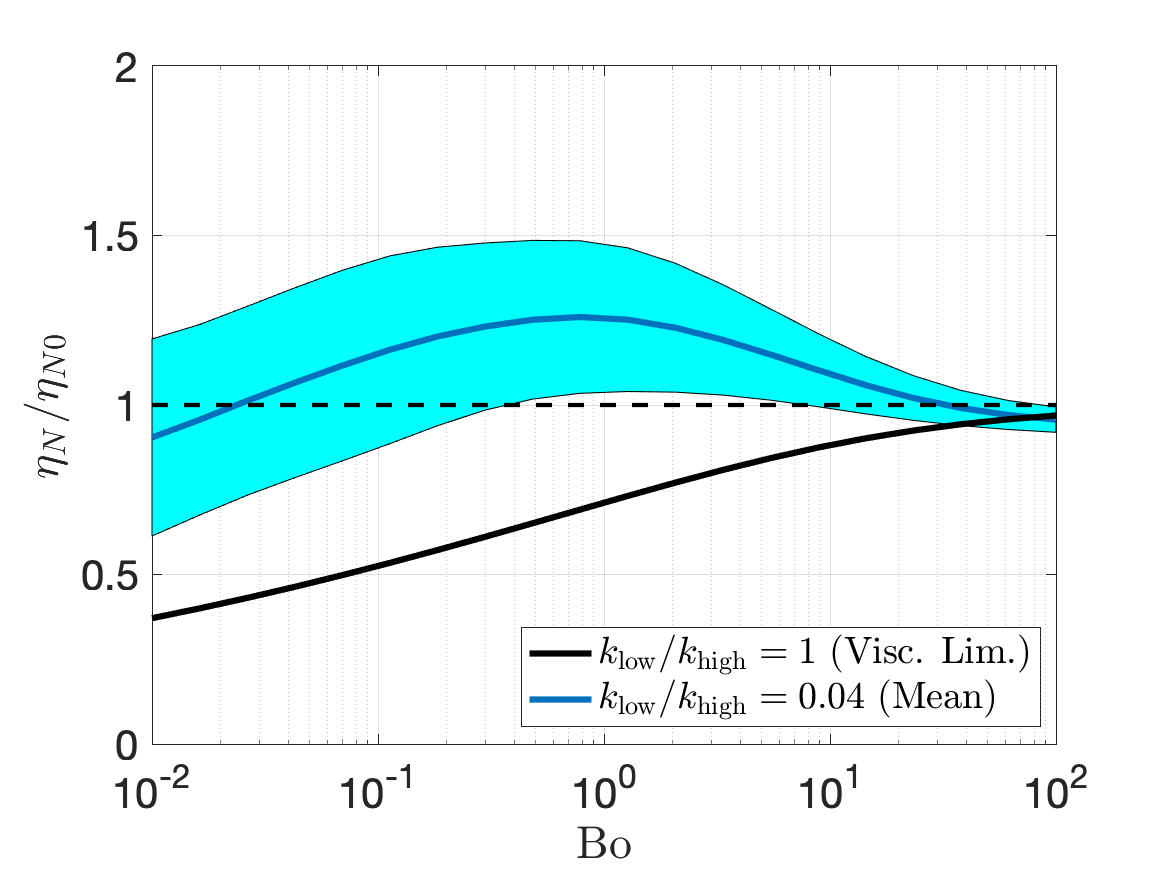}};
\node at (9,0) {\includegraphics[width=0.5\textwidth]{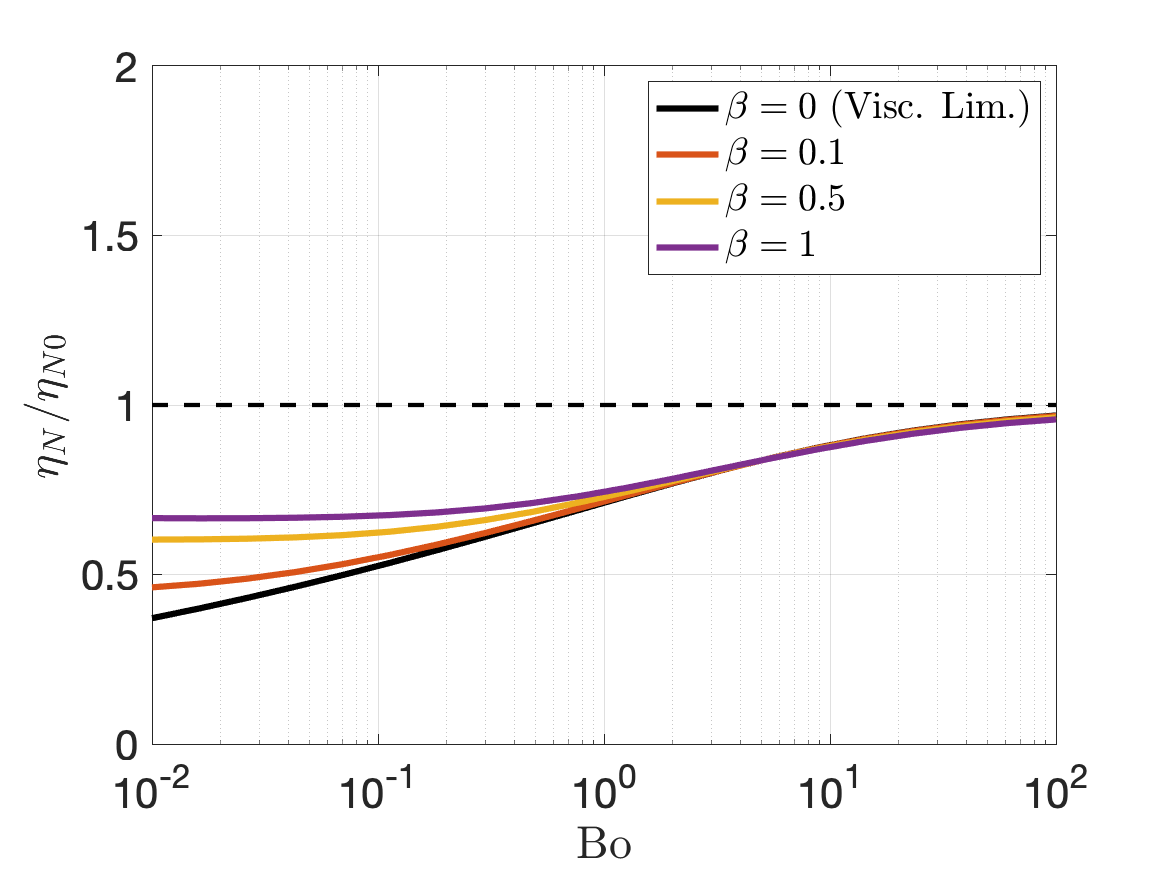}};
\node at (-4.5,3.2) {(c)};
\node at (5,3.2) {(d)};
\node at (0,3.5) {\bf \large Spectrum};
\node at (9,3.5) {\bf \large Compacted};
\node at (2.5,1.8) {\includegraphics[height=0.09\textwidth]{perm4}};
\draw[line width=1,->] (2,2.5) -- (2,1.0);
\draw[line width=1,->] (2,2.5) -- (3.3,2.5);
\node at (8.0,1.8) {\includegraphics[height=0.09\textwidth]{perm3}};
\draw[line width=1,->] (7.5,2.5) -- (7.5,1.0);
\draw[line width=1,->] (7.5,2.5) -- (8.8,2.5);
\end{tikzpicture}
\caption{Nose growth prefactor $\eta_N$ given in terms of the single phase limit $\eta_{N_0}=1.155$ for all heterogeneity types, parameterised by the permeability ratio $k_\mathrm{low}/k_\mathrm{high}$, the width ratio $H_\mathrm{low}/H_\mathrm{high}$, and the compaction power law $\beta$. 
In the case of the permeability spectrum, we show the mean result alongside one standard deviation above and below.
Limiting behaviours are illustrated with dashed lines. Solid black curves correspond to the homogeneous case, which is equivalent to the viscous limit, whereas all other curves correspond to the capillary limit. \label{lims}}
\end{figure}

Let us first focus on the non-compacted cases in the capillary limit (figures \ref{het1} and \ref{het2}a,c,e). For small Bond number Bo$\ll1$, the saturation becomes near-zero $s\approx 0$, but with spikes of linearly increasing magnitude that represent the thin regions where the permeability is near its maximum value $k\approx k_\mathrm{high}$. As we increase the Bond number towards unity, the saturation becomes larger, with an overall curved profile and significant oscillations. At high Bond number, the saturation tends towards $s\approx 1$ everywhere except very close to $z=h$, where it rapidly drops to $s=0$.
The shape of the gravity current changes from having a sharp nose at high Bond number to having a rounded blunt nose at low Bond number. There is not a  noticeable difference in the shape of the gravity current between the different types of heterogeneity.

\begin{figure}
\centering
\begin{tikzpicture}[scale=0.75]
\node at (9,0) {\includegraphics[width=0.5\textwidth]{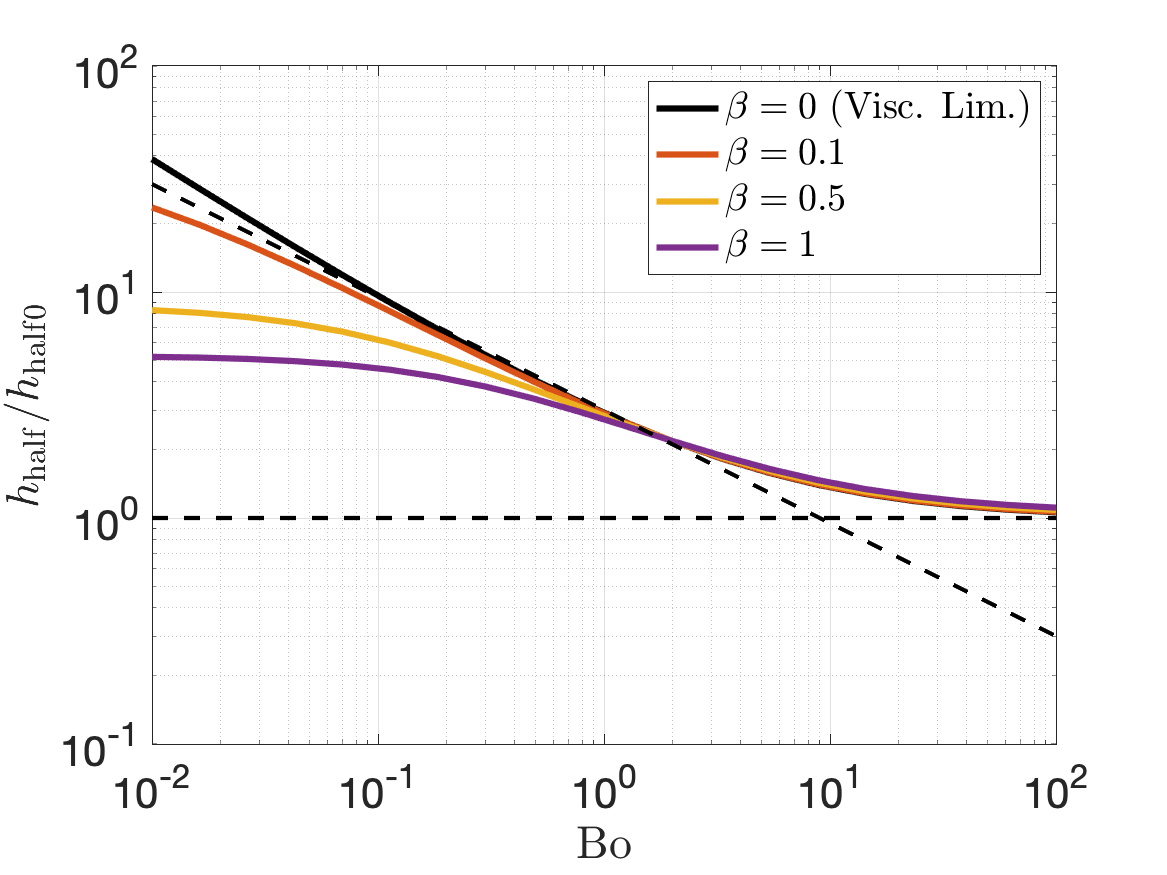}};
\node at (0,0) {\includegraphics[width=0.5\textwidth]{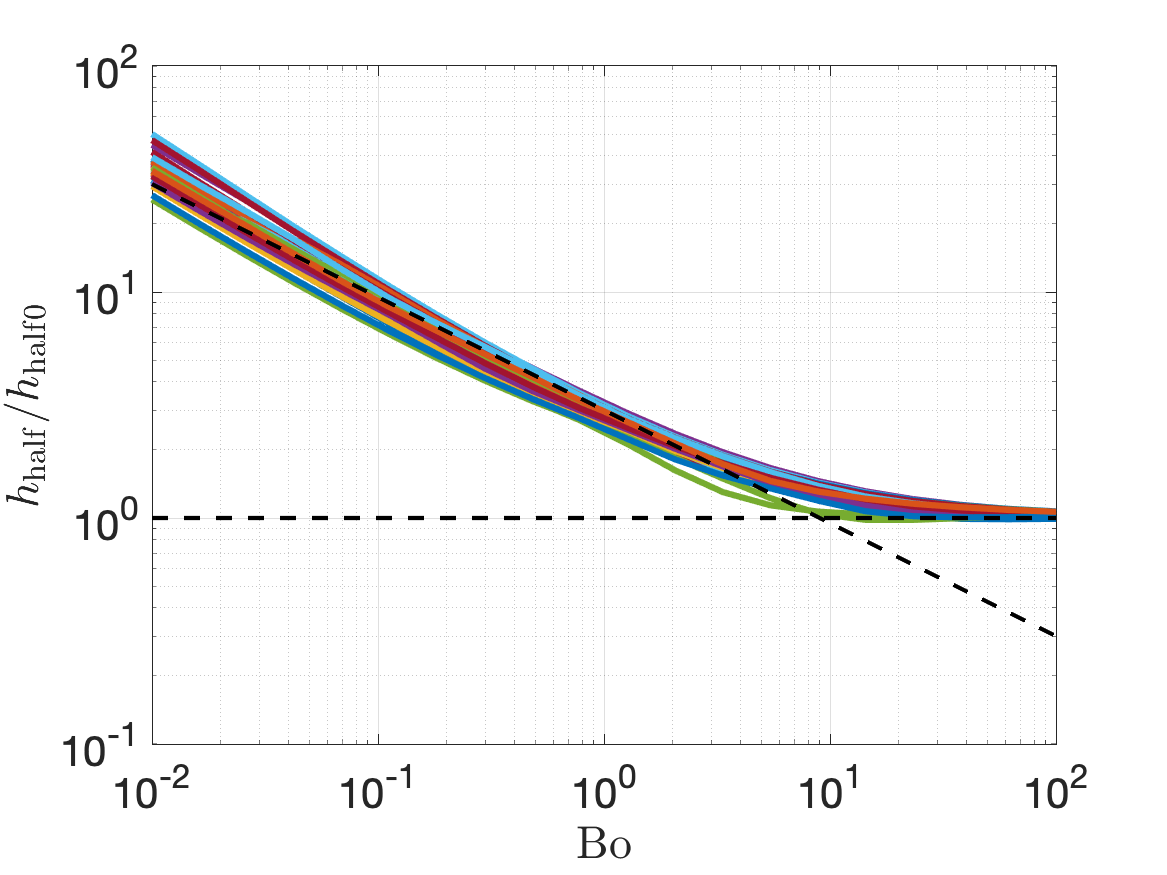}};
\node at (-4.5,3.2) {(a)};
\node at (5,3.2) {(b)};
\node at (9,3.5) {\bf \large Compacted};
\node at (0,3.5) {\bf \large Non-compacted};
\end{tikzpicture}
\caption{Mid-range thickness of the gravity current $h_\mathrm{half}=h(\eta_N/2)$, given in terms of the single phase limit $h_{\mathrm{half}_0}=0.348H$ for the non-compacted cases (a) and the compacted case (b). Limiting behaviours are illustrated with dashed lines. \label{lims2}}
\end{figure}
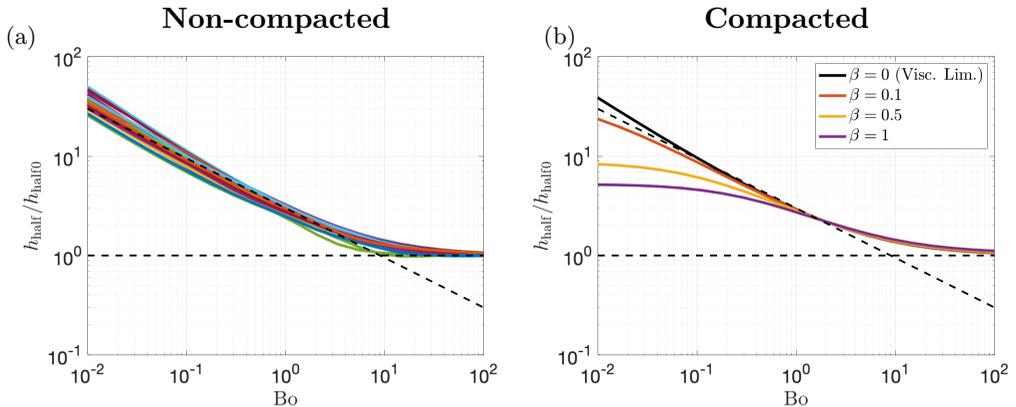

Apart from the shape and the saturation distribution, there are two other important metrics which are useful for describing the current. Firstly, the prefactor $\eta_N$ relates to the speed of the advancing nose, and secondly the mid-range thickness $h_\mathrm{half}=h(r_N(t)/2,t)$ indicates the approximate size of the current. Following \citet{golding2013effects}, we use the classic single phase limit values as a useful reference. Using a subscript notation, these are given by $\eta_{N_0}=1.155$ and $h_{\mathrm{half}_0}=0.348H$ \citep{lyle2005axisymmetric}.
In figures \ref{lims}a,b,c, \ref{lims2}a we plot these quantities for different values of the Bond number. In the limit Bo$\gg1$ all cases converge to the single phase limits, which is expected due to \eqref{sateqdimless}. Likewise, in every case the flux converges to a linear power law, corresponding to a uniform velocity profile, as can be seen in figures \ref{allfluxes0}c and \ref{allfluxes}f.

In the limit Bo$\ll1$ the mid-range thickness $h_\mathrm{half}$ behaves similarly for all three layered cases, growing approximately like $h_\mathrm{half}\sim$Bo$^{-1/2}$, as described by \citet{golding2013effects} for the homogeneous case. On the other hand, the prefactor $\eta_N$ behaves rather differently. 
In all cases, we see an increase in $\eta_N$ for stronger heterogeneities, indicating that capillary forces accelerate the gravity current. However, each heterogeneity affects the prefactor $\eta_N$ differently, as can be seen in the different shaped curves in figure \ref{lims}.
This reflects the complex nature of the velocity distributions and flux functions depicted in figures \ref{allfluxes0} and \ref{allfluxes}.
It is interesting to note that despite having a permeability profile with the same mean value, the different variations within each layer for each heterogeneity type are sufficient to alter the flux of saturation, and hence modify the speed of propagation of the gravity current. This sheds light on both the need for detailed bore hole measurements to infer as much information about the heterogeneities as possible, as well as the usefulness of such an upscaling approach as we have taken here.

For each of the different types of strata, we compare the capillary limit curves against the homogeneous case (solid black line), which is equivalent to the viscous limit. This allows us to quantify the effect of the heterogeneities on the prefactor more clearly. The strongest effect on the prefactor occurs when there are thin regions of high permeability $(H_\mathrm{low}/H_\mathrm{high}=10)$ in which the non-wetting saturation concentrates. This focussing of the saturation feeds into the nonlinearity of the relative permeability, thereby amplifying the effect on the flux function $\mathcal{F}$. By contrast, thin regions of low permeability  $(H_\mathrm{low}/H_\mathrm{high}=0.1)$ produce results which are very close to the homogeneous case, so we do not display them here.

In the case of the permeability spectrum, we choose a permeability distribution whose standard deviation divided by the mean is $\sigma(k)/k_0=1$, and whose mean permeability ratio (between lowest and highest values) is $\mu(k_\mathrm{low}/k_\mathrm{high})=0.04$. We calculate the prefactor $\eta_N$ for 50 different realisations of this distribution and plot the results in figure \ref{lims}c. For each Bond number we display the mean value, as well as one standard deviation on either side $\mu(\eta_N)\pm\sigma(\eta_N)$. The mean result is reminiscent of the previous sedimentary strata cases. However, it is interesting to note that the standard deviation is largest for low-medium Bond number and shrinks as the Bond number gets larger. Hence, predictions are particularly prone to uncertainty if the Bond number is less than order unity. For CO$_2$ sequestration applications, this indicates that particular attention towards measurements of heterogeneities should be paid for injection sites with low flow rates.

Next, we move on to describe the compacted case in the capillary limit (figure \ref{het2}b,d,f). The presence of compaction is most noticeable in the small Bond number cases. By comparing figures \ref{het2}b and \ref{het2}a, we see that compaction significantly increases the saturation within the gravity current, which is due to the permeability gradient forcing the non-wetting phase upwards. 
This is accompanied by an increase in the prefactor $\eta_N$ and a decrease in $h_\mathrm{half}$, as seen in figures \ref{lims}d, \ref{lims2}b. 
This is expected since, if a larger saturation is maintained, volume conservation indicates that the gravity current must be thinner.
By varying the compaction power law $\beta$ from 0 to 1, we illustrate a fairly uniform transition of the values of $\eta_N$ and $h_\mathrm{half}$ between those of a uniform rock and those of a strongly compacted rock.

For $\mathrm{Bo}\gg1$ the saturation becomes near uniform $s\approx1$, as before, and this is accompanied by both $\eta_N$ and $h_\mathrm{half}$ converging to their single phase limits. Hence, at such large Bond numbers the effect of compaction on the saturation distribution and gravity current shape is not particularly noticeable. This is expected, since compaction forces the saturation upwards, just like gravity.

\begin{figure}
\centering
\begin{tikzpicture}[scale=1]
\node at (0.5,-0.2) {\includegraphics[width=0.7\textwidth]{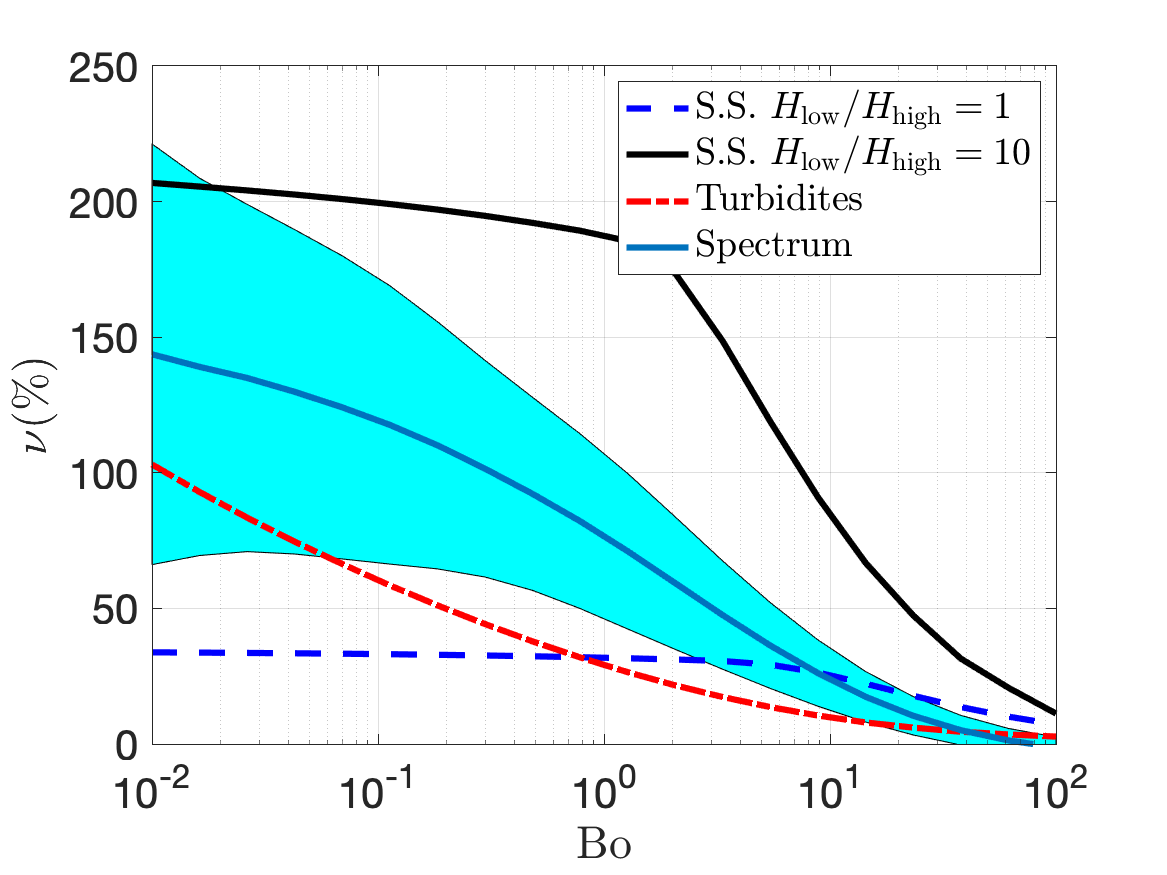}};
\draw[line width=2,cyan,->] (1.2,-1) .. controls (-0.7,0.1) .. (-2,0.5);
\node[cyan] at (0.5,0.3) {\rotatebox{320}{\bf Post-injection}};
\end{tikzpicture}
\caption{Heterogeneity efficiency $\nu$ \eqref{efficdef}, describing the relative increase in prefactor value $\eta_N$ due to heterogeneities, given as a ratio of the prefactor value for the homogeneous case. 
Here we focus on the layered cases (S.S. stands for sedimentary strata), ignoring compaction.
We illustrate how in post-injection scenarios the Bond number typically decreases over time, such that heterogeneities play an increasingly important role. In the case of the permeability spectrum we plot the mean value as well as one standard deviation on either side. The permeability ratio for all cases is $k_\mathrm{low}/k_\mathrm{high}=0.04$. \label{effic}}
\end{figure}

For applications to CO$_2$ storage, it is useful to infer from the above results how much we can expect heterogeneities to affect the speed of propagation of a gravity current.
Such information allows one to make efficiency predictions for CO$_2$ storage that help pinpoint the best sites for injection, as well as safety predictions that ensure the CO$_2$ does not spread beyond the desired perimeter.
Using the homogeneous case, $k_\mathrm{low}/k_\mathrm{high}=1$ (which is equivalent to the viscous limit), as the base case, we define the efficiency parameter $\nu$ as the relative difference we can expect heterogeneities to make on the prefactor $\eta_N$, such that
\beq
\nu(\mathrm{Bo},k(z))= {\eta_{N_\mathrm{het}}}/{\eta_{N_\mathrm{hom}}}-1.\label{efficdef}
\eeq
Clearly $\nu$ depends on a number of parameters, but here we focus on the different types of heterogeneity $k(z)$ and the Bond number. 
Restricting our attention to the layered cases (ignoring compaction), we plot the heterogeneity efficiency $\nu$ in figure \ref{effic} for different values of the Bond number.
The largest heterogeneity efficiency is observed for the sedimentary strata with thin bands of high permeability ($H_\mathrm{low}/H_\mathrm{high}=10$). As we mentioned earlier, this can be explained by a nonlinear focussing of the saturation into these high-speed bands. The most we can expect heterogeneities to accelerate the gravity current at high Bond number is around $\nu=10-50\%$, whereas at low Bond number the speedup can be as much as $\nu=100-200\%$. In the case of the permeability spectrum we also illustrate the standard deviation of the predictions, indicating that the results are particularly sensitive to uncertainty at low Bond number.

It is interesting to note that whilst fluid is injected at constant flow rate $Q$, the Bond number is held constant, but if the flow were to stop suddenly this would no longer be the case. In such situations, the buoyancy length scale $H=\sqrt{Q/u_b}$, which was previously used to define the Bond number, would be rendered meaningless. Instead, the appropriate length scale for the flow would be the gravity current thickness itself $h$ which, after the cessation of $Q$, would gradually decrease towards zero. Hence, the Bond number of the flow would  decrease accordingly, as illustrated in figure \ref{effic}, causing the effect of the heterogeneities to be increasingly amplified with time. 

This is particularly relevant for CO$_2$ storage applications, where the injection of gas is switched off once the aquifer is deemed to have reached maximum safe capacity. Hence, in such situations, it is clear that modelling heterogeneities is essential for understanding the post-injection spread of the CO$_2$. However, it is important to note that such situations involve imbibition flows, as opposed to drainage flows, as we have studied here \citep{woods2015flow}. Imbibition flows typically have different capillary pressure and relative permeability curves than drainage flows, though the approach studied here is still applicable.

In any case, we have shown here that heterogeneities have the potential to significantly alter the growth of the gravity current, and consequently careful upscaled modelling is required. In the next section, we will show that some of these limiting behaviours can be explained using asymptotic analysis.

\subsection{Limiting cases and analytical solutions}\label{limcasesec}

Some simplifications can be made in the limits of strong and weak capillary forces (i.e. small and large Bond number). Here we address these and derive analytical solutions which we use to explain some of our earlier numerical results. We split the analysis into situations without compaction $\beta=0$ and with compaction $\beta>0$. As in the previous section, here we restrict our attention to viscous limit and capillary limit behaviour only.

\subsubsection{Weak capillary forces without compaction}

We already showed earlier that in the limit of large Bond number the saturation distribution \eqref{sateqdimless} is approximately uniform $s\approx 1$. Inserting this into the integrals \eqref{Seqn},\eqref{Feqn1}, we see that
\begin{align}
\mathcal{S}&\approx\varphi h,\label{Sweak}\\
\mathcal{K}&\approx k_0 k_{rn0} h,
\end{align}
which allows us to calculate the dimensionless flux 
\beq
\hat{\mathcal{F}}=\hat{f}.\label{simgov2}
\eeq
This linear power law matches with our numerical observations in figure \ref{allfluxes0}c. Comparing \eqref{Sweak},\eqref{simgov2} with the study by \citet{lyle2005axisymmetric}, we see that the limit of large Bond number is identical to the classic single phase limit. 
This is of course expected, since in the limit of weak capillary forces the flow of phases decouples, such that a single phase model becomes appropriate.
This explains the convergence behaviour for both $\eta_N$ and $h_\mathrm{half}$ for Bo$\gg1$, as we illustrate with dashed lines in figures \ref{lims} and \ref{lims2}.

\subsubsection{Strong capillary forces without compaction}

To address the limit of strong capillary forces, we first consider the homogeneous case $k_\mathrm{low}/k_\mathrm{high}=1$ (equivalent to the viscous limit) since this makes the analysis for the subsequent heterogeneous cases easier. Hence, in the limit of small $\mathrm{Bo}\ll1$ in the homogeneous case, the saturation distribution \eqref{sateq2} approximates to a linear function
\beq
s\approx\lambda \mathrm{Bo} \frac{(h-z)}{H}.\label{seasy}
\eeq
Inserting this into the integrals \eqref{Seqn},\eqref{Feqn1} we get
\begin{align}
\mathcal{S}&\approx\frac{\varphi \lambda  \mathrm{Bo} }{2H}h^2\label{SVCG},\\
\mathcal{K}&\approx \frac{k_0k_{rn0}H\lambda^\alpha\mathrm{Bo}^\alpha }{\alpha+1}\lb\frac{h}{H}\rb^{\alpha+1},
\end{align}
from which we calculate the dimensionless flux
\beq
\hat{\mathcal{F}}=\left[\frac{2^{\alpha/2}(\lambda \mathrm{Bo})^{\alpha/2-1}}{\alpha+1}\right]\hat{f}^{\alpha/2} .\label{simgov3}
\eeq
In this case, the flux has an $\alpha/2$ power law, which matches with our numerical calculations in figure \ref{allfluxes0}a (for which $\alpha=4$). The gravity current thickness is given by inverting \eqref{SVCG}, such that
\beq
h= H\lb{2\hat{f}}/{\lambda\mathrm{Bo}}\rb^{1/2}.\label{hinvert}
\eeq
Clearly, the thickness grows like $h\sim\mathrm{Bo}^{-1/2}$ as Bo$\rightarrow0$, which we illustrate in figure \ref{lims2} with dashed lines. Our numerical results show good agreement, indicating their robustness. 
We also note that the square root in \eqref{hinvert} explains why we see a blunting of the gravity current nose at low Bond number in figures \ref{het1},\ref{het2}.

If we now consider a finite heterogeneity $k_\mathrm{low}/k_\mathrm{high}<1$, then in the case of sedimentary strata, the saturation distribution takes one of two possible values
\beq
s\approx \begin{cases} 0:&k=k_\mathrm{low}\\
\lambda \mathrm{Bo} {(h-z)}/{H}:&k=k_\mathrm{high}
\end{cases}.\label{stricky}
\eeq
Since the low/high permeability layers are distributed according to the ratio $H_\mathrm{low}/H_\mathrm{high}$, the integrals \eqref{Seqn},\eqref{Feqn1} approximate to
\begin{align}
\mathcal{S}&\approx\lb\frac{H_\mathrm{high}}{H_\mathrm{high}+{H_\mathrm{low}}}\rb\lb\frac{k_\mathrm{high}}{k_0}\rb^a\frac{\varphi \lambda  \mathrm{Bo} }{2H}h^2,\label{SCCG}\\
\mathcal{K}&\approx \lb\frac{H_\mathrm{high}}{H_\mathrm{high}+{H_\mathrm{low}}}\rb\frac{k_\mathrm{high}k_{rn0}H\lambda^\alpha\mathrm{Bo}^\alpha }{\alpha+1}\lb\frac{h}{H}\rb^{\alpha+1}.\label{linearF2}
\end{align}
Hence, the dimensionless flux is
\beq
\hat{\mathcal{F}}=\left[\lb\frac{H_\mathrm{high}}{H_\mathrm{high}+{H_\mathrm{low}}}\rb^{\alpha/2}\lb\frac{k_\mathrm{high}}{k_0}\rb^{1-a(1+\alpha/2)}\frac{2^{\alpha/2}(\lambda \mathrm{Bo})^{\alpha/2-1}}{\alpha+1} \right]\hat{f}^{\alpha/2}.\label{simgov4}
\eeq
Like the homogeneous case, the heterogeneous flux has an $\alpha/2$ power law, which also matches with our numerical calculations for sedimentary strata in figure \ref{allfluxes0}a (for which $\alpha=4$). 
We note that in the above analysis, the saturation approximation \eqref{stricky}, and consequently the flux \eqref{simgov4}, are only valid for heterogeneities which have a small enough permeability ratio that $(1-k_\mathrm{low}/k_\mathrm{high})/(1+k_\mathrm{low}/k_\mathrm{high})\gg\mathrm{Bo}$. However, for this study we restrict our attention to significantly heterogeneous media (rather than weakly heterogeneous).

We also note that in both the homogeneous case and the heterogeneous case, the coefficients in \eqref{simgov3},\eqref{simgov4} depend on the Bond number itself. Therefore, we do not expect a constant value asymptote for the prefactor $\eta_N$ in the limit Bo$\rightarrow 0$ (except in the specific case $\alpha=2$), which is consistent with our numerical observations in figure \ref{lims}.

In the case of the permeability spectrum, a similar analysis is possible since the only contribution to the integrals will come from the regions with the largest permeability value $k=k_\mathrm{high}$. However, since there are potentially many more than just two permeability values in the spectrum, one needs to replace the factor $H_\mathrm{high}/(H_\mathrm{high}+{H_\mathrm{low}})$ in \eqref{SCCG}-\eqref{simgov4} with the fraction of the aquifer occupied by such high permeability layers, which we denote $H_\mathrm{high}/H_\mathrm{total}$. In the case of the turbidites, a similar analysis to the above is much more difficult, since we cannot approximate the saturation distribution as simply as \eqref{stricky}.

\subsubsection{Weak capillary forces with compaction}

In the case where the rock is compacted, the saturation \eqref{sateqdimless} must be approximately uniform $s\approx 1$ in the limit of large Bond number, as before. Inserting this into the integrals \eqref{Seqn},\eqref{Feqn1}, we see that
\begin{align}
\mathcal{S}&\approx\varphi H \frac{(1 + h/H)^{1-a \beta}  - 1}{1 - a \beta},\label{Scompact1}\\
\mathcal{K}&\approx k_0 k_{rn0} H\frac{(1 + h/H)^{1- \beta}  - 1}{1 -  \beta}.\label{Kcompact1}
\end{align}
Note that if either $a$ or $a\beta$ equals unity, we get analytical expressions with logarithms instead of \eqref{Scompact1},\eqref{Kcompact1}. Assuming $\beta\neq1$, $a\beta\neq 1$, we then calculate the dimensionless flux, which we write in terms of $h$ for now:
\beq
\hat{\mathcal{F}}=\frac{(1 + h/H)^{\beta(a-1)+1} -(1+h/H)^{ a\beta}}{1 -  \beta}.\label{fluxcompact1}
\eeq
Clearly the flux \eqref{fluxcompact1} is not a linear function of $\mathcal{S}$ \eqref{Scompact1} as in the homogeneous case. However, we note that for weak porosity-permeability power laws $a\ll1$ (e.g. $a=0.14$ for the Salt Creek case study \citep{bickle2017rapid}) the integrated saturation \eqref{Scompact1} approximates to a linear function $\mathcal{S}\approx \varphi h$, as we observe in our numerical calculations in figure \ref{allfluxes}e. Furthermore, in situations where the compaction law is weak $\beta\ll1$, or alternatively where the gravity current thickness is small ${h}/H\ll1$, the flux \eqref{fluxcompact1} reduces to $\hat{\mathcal{F}}\approx {h}/H$, thereby recovering the single phase limit \eqref{simgov2}. This is in accordance with our numerical observations in figure \ref{allfluxes0}c.

However, we note that by choosing sufficiently strong compaction/porosity power laws $a,\beta$, the single phase limit is no longer recovered in the limit of large Bond number.
Physically speaking, this is because even with weak capillary forces, if the medium is sufficiently compacted the velocity distribution within the gravity current becomes dominated by the permeability variation, which upon integration creates a flux power law that is not equal to one.

\subsubsection{Strong capillary forces with compaction}

To address the compacted case at small Bond number, we first consider the form of the saturation \eqref{sateq}. In particular, we note that since $k$ is monotone decreasing with depth, the saturation $s$ is also monotone decreasing. In particular, we can calculate that $s$ will intercept zero at some depth $z^*$ which satisfies $\Delta\rho g (h-z^*)+p_0=p_e(z^*)$, and will remain zero for all larger depths than this $z>z^*$. Therefore, without loss of generality, we take $p_0=p_e(h)$, such that there are no regions of zero saturation within the gravity current. In this way, for small Bond number \eqref{sateq} approximates to a finite expression which is independent of Bo,
\beq
s\approx1-\lb\frac{1+z}{1+h}\rb^{\lambda \beta b  }.\label{seasycomp}
\eeq
This explains why the properties of the gravity current (e.g. $\eta_N,h_\mathrm{half}$) asymptote to constant values for Bo$\rightarrow 0$ in figures \ref{lims}d,\ref{lims2}b. 
For general power laws $a,b,\lambda,\alpha,\beta$, inserting \eqref{seasycomp}  into the integrals \eqref{Seqn},\eqref{Feqn1} leads to very complicated analytical expressions, which we do not display here. As we can see in figure \ref{allfluxes0}a, the flux for this case does not obey a fixed power law, but the exponent varies roughly between 1 and 2.

\subsection{Viscous-capillary transition }\label{transsec}

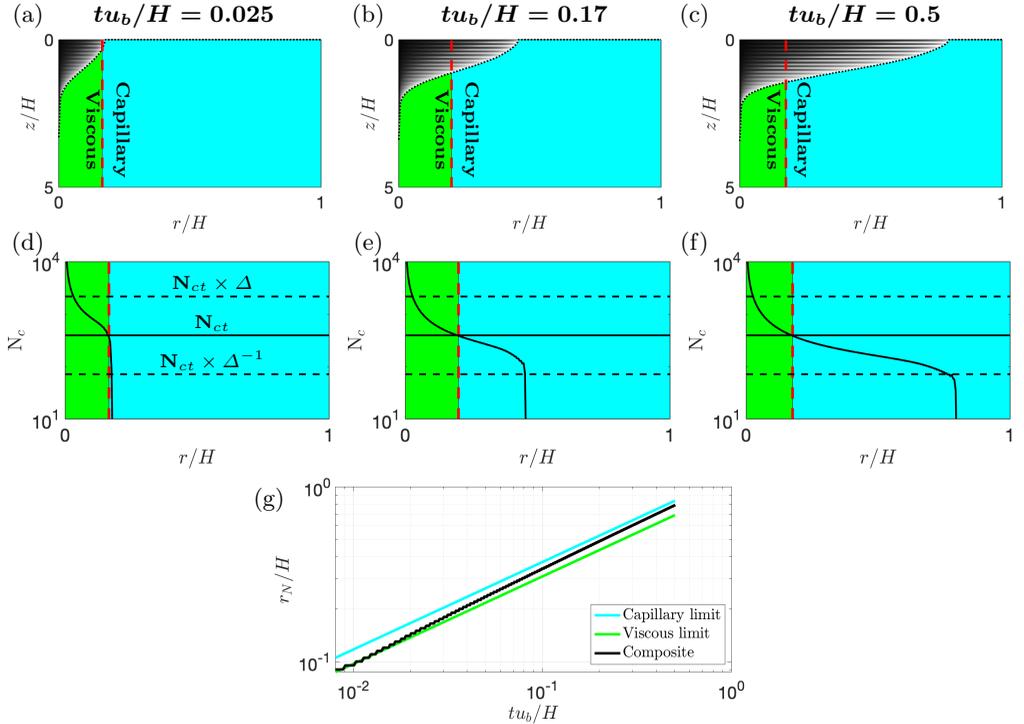
\begin{figure}
\centering
\begin{tikzpicture}[scale=0.75]
\node at (0,0) {\includegraphics[width=0.33\textwidth]{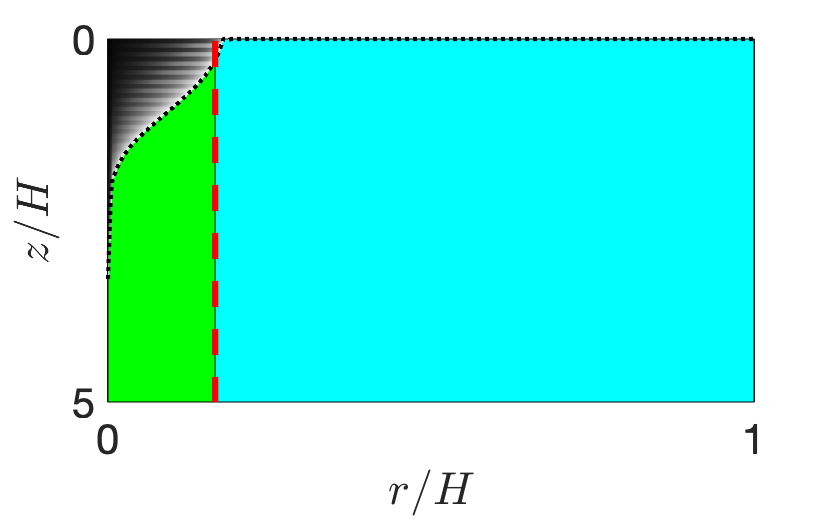}};
\node at (6,0) {\includegraphics[width=0.33\textwidth]{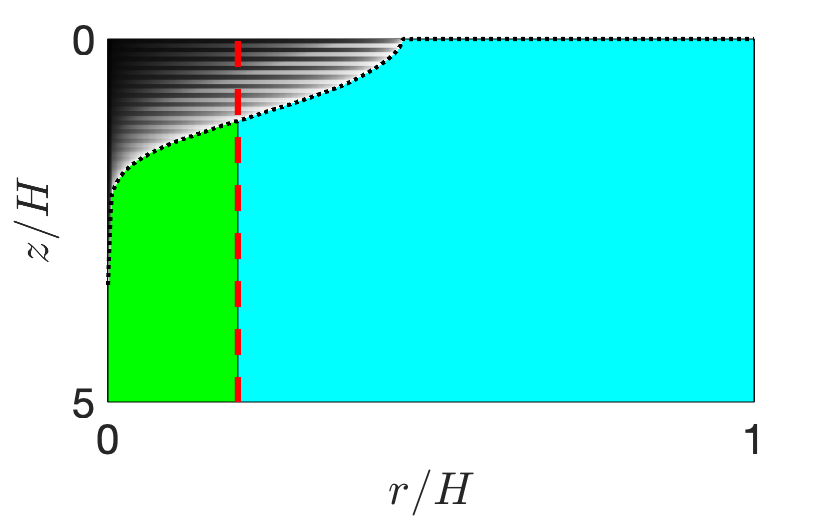}};
\node at (12,0) {\includegraphics[width=0.33\textwidth]{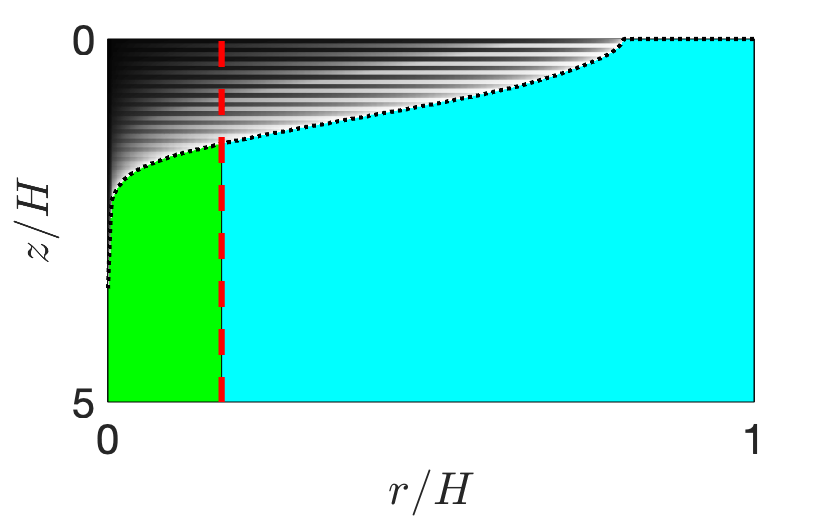}};
\node at (0,-4) {\includegraphics[width=0.35\textwidth]{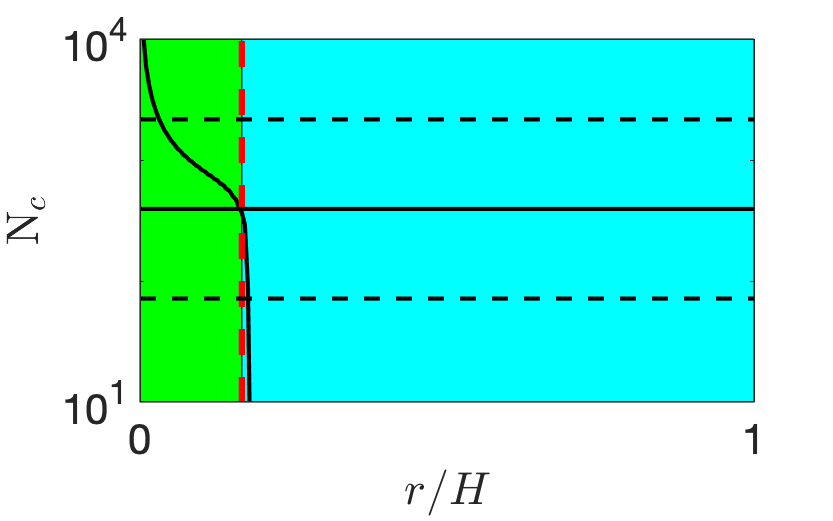}};
\node at (6,-4) {\includegraphics[width=0.35\textwidth]{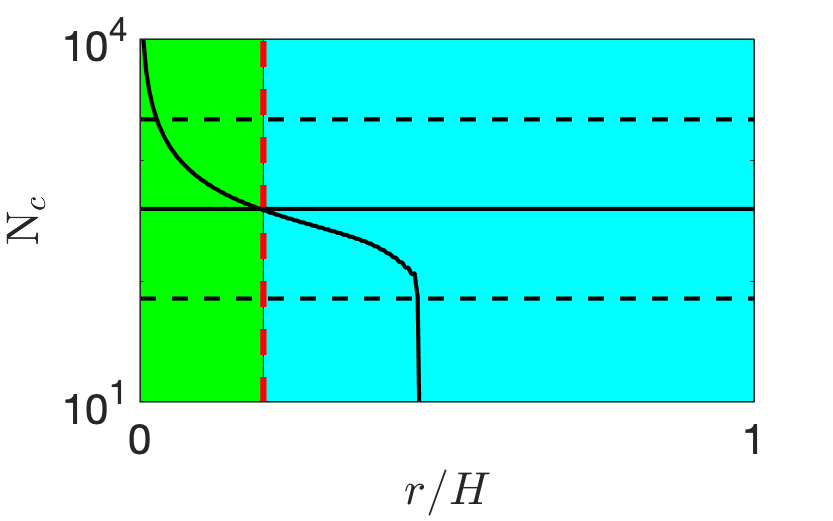}};
\node at (12,-4) {\includegraphics[width=0.35\textwidth]{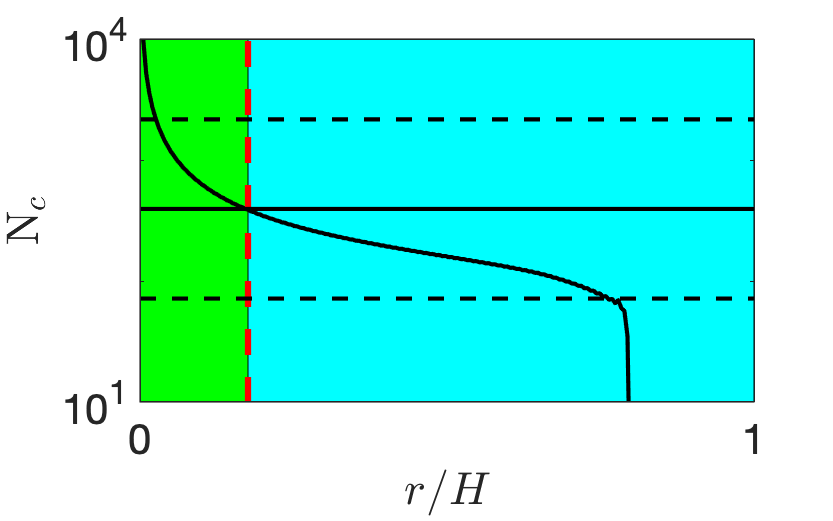}};
\node at (-2.75,2) {(a)};
\node at (3.25,2) {(b)};
\node at (9,2) {(c)};
\node at (-2.75,-2) {(d)};
\node at (3.25,-2) {(e)};
\node at (9,-2) {(f)};
\node at (0,2) {$\boldsymbol{tu_b/H=0.025}$};
\node at (6,2) {$\boldsymbol{tu_b/H=0.17}$};
\node at (12,2) {$\boldsymbol{tu_b/H=0.5}$};
\node at (-1.6,0) {\rotatebox{270}{\bf \scriptsize Viscous}};
\node at (-1.1,0) {\rotatebox{270}{\bf  \scriptsize Capillary}};
\node at (4.55,0) {\rotatebox{270}{\bf \scriptsize Viscous}};
\node at (5.05,0) {\rotatebox{270}{\bf  \scriptsize Capillary}};
\node at (10.4,0) {\rotatebox{270}{\bf \scriptsize Viscous}};
\node at (10.9,0) {\rotatebox{270}{\bf  \scriptsize Capillary}};
\node at (0.5,-2.7) {\bf \scriptsize N$_{ct}\times \Delta$};
\node at (0.5,-3.4) {\bf \scriptsize N$_{ct} $};
\node at (0.5,-4.05) {\bf \scriptsize N$_{ct}\times \Delta^{-1}$};
\node at (6,-8.2)  {\includegraphics[width=0.5\textwidth]{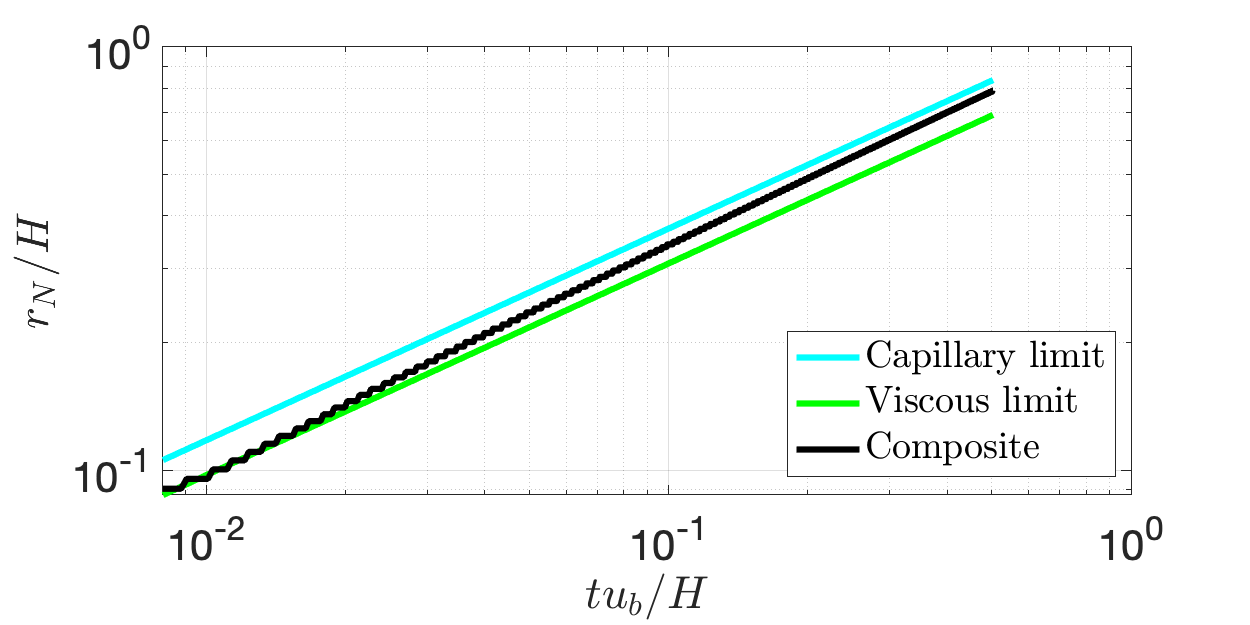}};
\node at (1.5,-6.5) {(g)};
\end{tikzpicture}
\caption{Numerical solution of the evolution of the gravity current \eqref{Geqn}, accounting for transition behaviour between viscous and capillary limits using composite expressions \eqref{transfun} for the upscaled flow properties, where the capillary number is given implicitly by \eqref{capdef3} (with Bo$=1$). The gravity current shape, shaded to illustrate the saturation distribution (using the same colour scale as in figures \ref{het1},\ref{het2}), is illustrated in (a,b,c), whereas the local capillary number N$_c$ is illustrated in (d,e,f). For all plots we shade regions with capillary number larger than the transition value N$_c>$N$_{c_t}=394$ in green, and regions with N$_c<$N$_{c_t}$ in blue. We also illustrate one folding scale $\Delta$ on either side of N$_{c_t}$ with dashed lines in (d,e,f). The evolution of the gravity current nose position $r_N(t)$ is shown on a log-log plot in (g). \label{transplots}}
\end{figure}

Up until now we have only discussed situations where the capillary number \eqref{capdef2} is either very large (viscous limit), in which capillary forces due to heterogeneities can be ignored (i.e. effectively the homogeneous case), or very small (capillary limit), in which the heterogeneities play a dominant role on the flow behaviour (i.e. the heterogeneous cases studied above). However, in general the capillary number may vary between small and large values throughout the aquifer. In this case, neither the viscous nor the capillary limit can be applied to the flux function in \eqref{Geqn}, and instead the flux must depend on the local capillary number, which is effectively a measure of the local pressure gradients within the gravity current. In this section, we discuss how to model this using numerical simulations of the full PDE \eqref{Geqn} coupled to the transcendental equation \eqref{capdef2}, thereby determining which regions of the gravity current are within the viscous and capillary limits, and which regions lie in between these limits.

Following the same approach as \citet{benham2020upscaling}, we formulate composite functions for the upscaled properties of the flow, which capture both the viscous and capillary limit regimes, as well as the transition between these limits. The two upscaled quantities of interest are the integrated saturation $\mathcal{S}$ and the flux $\mathcal{F}$. For each of these upscaled quantities $\left\{\mathcal{S},\mathcal{F}\right\}$, the transition behaviour is given in terms of the mean saturation \eqref{meansatdef} and capillary number \eqref{capdef2} by the formula
\beq
\left\{\mathcal{S},\mathcal{F}\right\}_{\mathrm{trans}}=\frac{1}{2}\left[ \left\{\mathcal{S}(\bar{s}),\mathcal{F}(\bar{s})\right\}_{-} \tanh \lb\frac{\log{\mathrm{N}_c}-\log{\mathrm{N}_{c_t}}}{\log \Delta}\rb+\left\{\mathcal{S}(\bar{s}),\mathcal{F}(\bar{s})\right\}_{+}\right],\label{transfun}
\eeq
where $\left\{\mathcal{S},\mathcal{F}\right\}_{\pm}=\left\{\mathcal{S},\mathcal{F}\right\}_{\mathrm{visc}}\pm \left\{\mathcal{S},\mathcal{F}\right\}_{\mathrm{cap}}$ is given in terms of the viscous limit (homogeneous) and capillary limit (heterogeneous) upscaled properties derived earlier, and N$_{c_t}$, $\Delta$, are two fitting parameters which represent the transition capillary number and folding scale, respectively. The precise values of these fitting parameters depend on numerous factors, including the type of heterogeneity and the specific power laws used, but here we use the same values as calculated by \citet{benham2020upscaling}, which are N$_{c_t}= 394$, $\Delta=5.5$, and these were shown to give good comparison with numerical simulations across a large range of capillary numbers with mean relative error of $\sim1\%$.

By accounting for the dependence of the upscaled flux on the capillary number, the system loses its self-similarity. In mathematical terms, this can be seen by noticing that the capillary number \eqref{capdef2} in the governing equation \eqref{Geqn} contains derivatives of $\mathcal{S}$ with respect to $r$, compromising the self-similar structure of the equations. Therefore, since we are no longer able to convert to a single governing similarity ODE, we must instead solve the full PDE \eqref{Geqn} numerically, in conjunction with the algebraic equation for the capillary number \eqref{capdef2}.  
Written in dimensionless terms, we observe that the capillary number is related to the Bond number according to
\beq
\mathrm{N}_c=\frac{\mathrm{Bo}p_0}{\Delta p_e}\left|\frac{h(\mathcal{S},\mathrm{N}_c)/H}{\mathcal{S}_h(h(\mathcal{S},\mathrm{N}_c),\mathrm{N}_c)}\frac{\partial \mathcal{S}}{\partial r}\right|.\label{capdef3}
\eeq
For the sake of the calculations in this section, we use a mid-range value of Bo$=1$ for the Bond number, since this represents an even balance between capillary and gravity forces.

In figure \ref{transplots} we display the numerical solution in the case of sedimentary strata ($k_\mathrm{low}/k_\mathrm{high}=1/3,H_\mathrm{low}/H_\mathrm{high}=1$), plotted at three different times after injection, illustrating the evolution of the gravity current, as well as the spatial variation in the local capillary number \eqref{capdef3}. At early times the majority of the gravity current lies within the viscous limit, except the very tip of the gravity current nose. Indeed, the nose of the gravity current lies in the capillary limit for all times, since the flux vanishes there (since N$_c\sim h\partial p_n/\partial r\sim h\partial h/\partial r\rightarrow 0$ as $r\rightarrow r_N(t)$ due to \eqref{zerofluxbc}). At later times, a substantial portion of the gravity current lies within the capillary limit, whilst the viscous-capillary transition (N$_c=$N$_{c_t}$) tends to a constant position of ${r}/H\approx 0.2$. The position of the nose of the gravity current $r_N(t)$ is plotted on a logarithmic scale in figure \ref{transplots}g, also indicating the viscous and capillary limit curves.

We make several observations. 
Firstly, we note the dynamic transition between viscous-like behaviour at early times and capillary like-behaviour at late times. This is evident from figure \ref{transplots}g, where we see that the nose position initially grows like $\sim t^{1/2}$ with prefactor corresponding to the viscous limit, and later grows like $\sim t^{1/2}$ with a capillary limit prefactor. 
There is a transition period at intermediate times where the behaviour does not obey a power law, as observed by \citet{benham2020upscaling} in the absence of gravity. 
This can be explained by the fact that, as the gravity current spreads out radially, the capillary number reduces everywhere except near the origin (where N$_c\rightarrow\infty$).
In this way, the composite expressions \eqref{transfun} in the majority of the gravity current switch between viscous limit behaviour to capillary limit behaviour over time.
Secondly, we note that since the viscous-capillary transition (N$_c=$N$_{c_t}$) tends to a constant position, if we continue the simulation indefinitely, the fraction of the gravity current in the capillary limit tends towards unity, indicating that at late times a capillary limit model is a good approximation for the whole aquifer.

\section{Discussion of applications to CO$_2$ sequestration}
\label{sleipsec}

In this section we discuss the implications of the upscaled description of gravity currents in the context of CO$_2$ sequestration. There are many different sites that we could choose as case studies, but probably the one with the most available data is at the Sleipner project in the North Sea \citep{bickle2007modelling}. In particular, we focus on the top layer, which has been investigated by numerous authors \citep{chadwick2009flow,williams2017improved}. Whilst various attempts have been made to model the CO$_2$ migration at Sleipner, and comparisons have been made with seismic measurements of the extent of the plume \citep{bickle2007modelling,golding2013effects,cowton2016inverse,cowton2018benchmarking}, 
these attempts have often fitted certain parameters (e.g. the permeability) to match observations, without accounting for the possible effect of heterogeneities. Therefore, here we use our previous upscaling procedure to perform this analysis, calculating the effect that different types of heterogeneities could have had on the speed of the gravity current. Since little is known about the geological heterogeneities at Sleipner, we investigate all the different types of layering we studied earlier, and we account for how uncertainty in such layering may propagate to uncertainty in the modelling predictions.

As described earlier in Section \ref{sechet}, there are 8 non-dimensional parameters in our model (excluding the capillary number). Of these parameters, 5 relate to type of heterogeneities, about which little is known for Sleipner. Therefore, instead we take the following values from other similar studies: the porosity-permeability power law used by \citet{bickle2017rapid} $a=0.14$;  the pore-entry pressure power law taken from the scaling proposed by \cite{leverett1941capillary} $b=0.43$; the permeability ratio taken from \citet{bickle2017rapid} $k_\mathrm{low}/k_\mathrm{high}=0.01$, no compaction $\beta=0$, and we vary the layer ratio $H_\mathrm{low}/H_\mathrm{high}$ between $1$ and $10$. Since it is unknown whether sedimentary strata, turbidites or a permeability spectrum is most appropriate, we investigate all of these different heterogeneity types.

There are 2 parameters relating to the multiphase flow properties $\lambda$ and $\alpha$, which define the capillary pressure and relative permeability. We note that not all relative permeability curves are parameterised as simply as \eqref{krneq}. For example, other more nonlinear functional forms have been proposed by \citet{chierici1984novel}. However, we note that \cite{chadwick2009flow,williams2017improved} use the \textit{Brooks-Corey} formulation to model Sleipner, which is given by
\beq
k_{rn}=k_{rn0}s^2\left[1-(1-s)^2\right].\label{altkrn}
\eeq
We could change our formulation to account for this modified parameterisation, but instead we notice that \eqref{altkrn} can be approximated by \eqref{krneq} with mean relative error $3\%$ using a power law value of $\alpha=2.32$. Therefore, we stick with our original formulation \eqref{krneq} for the sake of consistency, without sacrificing much accuracy. Meanwhile, the pore size distribution is estimated by \citet{chadwick2009flow} as $\lambda=2/3$.

The final parameter we need to describe the problem is the Bond number, which is defined by \eqref{Bodef} in terms of 6 other physical parameters (excluding gravity, $g=9.81$ m/s$^2$). 
For the top layer at Sleipner, \citet{williams2017improved} estimate the temperature between $28$-$31^\circ$C and pressure between $8.2$-$8.9$ MPa, which gives a density difference of $\Delta \rho=232$-$309$ kg/m$^3$. 
Meanwhile the viscosity of CO$_2$ is taken as $\mu_n=54.7$-$65.5\times10^{-6}$ Pa$\cdot$s.
The input flux is best estimated by \citet{cowton2016inverse}, which for the first few years of injection is $Q=1.5$-$3\times 10^{-3}$ m$^3$/s. 
The mean permeability is estimated by \citet{bickle2007modelling} as $k_0=1.1-5\times 10^{-12}$ m$^2$.
Finally, the pore entry pressure and end-point relative permeability are given by \citet{williams2017improved} as $p_0=1.3$ kPa and $k_{rn0}=0.28$. Putting these all together, we estimate the Bond number as Bo$=8.9-35.5$.

Now that we have determined all the parameter values (up to a given uncertainty), we follow the procedure outlined earlier to calculate the prefactor $\eta_N$  for the gravity current using the various different heterogeneity types. In this way, we can measure the heterogeneity efficiency $\nu$ \eqref{efficdef}, which tells us how much we can expect the heterogeneities to modify the speed of propagation. 
In the low/high Bond number estimate Bo=$8.9/35.5$, we find $\nu=36\%/31\%$ for equally distributed sedimentary strata ($H_\mathrm{low}/H_\mathrm{high}=1$),  $\nu=147\%/108\%$ for sedimentary strata with thin streaks of high permeability ($H_\mathrm{low}/H_\mathrm{high}=10$), $\nu=11\%/6\%$ for turbidites, and $\nu=23\pm11\%/9\pm8\%$ for a permeability spectrum (mean value). In the latter case, we include the standard deviation values to indicate how these predictions vary due to the uncertainty of the heterogeneity measurements. Indeed, the large degree of uncertainty in these predictions not only illustrates the need for more detailed bore hole measurements at Sleipner, but also demonstrates the importance of accounting for such uncertainty in any modelling approach.

We note that the permeability ratio in the Sleipner field may not be as small as the value we have taken from Salt Creek, $k_\mathrm{low}/k_\mathrm{high}=0.01$ \citep{bickle2017rapid}. Therefore we also perform the same analysis as above using a permeability ratio ten times larger. We find that the efficiencies $\nu$ calculated for $k_\mathrm{low}/k_\mathrm{high}=0.1$ are between $1/5$ and $3/5$ of their  values for $k_\mathrm{low}/k_\mathrm{high}=0.01$. This indicates that, even if we have vastly overestimated the permeability ratio, the effect of heterogeneities is still likely to be significant.

From this analysis, it is clear that the possible effect that heterogeneities may have had on the injection of CO$_2$ at Sleipner largely depends on the type of heterogeneities present. In particular, thin sedimentary strata with very high permeability could have caused a potential speedup of more than $100\%$. 
However, for more moderate permeability profiles, such as evenly distributed strata or turbidites, these heterogeneities may have only caused a $10-30\%$ speedup.

\section{Concluding remarks}\label{concsec}

We have studied the upscaled effect of several different types of heterogeneity on the evolution of an axisymmetric gravity current under an impermeable cap rock. The four heterogeneity types considered, which were all given in terms of vertical variations in the rock properties, were each motivated by different physical mechanisms for the non-uniform deposition or compaction of sediments. 
We developed a general method for calculating the gravity current shape and growth rate in either the viscous or capillary limits, which involves solving a single ordinary differential equation that depends on an upscaled flux term evaluated either via numerical integration, or using analytical expressions which we derived in certain limiting cases. This simplified approach not only reduces the computation time significantly, but also provides key insights into the role of small-scale heterogeneities on the propagation of the large-scale flow.

In particular, we showed that heterogeneities have the ability to speed up the growth rate of the gravity current by means of a nonlinear focussing mechanism (via the relative permeability and capillary pressure) into high permeability layers. The degree of speedup depends on the type of heterogeneity, and most importantly the fraction of high/low permeability regions within the medium. The largest effect was seen in the case of sedimentary strata with thin regions of high permeability. 
Using a permeability profile composed of randomly sampled layers, we demonstrated how uncertainty in heterogeneity measurements can propagate to uncertainty in field predictions, an effect which is particularly pronounced for small values of the Bond number.
We also investigated modelling the transition from the viscous limit regime to the capillary limit regime, shedding light into which regions of the gravity current are most affected by heterogeneities, and when.

The main motivation for this study was to create an upscaled description of how small-scale heterogeneities affect large-scale CO$_2$ migration, for safe and efficient sequestration in porous aquifers.
To assess the risks associated with any CO$_2$ storage scheme, examining the effect of heterogeneities is essential. 
To illustrate this, we used the case study of CO$_2$ injection at the Sleipner project in the North Sea. In this case, for realistic parameter values, we demonstrated that heterogeneities may have sped up the gravity current by more than 100$\%$ during injection. However, we also illustrated that this figure depends greatly on the heterogeneity type, indicating the need for detailed core measurements from bore holes. 

For future work, variations in the heterogeneities along the length of the aquifer could be studied (in addition to the vertical heterogeneities investigated here), similarly to \citet{jackson2020small}. In such cases, the upscaled flow properties would depend on both the horizontal and vertical correlation length scales of the heterogeneities. In addition, using the upscaled results presented here, predictions of trapping efficiencies could be calculated for various sequestration sites. This could shed light onto which aquifers have heterogeneities that could potentially enhance their capability for CO$_2$ storage.
Furthermore, as we discussed earlier, it would be interesting to investigate the evolution of the gravity current after injection has stopped (see figure \ref{effic}). However, in this case both imbibition and drainage relative permeability/capillary pressure curves must be considered, depending on whether the CO$_2$ is  invading or being withdrawn from different regions of the aquifer, similarly to \citet{golding2017two}.

\acknowledgements{
This research is funded in part by the GeoCquest consortium, a BHP-supported collaborative project between Cambridge, Stanford and Melbourne Universities, and by a NERC consortium grant ``Migration of CO$_2$ through North Sea Geological Carbon Storage Sites'' (grant no. NE/N016084/1).
}\\

Declaration of Interests. The authors report no conflict of interest.\\

Code written to numerically evaluate the flux integrals and calculate the similarity  solution can be found on the
personal website of G.P. Benham: \url{https://yakari.polytechnique.fr/people/benham/gravity_current/upscale.m}

\appendix
\section{Extra plots}\label{appA}

\begin{figure}
\centering
\begin{tikzpicture}[scale=0.8]
\node at (0,0) {\includegraphics[width=0.48\textwidth]{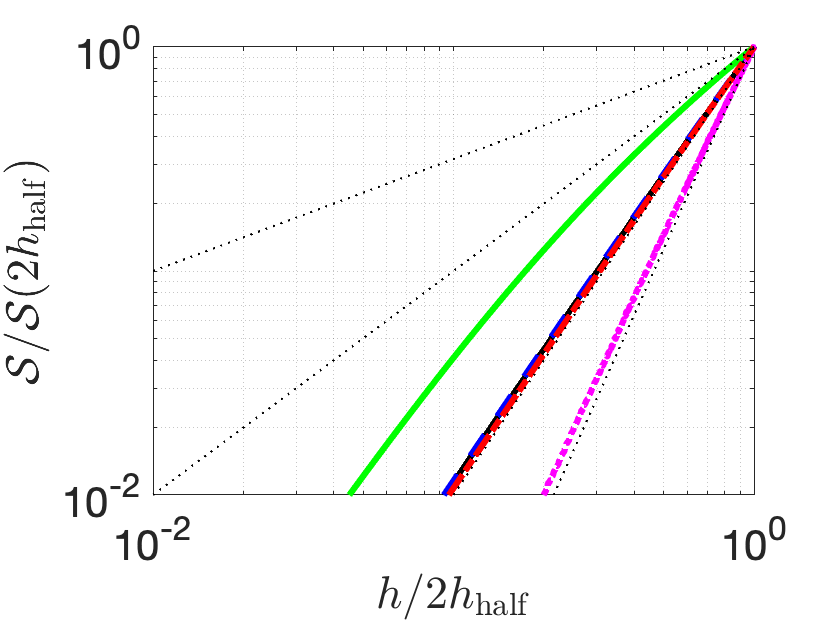}};
\node at (0,-6) {\includegraphics[width=0.48\textwidth]{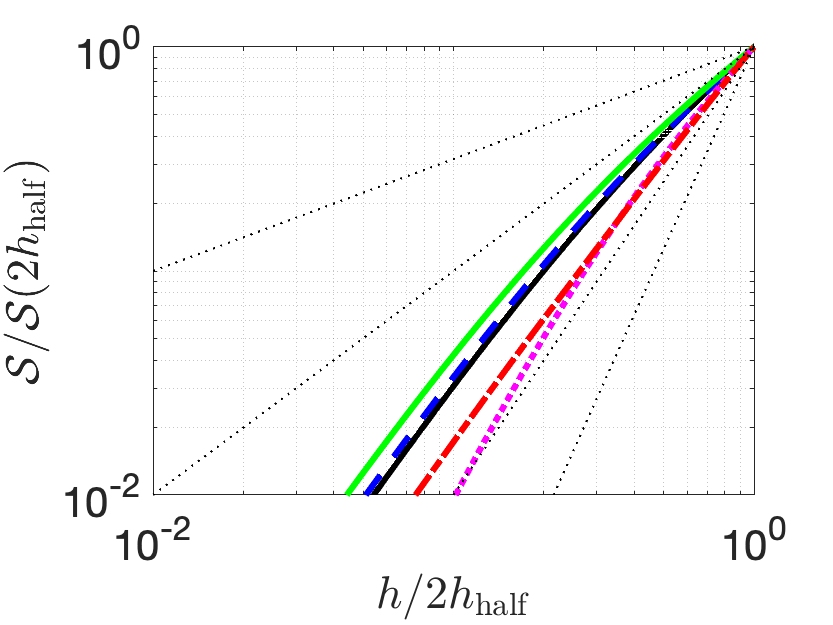}};
\node at (0,-12) {\includegraphics[width=0.48\textwidth]{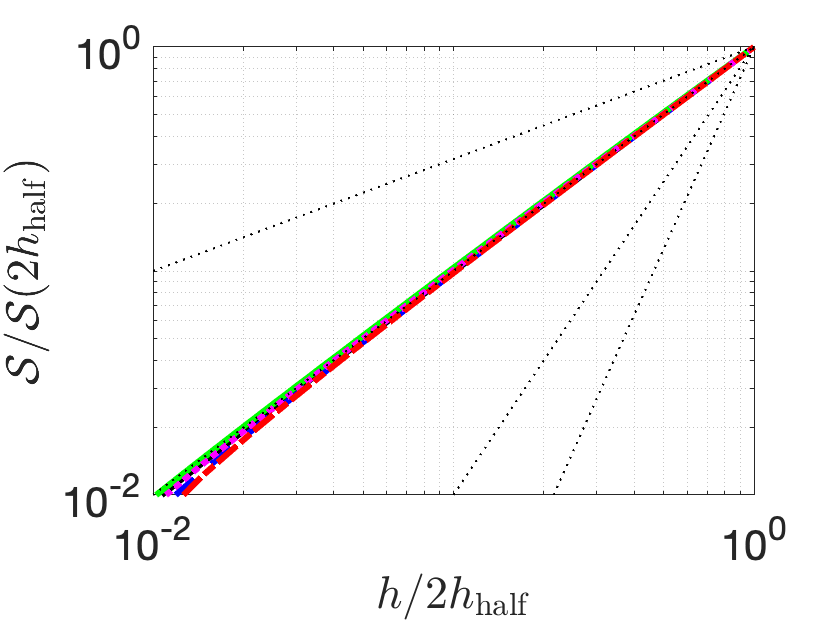}};
\node at (8,0) {\includegraphics[width=0.48\textwidth]{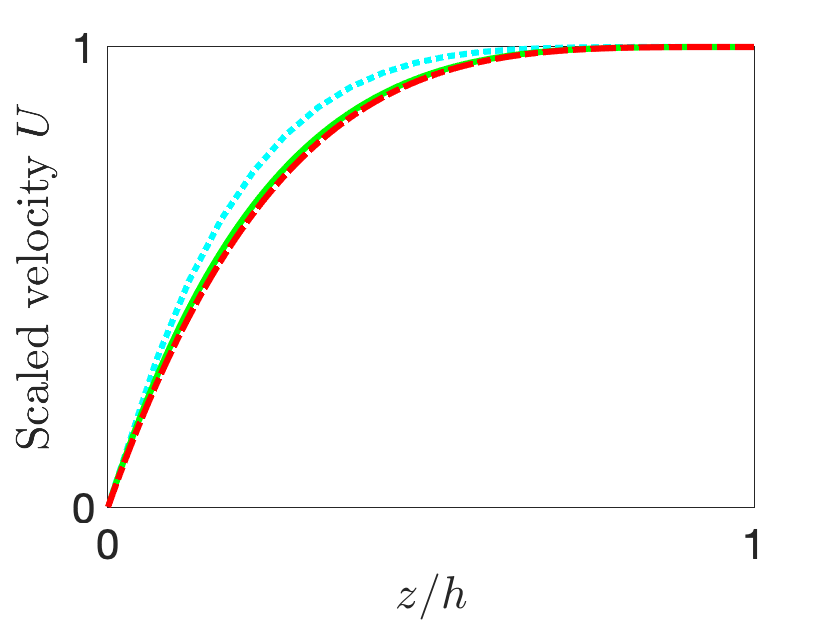}};
\node at (8,-6) {\includegraphics[width=0.48\textwidth]{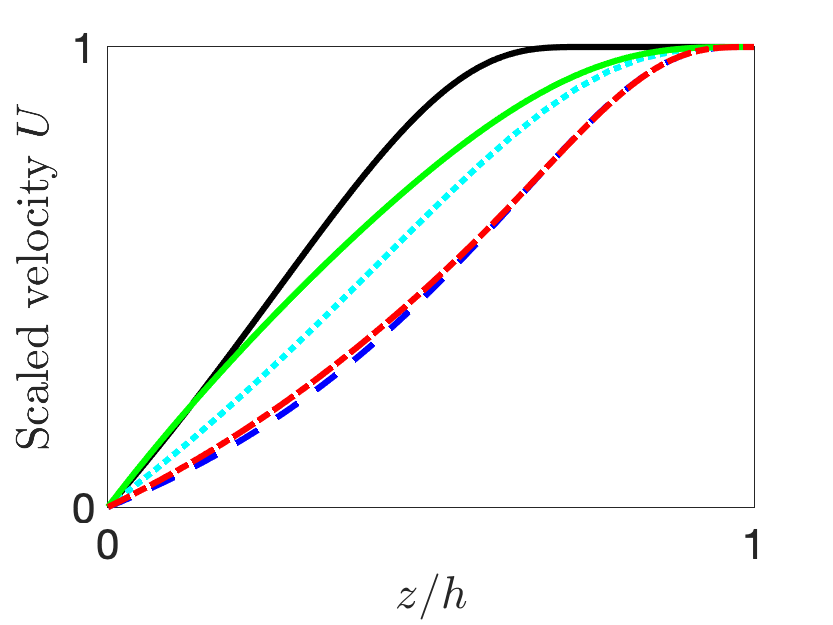}};
\node at (8,-12) {\includegraphics[width=0.48\textwidth]{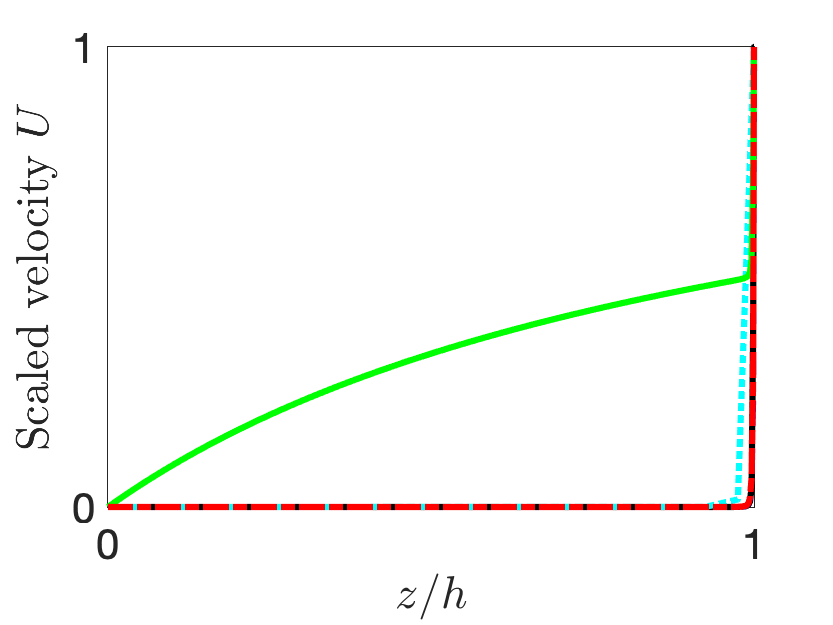}};
\node at (4,-16) {\includegraphics[width=0.9\textwidth]{legend}};
\node at (-4,3) {(a)};
\node at (4,3) {(b)};
\node at (-4,-3) {(c)};
\node at (4,-3) {(d)};
\node at (-4,-9) {(e)};
\node at (4,-9) {(f)};
\node at (9,-1) {\large \bf Bo$\boldsymbol{=10^{-3}}$};
\node at (9,-7) {\large \bf Bo$\boldsymbol{=1}$};
\node at (9,-11) {\large \bf Bo$\boldsymbol{=10^{3}}$};
\node[red] at (0.5,-10) {\large \bf 1/2};
\node[red] at (-0.5,-11.7) {\large \bf 1};
\node[red] at (1.0,-12) {\large \bf 2};
\node[red] at (2.5,-12) {\large \bf 3};
\end{tikzpicture}
\caption{(a,c,e) Variation of the integrated saturation $\mathcal{S}$ \eqref{Seqn} for different values of the Bond number Bo and different types of heterogeneity. Both $\mathcal{S}$ and $h$ are normalised by reference values (at  $h=2h_\mathrm{half}$) for illustration purposes. In each plot we indicate power law values of $1/2$, $1$, $2$ and $3$ with dotted lines for comparison. (b,d,f) Corresponding scaled velocity profiles $U=(u_n(0)-u_n(z))/(u_n(0)-u_n(h))$ where $u_n\propto \Delta \rho g k(z)k_{rn}(s)/\mu_n$. We plot the ensemble average of the velocity, rather than the velocity within each layer, so as not to display oscillatory behaviour between layers.  \label{allfluxes}}
\end{figure}

In this Appendix we present extra plots in figure \ref{allfluxes} for the integrated saturation $\mathcal{S}$ \eqref{Seqn} and the Darcy velocity of the non-wetting phase $u_n\propto \Delta \rho g k(z)k_{rn}(s)/\mu_n$ at different Bond numbers and for different types of heterogeneity. 

The relationship between the integrated saturation $\mathcal{S}$ and the gravity thickness $h$, as shown in figure \ref{allfluxes}a,c,e, is useful for understanding how to invert the solution of the governing equation \eqref{Geqn}. In all cases $\mathcal{S}$ has approximate power law dependence on $h$, where the power is between linear and cubic, as illustrated with dotted lines.

The velocity profiles in figure \ref{allfluxes}b,d,f, are useful for understanding the form of the flux function $\mathcal{F}$ (which is the integrated velocity), as displayed in figure \ref{allfluxes0}a,b,c. We present the scaled velocity $U=(u_n(0)-u_n(z))/(u_n(0)-u_n(h))$ so that $U$ varies between $U=0$ at $z=0$ and $U=1$ at $z=h$. In this way we can see the functional form of $U$ more clearly. For example, if $U$ is constant (as in figure \ref{allfluxes}f), when integrated this will give a linear dependence between the flux $\mathcal{F}$ and the gravity current thickness $h$.

\section{Derivation of the governing equation \eqref{Geqn}}\label{appB}

In this Appendix we provide the details for the derivation of the governing equation for the integrated saturation $\mathcal{S}$, which is given by \eqref{Geqn}. We start by considering the governing equation for the gravity current thickness \eqref{goveheq}. To re-write this in terms of the the integrated saturation  \eqref{Seqn}, we first need to transform the derivatives of $h$ into derivatives of $\mathcal{S}$. As illustrated by figure \ref{allfluxes}a,c,e, the integrated saturation is always a monotone increasing function of the gravity current thickness, so that we can define a unique inverse function
\beq
h=\mathcal{S}^{-1}(\mathcal{S}).\label{hinverse}
\eeq
Then, using the chain rule, the gradient is given by
\beq
\frac{\partial h}{\partial r} = \frac{\partial h}{\partial \mathcal{S}} 
\frac{\partial \mathcal{S}}{\partial r}.
\eeq
We use the inverse derivative identity to calculate derivatives of \eqref{hinverse}, such that
\beq
\frac{\partial h}{\partial \mathcal{S}} =\lb\frac{\partial \mathcal{S}}{\partial h} 
\rb^{-1}.
\eeq
Finally, by defining the flux function $\mathcal{K}$ as \eqref{Feqn1}, we arrive at the governing equation for $\mathcal{S}$, which is
\beq
\frac{\partial \mathcal{S}}{\partial t} =\frac{u_b}{k_0k_{rn0}}\frac{1}{r}\frac{\partial}{\partial r}\lb r \left[\frac{\mathcal{K}}{\partial\mathcal{S}/\partial h}\right] \frac{\partial \mathcal{S}}{\partial r}  \rb.
\eeq

\bibliographystyle{jfm}
\bibliography{bibfile.bib}

\begin{thebibliography}{40}
\expandafter\ifx\csname natexlab\endcsname\relax\def\natexlab#1{#1}\fi
\def\au#1{#1} \def\ed#1{#1} \def\yr#1{#1}\def\at#1{#1}\def\jt#1{\textit{#1}}
  \def\bt#1{#1}\def\bvol#1{\textbf{#1}} \def\vol#1{#1} \def\pg#1{#1}
  \def\publ#1{#1}\def\arxiv#1{#1}\def\org#1{#1}\def\st#1{\textit{#1}}

\bibitem[Bear(2013)]{bear2013dynamics}
{\sc \au{Bear, J}} \yr{2013} {\em Dynamics of fluids in porous media\/}.
  \publ{Courier Corporation}.

\bibitem[Benham {\em et~al.\/}(2021)Benham, Bickle \&
  Neufeld]{benham2020upscaling}
{\sc \au{Benham, GP}, \au{Bickle, MJ} \& \au{Neufeld, JA}} \yr{2021}
  \at{Upscaling multiphase viscous-to-capillary transitions in heterogeneous
  porous media}.  \jt{J. Fluid Mech.}  \bvol{(in press)}.

\bibitem[Berg {\em et~al.\/}(2013)Berg, Oedai \& Ott]{berg2013displacement}
{\sc \au{Berg, S}, \au{Oedai, S} \& \au{Ott, H}} \yr{2013}  \at{Displacement
  and mass transfer between saturated and unsaturated {CO}{$_2$}--brine systems
  in sandstone}.  \jt{Int. J. Greenh. Gas Con.}  \bvol{12},  \pg{478--492}.

\bibitem[Bickle(2009)]{bickle2009geological}
{\sc \au{Bickle, MJ}} \yr{2009}  \at{Geological carbon storage}.  \jt{Nat.
  Geosci.}  \bvol{2}~(12),  \pg{815--818}.

\bibitem[Bickle {\em et~al.\/}(2007)Bickle, Chadwick, Huppert, Hallworth \&
  Lyle]{bickle2007modelling}
{\sc \au{Bickle, M}, \au{Chadwick, A}, \au{Huppert, HE}, \au{Hallworth, M} \&
  \au{Lyle, S}} \yr{2007}  \at{Modelling carbon dioxide accumulation at
  {S}leipner: Implications for underground carbon storage}.  \jt{Earth Planet.
  Sci. Lett.}  \bvol{255}~(1-2),  \pg{164--176}.

\bibitem[Bickle {\em et~al.\/}(2017)Bickle, Kampman, Chapman, Ballentine,
  Dubacq, Galy, Sirikitputtisak, Warr, Wigley \& Zhou]{bickle2017rapid}
{\sc \au{Bickle, M}, \au{Kampman, N}, \au{Chapman, H}, \au{Ballentine, C},
  \au{Dubacq, B}, \au{Galy, A}, \au{Sirikitputtisak, T}, \au{Warr, O},
  \au{Wigley, M} \& \au{Zhou, Z}} \yr{2017}  \at{Rapid reactions between
  {{CO}{$_2$}}, brine and silicate minerals during geological carbon storage:
  Modelling based on a field {{CO}{$_2$}} injection experiment}.  \jt{Chem.
  Geol.}  \bvol{468},  \pg{17--31}.

\bibitem[Boon \& Benson(2021)]{boon2021}
{\sc \au{Boon, M} \& \au{Benson, SM}} \yr{2021}  \at{A physics-based model to
  predict the impact of horizontal lamination on {CO}{$_2$} plume migration}.
  \jt{Sub Judice} .

\bibitem[Braun {\em et~al.\/}(2005)Braun, Helmig \& Manthey]{braun2005macro}
{\sc \au{Braun, C}, \au{Helmig, R} \& \au{Manthey, S}} \yr{2005}
  \at{Macro-scale effective constitutive relationships for two-phase flow
  processes in heterogeneous porous media with emphasis on the relative
  permeability--saturation relationship}.  \jt{J. Contam. Hydrol.}
  \bvol{76}~(1-2),  \pg{47--85}.

\bibitem[Brooks \& Corey(1964)]{brooks1964hydrau}
{\sc \au{Brooks, R} \& \au{Corey, T}} \yr{1964}  \at{Hydraulic properties of
  porous media}.  \jt{Hydrology Papers, Colorado State University}  \bvol{24},
  \pg{37}.

\bibitem[Bui {\em et~al.\/}(2018)Bui, Adjiman, Bardow, Anthony, Boston, Brown,
  Fennell, Fuss, Galindo \& Hackett]{bui2018carbon}
{\sc \au{Bui, M}, \au{Adjiman, CS}, \au{Bardow, A}, \au{Anthony, EJ},
  \au{Boston, A}, \au{Brown, S}, \au{Fennell, PS}, \au{Fuss, S}, \au{Galindo,
  A} \& \au{Hackett, LA}} \yr{2018}  \at{Carbon capture and storage ({CCS}):
  the way forward}.  \jt{Energy Environ. Sci.}  \bvol{11}~(5),
  \pg{1062--1176}.

\bibitem[Cavanagh \& Haszeldine(2014)]{cavanagh2014sleipner}
{\sc \au{Cavanagh, AJ} \& \au{Haszeldine, RS}} \yr{2014}  \at{The sleipner
  storage site: Capillary flow modeling of a layered {CO}{$_2$} plume requires
  fractured shale barriers within the utsira formation}.  \jt{Int. J. Greenh.
  Gas Con.}  \bvol{21},  \pg{101--112}.

\bibitem[Chadwick {\em et~al.\/}(2009)Chadwick, Noy \&
  Holloway]{chadwick2009flow}
{\sc \au{Chadwick, RA}, \au{Noy, DJ} \& \au{Holloway, S}} \yr{2009}  \at{Flow
  processes and pressure evolution in aquifers during the injection of
  supercritical {CO}{$_2$} as a greenhouse gas mitigation measure}.  \jt{Pet.
  Geosci.}  \bvol{15}~(1),  \pg{59--73}.

\bibitem[Chierici(1984)]{chierici1984novel}
{\sc \au{Chierici, GL}} \yr{1984}  \at{Novel relations for drainage and
  imbibition relative permeabilities}.  \jt{Soc. Petrol. Eng. J.}
  \bvol{24}~(03),  \pg{275--276}.

\bibitem[Cloud(1941)]{cloud1941effects}
{\sc \au{Cloud, WF}} \yr{1941}  \at{Effects of sand grain size distribution
  upon porosity and permeability}.  \jt{Oil Weekly}  \bvol{103}~(8),  \pg{26}.

\bibitem[Corey(1954)]{corey1954interrelation}
{\sc \au{Corey, AT}} \yr{1954}  \at{The interrelation between gas and oil
  relative permeabilities}.  \jt{Prod. Monthly}  \bvol{19}~(1),  \pg{38--41}.

\bibitem[Cowton {\em et~al.\/}(2016)Cowton, Neufeld, White, Bickle, White \&
  Chadwick]{cowton2016inverse}
{\sc \au{Cowton, LR}, \au{Neufeld, JA}, \au{White, NJ}, \au{Bickle, MJ},
  \au{White, JC} \& \au{Chadwick, RA}} \yr{2016}  \at{An inverse method for
  estimating thickness and volume with time of a thin {CO}{$_2$}-filled layer
  at the {S}leipner field, {N}orth {S}ea}.  \jt{J. Geophys. Res. Solid Earth}
  \bvol{121}~(7),  \pg{5068--5085}.

\bibitem[Cowton {\em et~al.\/}(2018)Cowton, Neufeld, White, Bickle, Williams,
  White \& Chadwick]{cowton2018benchmarking}
{\sc \au{Cowton, LR}, \au{Neufeld, JA}, \au{White, NJ}, \au{Bickle, MJ},
  \au{Williams, GA}, \au{White, JC} \& \au{Chadwick, RA}} \yr{2018}
  \at{Benchmarking of vertically-integrated {CO}{$_2$} flow simulations at the
  {S}leipner field, {N}orth {S}ea}.  \jt{Earth Planet. Sci. Lett.}  \bvol{491},
   \pg{121--133}.

\bibitem[Golding {\em et~al.\/}(2013)Golding, Huppert \&
  Neufeld]{golding2013effects}
{\sc \au{Golding, MJ}, \au{Huppert, HE} \& \au{Neufeld, JA}} \yr{2013}  \at{The
  effects of capillary forces on the axisymmetric propagation of two-phase,
  constant-flux gravity currents in porous media}.  \jt{Phys. Fluids}
  \bvol{25}~(3),  \pg{036602}.

\bibitem[Golding {\em et~al.\/}(2017)Golding, Huppert \&
  Neufeld]{golding2017two}
{\sc \au{Golding, MJ}, \au{Huppert, HE} \& \au{Neufeld, JA}} \yr{2017}
  \at{Two-phase gravity currents resulting from the release of a fixed volume
  of fluid in a porous medium}.  \jt{J. Fluid Mech.}  \bvol{832},
  \pg{550--577}.

\bibitem[Golding {\em et~al.\/}(2011)Golding, Neufeld, Hesse \&
  Huppert]{golding2011two}
{\sc \au{Golding, MJ}, \au{Neufeld, JA}, \au{Hesse, MA} \& \au{Huppert, HE}}
  \yr{2011}  \at{Two-phase gravity currents in porous media}.  \jt{J. Fluid
  Mech.}  \bvol{678},  \pg{248--270}.

\bibitem[Harper {\em et~al.\/}(2021)Harper, Liu, Tavener \&
  Wildey]{harper2021coupling}
{\sc \au{Harper, G}, \au{Liu, J}, \au{Tavener, S} \& \au{Wildey, T}} \yr{2021}
  \at{Coupling {A}rbogast--{C}orrea and {B}ernardi--{R}augel elements to
  resolve coupled {S}tokes--{D}arcy flow problems}.  \jt{Comput. Methods in
  Appl. Mech. Eng.}  \bvol{373},  \pg{113469}.

\bibitem[Hinton \& Woods(2018)]{hinton2018buoyancy}
{\sc \au{Hinton, EM} \& \au{Woods, AW}} \yr{2018}  \at{Buoyancy-driven flow in
  a confined aquifer with a vertical gradient of permeability}.  \jt{J. Fluid
  Mech.}  \bvol{848},  \pg{411--429}.

\bibitem[Hinton \& Woods(2020{\natexlab{{\em a\/}}})]{hinton2020shear2}
{\sc \au{Hinton, EM} \& \au{Woods, AW}} \yr{2020{\natexlab{{\em a\/}}}}
  \at{Shear dispersion in a porous medium. {P}art 1. an intrusion with a steady
  shape}.  \jt{J. Fluid Mech.}  \bvol{899}.

\bibitem[Hinton \& Woods(2020{\natexlab{{\em b\/}}})]{hinton2020shear1}
{\sc \au{Hinton, EM} \& \au{Woods, AW}} \yr{2020{\natexlab{{\em b\/}}}}
  \at{Shear dispersion in a porous medium. {P}art 2. an intrusion with a
  growing shape}.  \jt{J. Fluid Mech.}  \bvol{899}.

\bibitem[Huppert(1982)]{huppert1982propagation}
{\sc \au{Huppert, HE}} \yr{1982}  \at{The propagation of two-dimensional and
  axisymmetric viscous gravity currents over a rigid horizontal surface}.
  \jt{J. Fluid Mech.}  \bvol{121},  \pg{43--58}.

\bibitem[Huppert \& Woods(1995)]{huppert1995gravity}
{\sc \au{Huppert, HE} \& \au{Woods, AW}} \yr{1995}  \at{Gravity-driven flows in
  porous layers}.  \jt{J. Fluid Mech.}  \bvol{292},  \pg{55--69}.

\bibitem[Jackson {\em et~al.\/}(2018)Jackson, Agada, Reynolds \&
  Krevor]{jackson2018characterizing}
{\sc \au{Jackson, SJ}, \au{Agada, S}, \au{Reynolds, CA} \& \au{Krevor, S}}
  \yr{2018}  \at{Characterizing drainage multiphase flow in heterogeneous
  sandstones}.  \jt{Water Resour. Res.}  \bvol{54}~(4),  \pg{3139--3161}.

\bibitem[Jackson \& Krevor(2020)]{jackson2020small}
{\sc \au{Jackson, SJ} \& \au{Krevor, S}} \yr{2020}  \at{Small-scale capillary
  heterogeneity linked to rapid plume migration during {CO}{$_2$} storage}.
  \jt{Geophys. Res. Lett.}  \pg{p. e2020GL088616}.

\bibitem[Krevor {\em et~al.\/}(2015)Krevor, Blunt, Benson, Pentland, Reynolds,
  Al-Menhali \& Niu]{krevor2015capillary}
{\sc \au{Krevor, S}, \au{Blunt, MJ}, \au{Benson, SM}, \au{Pentland, CH},
  \au{Reynolds, C}, \au{Al-Menhali, A} \& \au{Niu, B}} \yr{2015}  \at{Capillary
  trapping for geologic carbon dioxide storage--from pore scale physics to
  field scale implications}.  \jt{Int. J. Greenh. Gas Con.}  \bvol{40},
  \pg{221--237}.

\bibitem[Leverett(1941)]{leverett1941capillary}
{\sc \au{Leverett, MC}} \yr{1941}  \at{Capillary behavior in porous solids}.
  \jt{Trans. AIME}  \bvol{142}~(01),  \pg{152--169}.

\bibitem[Li \& Benson(2015)]{li2015influence}
{\sc \au{Li, B} \& \au{Benson, SM}} \yr{2015}  \at{Influence of small-scale
  heterogeneity on upward {CO}{$_2$} plume migration in storage aquifers}.
  \jt{Adv. Water Resour.}  \bvol{83},  \pg{389--404}.

\bibitem[Liu {\em et~al.\/}(2016)Liu, Sadre-Marandi \& Wang]{liu2016darcylite}
{\sc \au{Liu, J}, \au{Sadre-Marandi, F} \& \au{Wang, Z}} \yr{2016}
  \at{Darcy{L}ite: A {M}atlab toolbox for {D}arcy flow computation}.
  \jt{Procedia Comput. Sci.}  \bvol{80},  \pg{1301--1312}.

\bibitem[Lyle {\em et~al.\/}(2005)Lyle, Huppert, Hallworth, Bickle \&
  Chadwick]{lyle2005axisymmetric}
{\sc \au{Lyle, S}, \au{Huppert, HE}, \au{Hallworth, M}, \au{Bickle, M} \&
  \au{Chadwick, A}} \yr{2005}  \at{Axisymmetric gravity currents in a porous
  medium}.  \jt{J. Fluid Mech.}  \bvol{543},  \pg{293--302}.

\bibitem[Nelson(1994)]{nelson1994permeability}
{\sc \au{Nelson, PH}} \yr{1994}  \at{Permeability-porosity relationships in
  sedimentary rocks}.  \jt{The Log Analyst}  \bvol{35}~(03).

\bibitem[Nordbotten \& Celia(2011)]{nordbotten2011geological}
{\sc \au{Nordbotten, JM} \& \au{Celia, MA}} \yr{2011} {\em Geological storage
  of {CO}{$_2$}: modeling approaches for large-scale simulation\/}.  \publ{John
  Wiley \& Sons}.

\bibitem[Pegler {\em et~al.\/}(2014)Pegler, Huppert \&
  Neufeld]{pegler2014fluid}
{\sc \au{Pegler, SS}, \au{Huppert, HE} \& \au{Neufeld, JA}} \yr{2014}
  \at{Fluid injection into a confined porous layer}.  \jt{J. Fluid Mech.}
  \bvol{745},  \pg{592--620}.

\bibitem[Rabinovich {\em et~al.\/}(2016)Rabinovich, Li \&
  Durlofsky]{rabinovich2016analytical}
{\sc \au{Rabinovich, A}, \au{Li, B} \& \au{Durlofsky, LJ}} \yr{2016}
  \at{Analytical approximations for effective relative permeability in the
  capillary limit}.  \jt{Water Resour. Res.}  \bvol{52}~(10),  \pg{7645--7667}.

\bibitem[Trevisan {\em et~al.\/}(2017)Trevisan, Krishnamurthy \&
  Meckel]{trevisan2017impact}
{\sc \au{Trevisan, L}, \au{Krishnamurthy, PG} \& \au{Meckel, Tip~A}} \yr{2017}
  \at{Impact of 3{D} capillary heterogeneity and bedform architecture at the
  sub-meter scale on {CO}{$_2$} saturation for buoyant flow in clastic
  aquifers}.  \jt{Int. J. Greenh. Gas Con.}  \bvol{56},  \pg{237--249}.

\bibitem[Williams \& Chadwick(2017)]{williams2017improved}
{\sc \au{Williams, GA} \& \au{Chadwick, RA}} \yr{2017}  \at{An improved
  history-match for layer spreading within the {S}leipner plume including
  thermal propagation effects}.  \jt{Energy Procedia}  \bvol{114},
  \pg{2856--2870}.

\bibitem[Woods(2015)]{woods2015flow}
{\sc \au{Woods, AW}} \yr{2015} {\em Flow in porous rocks\/}.  \publ{Cambridge
  University Press}.

\end{thebibliography}

\end{document}